\documentclass[11pt]{article}
\usepackage{jcappub,amsmath,amssymb,hyperref,framed}
\usepackage{mathtools,mathrsfs}
\usepackage{bm}
\usepackage[table]{xcolor}
\usepackage{amsmath}
\usepackage{amssymb}
\usepackage{bbold}
\usepackage{pgfplots}
\usepackage[utf8]{inputenc}
\usepackage{comment}
\usepackage{graphics}
\usepackage{todonotes}
\usepackage{tikz}
\usetikzlibrary{arrows,positioning,decorations.markings,decorations.pathmorphing,calc}
\definecolor{hyperref}{RGB}{026,028,087}
\DeclarePairedDelimiter{\abs}{\lvert}{\rvert}

\def\gsim{ \lower .75ex \hbox{$\sim$} \llap{\raise .27ex \hbox{$>$}} }
\def\lsim{ \lower .75ex \hbox{$\sim$} \llap{\raise .27ex \hbox{$<$}} }
\def\be{\begin{equation}}
\def\ee{\end{equation}}
\def\bea{\begin{eqnarray}}
\def\eea{\end{eqnarray}}

\newcommand{\ba}{\begin{array}}
\newcommand{\ea}{\end{array}}

\newcommand{\commentout}[1]{}

\newcommand{\bs}{\begin{split}}

\def\ba{\begin{eqnarray}}
\def\ea{\end{eqnarray}}
\def\nn{\nonumber}

\def\({\left(}
\def\){\right)}

\definecolor{jn}{RGB}{10, 10, 200} 
\definecolor{js}{RGB}{204, 0, 0} 
\definecolor{pgf}{RGB}{10, 150, 10} 
\newcommand*{\mathcolor}{}
\def\mathcolor#1#{\mathcoloraux{#1}}
\newcommand*{\mathcoloraux}[3]{%
\protect\leavevmode
\begingroup
\color#1{#2}#3%
\endgroup
}
\newlength{\stheight}
\newcommand\textst[1][fu-grey]{
\ifmmode\setlength{\stheight}{+1.0ex}
\else\setlength{\stheight}{+0.5ex}
\fi
\bgroup\markoverwith{\textcolor{#1}{\rule[\the\stheight]{2pt}{1.0pt}}}\ULon
} 
\newcommand{\textins}[2][fu-grey]{
\ifmmode\mathcolor{#1}{#2}
\else\textcolor{#1}{#2}\@\,
\fi
}
\graphicspath{{./}}
\allowdisplaybreaks

\tikzstyle{vecArrow} = [thick, decoration={markings,mark=at position
1 with {\arrow[semithick]{open triangle 60}}},
double distance=1.4pt, shorten >= 5.5pt,
preaction = {decorate},
postaction = {draw,line width=1.4pt, white,shorten >= 4.5pt}]
\begin{document}

\title{A general theory of linear cosmological perturbations: stability conditions, the quasistatic limit and dynamics.}

\author[a]{Macarena Lagos,}
\author[b]{Emilio Bellini,}
\author[c]{Johannes Noller,}
\author[b]{Pedro G. Ferreira,}
\author[b]{Tessa Baker}
\affiliation[a]{Kavli Institute for Cosmological Physics, The University of Chicago, Chicago, IL 60637, USA}
\affiliation[b]{Astrophysics, University of Oxford, DWB, Keble road, Oxford OX1 3RH, UK}
\affiliation[c]{Institute for Theoretical Studies, ETH Zurich, Clausiusstrasse 47, 8092 Zurich, Switzerland}

\emailAdd{mlagos@kicp.uchicago.edu}
\emailAdd{emilio.bellini@physics.ox.ac.uk}
\emailAdd{johannes.noller@eth-its.ethz.ch}
\emailAdd{p.ferreira1@physics.ox.ac.uk}
\emailAdd{tessa.baker@physics.ox.ac.uk}

\abstract{
We analyse cosmological perturbations around a homogeneous and isotropic background for scalar-tensor, vector-tensor and bimetric theories of gravity. Building on previous results, we propose a unified view of the effective parameters of all these theories. Based on this structure, we explore the viable space of parameters for each family of models by imposing the absence of ghosts and gradient instabilities. We then focus on the quasistatic regime and confirm that all these theories can be approximated by the phenomenological two-parameter model described by an effective Newton's constant and the gravitational slip. Within the quasistatic regime we pinpoint signatures which can distinguish between the broad classes of models (scalar-tensor, vector-tensor or bimetric). Finally, we present the equations of motion for our unified approach in such a way that they can be implemented in Einstein-Boltzmann solvers.}


\maketitle

\setcounter{tocdepth}{2}

\section{Introduction}

There is now a complete and thorough understanding of linear cosmological perturbation theory in general relativity (GR). The tremendous progress of cosmology of the past twenty-five years has been primarily due to our ability to calculate the phenomenology of the large scale structure of the universe using linear theory. The most recent spectacular success of the Planck satellite in measuring anisotropies in the cosmic microwave background (CMB) and the resulting constraints on cosmological parameters \cite{Ade:2015xua} are possible due to fast and accurate Einstein-Boltzmann solvers that incorporate a bewildering number of different physical effects \cite{Lewis:1999bs,Blas:2011rf}.

Over the past few years, there has been tremendous progress in constructing a more general linear cosmological perturbation theory that can go beyond GR\footnote{To be specific, we mean beyond the theory that is described in terms of one metric, $g_{\alpha\beta}$, minimally coupled to other matter fields, and whose dynamics is exclusively given by the Einstein-Hilbert action} \cite{Clifton:2011jh}. Originally, this was done by assuming ad-hoc modifications to the equations of motion (the linearized Einstein field equations), introducing a general time varying Newton's constant and gravitational slip \cite{Bertschinger:2008zb,Caldwell:2007cw}. The approach is remarkably effective and widely used to approximate various theories as well as in cosmological parameter constraints. There are currently two Einstein-Boltzmann solvers, MGCAMB \cite{Zhao:2008bn,Hojjati:2011ix} and ISiTGR \cite{Dossett:2011tn} which are widely used in both theoretical and phenomenological work. 

A more systematic approach, known as Parametrized Post-Friedmann formalism (PPF), using the field equations was proposed in \cite{Baker:2011jy,Baker:2012zs} where it was shown that it was possible to encompass a very broad class of theories in a completely consistent way in one set of modified, linearized field equations. Crucial to this approach were the gauge transformation properties of any extra gravitational degrees of freedom. The resulting set of equations has a large number of free parameters which can be mapped onto a very large subset of the theories of gravity that have been proposed. Nevertheless, and although that number can be reduced by judicious use of the Bianchi identities, the parametrization is still sufficiently large to be impractical for a systematic study of the phenomenology of gravitational theories.

Building on the success of the effective field theory of inflation \cite{Cheung:2007st,Weinberg:2008hq}, an alternative action-based approach, known as Effective Field Theory (EFT) of dark energy, has been proposed that focuses solely on scalar-tensor theories \cite{Gubitosi:2012hu,Creminelli:2008wc}. Constructing a general quadratic action for the metric in the unitary gauge (i.e.~where fluctuations in the scalar field are set to zero), it is possible to recover the quadratic action for the most general scalar-tensor theory \cite{Gleyzes:2013ooa,Gleyzes:2014rba}. This approach has been successfully implemented in Einstein-Boltzmann solvers (EFTCAMB \cite{EFTCAMB1} and {\tt hi\_class} \cite{Zumalacarregui:2016pph}) and been used to obtain preliminary, weak, cosmological constraints from a selection of cosmological data sets. Furthermore, it has been used to understand the nature of scalar-tensor theories and how they can be extended beyond what was originally thought \cite{Gleyzes:2014dya}.

A more general approach to cosmological perturbation theory was proposed in \cite{Lagos:2016wyv,Lagos:2016gep}\footnote{A similar method was proposed in \cite{Battye:2012eu}}. There, the idea is to consider the most general, quadratic action for the gravitational degrees of freedom which satisfies a particular set of symmetry principles. By looking at the Noether symmetries and constraints of this action it is possible to identify the subset of functions which characterize scalar-tensor, vector-tensor and tensor-tensor (or bimetric) theories. Similar to the EFT approach, it is constructed at the level of the action but like the PPF approach it is completely general and systematic. 

While great strides have been taken at the level of linear cosmological perturbation theory, there has also been great progress in understanding a variety of theories of gravity at a more fundamental, non-linear level. For scalar-tensor theories, we highlight the rediscovery of \cite{Horndeski:1974wa,Deffayet:2009wt} and its extensions \cite{Gleyzes:2014dya,Zumalacarregui:2013pma,Langlois:2017mxy}. For vector-tensor theories, generalizations of the Einstein-Aether theory have been proposed \cite{Zlosnik:2006zu,Zuntz:2010jp} as well as the systematic construction of a more general vector-tensor theory 
\cite{Tasinato:2014eka,Heisenberg:2014rta,Allys:2015sht,Jimenez:2016isa}, 
mirroring what has been done in the case of scalar-tensor theories. And, of course, the resurrection of massive gravity has led to a plethora of new and interesting results in theories involving more than one metric \cite{deRham:2010kj,Hassan:2011zd}. There has also been interest in other theories that explore more exotic extensions to the standard picture: non-local theories \cite{Deser:2007jk,Maggiore:2016gpx} and theories with higher derivatives \cite{Sotiriou:2008rp} are all being considered as possible extensions to GR.

The literature on cosmological linear perturbations of theories of gravity is rich yet heterogeneous. Each class of theories is approached in a distinct way and it is difficult to relate their phenomenology to each other in a systematic way. If we are to explore the realm of gravitational theories in a coherent fashion, it makes sense to try and understand their structure in a unified way. In particular, at the linear level, we expect each subclass of theories to be characterized by a finite (and hopefully) small subset of functions of time. A notable example is scalar-tensor theories where it has been shown \cite{Bellini:2014fua} that the dynamics of linear perturbations is completely governed by a small set of functions of time $\alpha_X(t)$ (where each subscript $X$ can be tied to a particular type of term in the quadratic action). Much like in the Paramaterized-Post-Newtonian (PPN) approach \cite{1993tegp.book.....W} where each of the parameters can be (roughly) associated to a specific physical effect, one hopes that something similar can be said of the parameters $\alpha_X$.

In this paper, we will take a unified view of the parametrised linear cosmological perturbations described in \cite{Lagos:2016wyv,Lagos:2016gep} for scalar-tensor (ST), vector-tensor (VT) and tensor-tensor (TT) theories of gravity, and we will cover the four cases summarised in Table \ref{SummaryTable}. We separate the family of VT theories into two distinct subclasses: general vector-tensor (with a general vector field) and Einstein-Aether (with a unitary vector field) theories. In this table we show the number of free parameters that describe each one of the models, in addition to one example of non-linear theory encompassed by the corresponding parametrisation. These four models are constructed in such a way that they all lead to second-order derivative equations of motion and are linearly diffeomorphism invariant. In addition, they all propagate at most one scalar gravitational degree of freedom (DoF), which comes from the scalar field in ST, the helicity-0 mode of vector field in VT, or the helicity-0 mode of a tensor in TT. 

\begin{table}[h!]
	\centering
	\begin{tabular}{| l || c || c || c |}
		\hline
		Model & Fields & Parameters & Example \\ \hline \hline
		Scalar-Tensor & $g_{\mu\nu}$, $\chi$ & 5 & Horndeski \cite{Horndeski:1974wa}\\ \hline
		General Vector-Tensor & $g_{\mu\nu}$, $A^{\mu}$ & 8 & Generalised Proca \cite{Heisenberg:2014rta}\\
		Einstein-Aether & $g_{\mu\nu}$, $A^{\mu}$ & 4 & Einstein-Aether \cite{Zlosnik:2006zu} \\ \hline
	  Tensor-Tensor & $g_{\mu\nu}$, $f_{\mu\nu}$ & 5 & Massive bigravity \cite{Hassan:2011zd} \\ \hline
	\end{tabular}
	\caption{\label{SummaryTable} Parametrised ST, general VT, Einstein-Aether, and TT quadratic theories of gravity. We show the field content of each model in the second column, while the third column shows the number of free functions of time parametrising the cosmological evolution of the background and scalar linear perturbations. The fourth column shows examples of non-linear theories that are encompassed by the corresponding parametrisations.}
\end{table}

This paper will be structured as follows. In Section \ref{Sec:Models} we will standardise the parametrisation scheme used for ST, VT and TT theories, and propose a set of functions that will be considered as a basis to classify linear cosmological perturbations in these families of gravity theories. We propose a basis such that the four models are parametrised by a mass scale $M(t)$ in addition to a finite number of dimensionless parameters $\alpha_X(t)$, and GR is recovered when $M$ is the Planck mass and all $\alpha_X$ vanish. In this section we also give the expressions for ($M$, $\alpha_X$) in the case of the examples of non-linear theories of Table \ref{SummaryTable}, thus showing explicitly how these theories are encompassed by the general parametrised models. Using our standardised parametrisation, in Section \ref{Sec:constraints} we proceed to explore theoretical constraints on the parameter space in order avoid ghost and gradient instabilities. In Section \ref{Sec:QSA} we analyse the quasistatic regime and discuss possible observational signatures from different regions of the theory space. In Section \ref{Sec:Phenomenology} we work towards a unified phenomenology for the four classes of theories. In particular, we present the field equations for cosmological perturbations in a unified form such that all of them can be implemented in Einstein-Boltzmann solvers. Finally, in Section \ref{Sec:discussion} we summarise and discuss the findings of this paper.


\section{Parametrised Actions}\label{Sec:Models}

In this section we will explore a broad range of gravitational theories, and generate a standardised description of linear scalar cosmological perturbations. We will look at scalar-tensor, vector-tensor and tensor-tensor gravity theories, which have all been studied in great detail over the past two decades. We will start by rewriting the parametrised quadratic actions derived in \cite{Lagos:2016wyv,Lagos:2016gep} for the four families of models in Table \ref{SummaryTable}. While in previous works certain sets of parameters were used to describe the cosmological background and linear perturbation evolution, here we will rewrite these actions in such a way that all models are parametrised in the same way: with one free function that determines the evolution of the background, the scale factor $a$, as well as a finite set of free functions --one mass scale $M(t)$ and a set of dimensionless functions $\alpha_X(t)$-- that determine the evolution of linear scalar cosmological perturbations. We name and define this set of $\alpha_X$ parameters according to their physical interpretation and how they affect the relevant DoFs present in each model. 

We will then consider some well-known non-linear theories that are encompassed by the four parametrised actions, and show the relation between the free functions of these non-linear theories and the set of parameters ($M$, $\alpha_X$), providing then a dictionary that can be straightforwardly used to translate future constraints on these free parameters into constraints on specific non-linear theories. In most cases, we will find that the four parametrised linear actions are more general than the examples of non-linear theories considered here, which raises the question of whether such non-linear models can be extended.

Finally, once we have standardised the way all the actions are parametrised, we will also address the possibility of constructing a unified local action that can encompass scalar-tensor, vector-tensor and bimetric theories at the same time. We will discuss the main obstacles that prevent us from doing so. 

Next, let us establish some terminology and notation. We consider modified gravity models that are all composed by one metric $g_{\mu\nu}$ and one additional field (scalar, vector or tensor). We parametrise our ignorance in the gravitational interactions, while always assuming that additional matter interactions are known. Specifically, we assume that $g_{\mu\nu}$ is the physical metric describing the spacetime, and couples minimally to any matter components. We will be using a scalar field $\hat\varphi$ to represent external matter components, with an action given by:
\begin{equation}\label{FullMatterAction}
S_m=-\int d^4x \sqrt{-g}\left[ \frac{1}{2}\partial_\mu \hat\varphi \partial^\mu \hat\varphi+V(\hat\varphi) \right],
\end{equation}
where $g$ is the determinant of $g_{\mu\nu}$, and $V(\varphi)$ is an arbitrary potential. 

We follow the standard scalar-vector-tensor decomposition of cosmological perturbations \cite{Kodama:1985bj} and focus on scalar perturbations only. We will express the linearly perturbed line element of the metric $g_{\mu\nu}$ in the following way:
\begin{equation}\label{Def4Pert}
ds^2_g=-\left(1+2\Phi\right)dt^2 +2 \partial_iB dt dx^i + a^2\left[\left(1-2\Psi\right)\delta_{ij}+2\partial_i\partial_jE\right]dx^idx^j,
\end{equation}
where $a$ is the scale factor and depends only on the time $t$, while there are four scalar perturbations $\Phi$, $\Psi$, $B$ and $E$, which depend on time and space in general. 
Similarly, the linearly perturbed matter scalar field will be written as $\hat\varphi=\varphi(t)+\delta \varphi(t, \vec{x})$. In addition, depending on the theories to be considered, there will be an extra scalar, vector or tensor field, whose linearly perturbed form will be given in the corresponding subsections. 

At the background level, the matter equation of motion will be fixed and given by:
\begin{eqnarray}\label{BackMatterEqn}
{\ddot \varphi}+3H{\dot \varphi}+V_{,\varphi}=0,
\end{eqnarray}
where $V_{,\varphi}$ is the derivative of the potential with respect to the field $\varphi$ evaluated at the background.
While this background equation will be assumed throughout this paper, the scale factor $a$ will be kept free in all parametrised models. Nevertheless, it is useful to note that, in the case of GR, the background metric equations reduce to:
\begin{eqnarray}
H^2&=&\frac{1}{3M_P^2}\left(\frac{1}{2}{\dot \varphi}^2+V(\varphi)\right), \\
{\dot H}&=&-\frac{1}{2M_P^2}{\dot \varphi}^2,
\end{eqnarray}
where $M_P$ is the Planck mass. It will sometimes be convenient to express the scalar field as a fluid, with energy density $\rho={\dot \varphi}^2/2+V(\varphi)$ and pressure $P={\dot \varphi}^2/2-V(\varphi)$. 
 
At the level of linear perturbations, in all models, we will be able to write the total parametrised quadratic actions for scalar perturbations as:
\begin{equation}
S^{(2)}=S_G^{(2)}+S_{m,\varphi}^{(2)}\, ,
\end{equation} 
where $S_G^{(2)}$ describes the gravitational action for the corresponding family of gravity theories, and $S_{m,\varphi}^{(2)}$ is the quadratic expansion of the matter action (\ref{FullMatterAction}) that involves terms with $\delta\varphi$ only (any quadratic term of the metric perturbations from $S_m$ will be included in $S^{(2)}_G$). Explicitly,
\begin{align}\label{Matter lagrangian}
S_{m,\varphi}^{(2)}=&\int dtd^3k\; a^3\left[ \frac{1}{2}\dot{\delta\varphi}^{2}-\frac{1}{2}\left(\frac{k^{2}}{a^{2}}+V_{,\varphi\varphi}\right)\delta\varphi^{2}-\Phi\left(\dot{\varphi}\dot{\delta\varphi}+\delta\varphi V_{,\varphi}\right)\right. \nonumber\\
 &\left. +\dot{\varphi}\left(3\dot{\Psi}+k^{2}\dot{E}-\frac{k^{2}}{a^{2}}B\right)\delta\varphi\right] ,
\end{align}
where $V_{,\varphi\varphi}$ represents the second derivative of the potential with respect to the scalar field evaluated at the background. For simplicity, we have expressed the matter Lagrangian in Fourier space and used the same notation to describe the field in phase and Fourier space. In what follows, we will present all the parametrised actions in Fourier space, and assume an implicit dependence on the wavenumber $k$ and time $t$ for all perturbation fields. All of these actions are constructed in such a way that they are linearly diffeomorphism invariant and lead to second-order derivative equations of motion.


\subsection{Scalar-Tensor Theories}
Using the methodology presented in \cite{Lagos:2016wyv} one can construct the most general linearly diffeomorphism-invariant scalar-tensor quadratic action around a spatially flat homogeneous and isotropic background, leading to second-order derivative equations of motion. This action is given by:
\begin{align}\label{ScalarActionFinal}
 S^{(2)} &= S_{m,\varphi}^{(2)} + \int d^3kdt\; a^3M_S^2\mathcal{L}_{ST},
\end{align}
where $\mathcal{L}_{ST}$ is the parametrised gravitational Lagrangian for scalar-tensor theories, given by:
\begin{align}
\mathcal{L}_{ST}= & -\left[3\dot{\Psi}+2\tilde{B}\right]\dot{\Psi}-\left(2-\hat{\alpha}_{\textrm{B}}\right)\left[3\dot{\Psi}+\tilde{B}\right]H\Phi-2\frac{k^{2}}{a^{2}}\Phi\Psi+\left(1+\hat{\alpha}_{\textrm{T}}\right)\frac{k^{2}}{a^{2}}\Psi^{2}\nonumber \\
 & -\frac{1}{2}\left[\left(6-6\hat{\alpha}_{\textrm{B}}-\hat{\alpha}_{\textrm{K}}\right)H^{2}-\frac{\dot{\varphi}^{2}}{M_{S}^{2}}\right]\Phi^{2}+\frac{1}{2}\hat{\alpha}_{\textrm{K}}H^{2}\dot{v}_{X}^{2}+\hat{\alpha}_{\textrm{B}}\left(3\dot{\Psi}+\tilde{B}\right)H\dot{v}_{X}\nonumber \\
 & +\left(3\hat{\alpha}_{\textrm{B}}+\hat{\alpha}_{\textrm{K}}\right)H^{2}\Phi\dot{v}_{X}+3\left(2\dot{H}+\frac{\dot{\varphi}^{2}}{M_{S}^{2}}\right)\dot{\Psi}v_{X}+\left(2\dot{H}+\frac{\dot{\varphi}^{2}}{M_{S}^{2}}\right)\tilde{B}v_{X}\nonumber \\
 & +2\left(\hat{\alpha}_{\textrm{M}}-\hat{\alpha}_{\textrm{T}}\right)H\frac{k^{2}}{a^{2}}\Psi v_{X}+\left[3\left(2-\hat{\alpha}_{\textrm{B}}\right)\dot{H}+3\frac{\dot{\varphi}^{2}}{M_{S}^{2}}+\hat{\alpha}_{\textrm{B}}\frac{k^{2}}{a^{2}}\right]H\Phi v_{X}\nonumber \\
 & +\frac{1}{2}\left[\left(2\dot{H}+\frac{\dot{\varphi}^{2}}{M_{S}^{2}}\right)-2\left(\hat{\alpha}_{\textrm{M}}-\hat{\alpha}_{\textrm{T}}\right)H^{2}-\frac{\left(aH\hat{\alpha}_{\textrm{B}}M_{S}^{2}\right)^{.}}{aM_{S}^{2}}\right]\frac{k^{2}}{a^{2}}v_{X}^{2}\nonumber \\
 & -\frac{3}{2}\left[\left(2\dot{H}+\frac{\dot{\varphi}^{2}}{M_{S}^{2}}\right)\dot{H}-\frac{\left(a^{3}H\dot{H}\hat{\alpha}_{\textrm{B}}M_{S}^{2}\right)^{.}}{a^{3}M_{S}^{2}}\right]v_{X}^{2}\,.\label{STperts}
\end{align}
Here $v_X$ describes the perturbations of the extra gravitational scalar field (whose background value does not appear in the action). We have introduced, to simplify our expressions, the following perturbation field:
\begin{equation}\label{Bredefinition}
 \tilde{B}=k^{2}\dot{E}-\frac{k^{2}}{a^{2}}B\,.
\end{equation}
This action depends on two background quantities: the scale factor and the matter field. Whereas the matter field is considered to be a dependent function through the matter eq.~(\ref{BackMatterEqn}), the scale factor is a free parameter of the model. In addition to the scale factor, the dynamics of linear perturbation fields is completely determined by a set of four free independent functions of time: one mass dimensional parameter $M_S$, and three dimensionless parameters $\hat{\alpha}_T$, $\hat{\alpha}_B$, and $\hat{\alpha}_K$. We note that for simplicity we have also introduced the function $\hat{\alpha}_M$, given by
\begin{equation}\label{STAlphaM}
\hat{\alpha}_M=\frac{d \ln M_S^2}{d\ln a},
\end{equation}
 which describes the running of the mass $M_S$ and thus is not independent from the other four parameters. As shown in \cite{Bellini:2014fua}, from the previous action we can find a physical interpretation for the free parameters: $M_S$ and $\hat{\alpha}_M$ characterise the time dependence of Newton's constant, $\hat{\alpha}_K$ (known as ``kineticity") extends the kinetic term of canonical scalar field models while $\hat{\alpha}_B$ (known as ``braiding") singles out the kinetic mixing between $v_X$ and metric perturbations. The remaining function, $\hat{\alpha}_T$ (known as ``tensor speed excess''), describes deviations of the speed of propagation of tensor modes from the speed of light, as well as the anisotropic stress in the scalar sector. We note that all these parameters describe deviations from GR. Indeed, GR is recovered when $M_S=M_P$, and $\hat{\alpha}_K=\hat{\alpha}_B=\hat{\alpha}_T=0$. 

Furthermore, as previously discussed in \cite{Gleyzes:2013ooa,Lagos:2016wyv}, the action in eq.~(\ref{STperts}) propagates only one physical scalar degree for any value of the parameters. This can explicitly be seen from the action, as $B$ and $\Phi$ appear as auxiliary variables (without time derivatives) that can be integrated out, leading to a gravitational action with only three explicit fields $E$, $\Psi$ and $v_X$, two of which can be gauge fixed leaving one physical propagating degree of freedom.

We should note that it is now customary to include an additional parameter, $\hat{\alpha}_H$ which arises if we include terms with higher derivatives in the action but which, nevertheless, do not lead to additional propagating degrees of freedom and do not violate Ostrogradski's theorem. In other words, they very much behave like theories with purely second order equations of motion \cite{Zumalacarregui:2013pma,Gleyzes:2014dya,Gleyzes:2014qga}. It has been conjectured that this approach can be extended to an almost arbitrary number of higher order derivatives subject to certain constraints (and leading, therefore, to an arbitrary number of extra $\alpha_X$ functions) and still only propagating an extra degree of freedom
(see \cite{Lagos:2016wyv,BenAchour:2016fzp} for a discussion on how this can be done).

The action previously presented is the most general quadratic action for scalar-tensor gravity theories that are linearly invariant under general coordinate transformations and have second-order derivative equations of motion. Therefore, such an action is expected to be related to quadratic expansions of non-linear scalar-tensor theories that are diffeomorphism invariant and also lead to second derivatives. Indeed, it has been previously found that one can write the most general diffeomorphism-invariant non-linear action from a metric, $g_{\mu\nu}$ and scalar field, $\phi$ \cite{Horndeski:1974wa,Deffayet:2009wt} with second-order derivative equations of motion as:
\begin{eqnarray}\label{eq:L_horndeski}
S=\int d^4x \sqrt{-g}\left\{\sum_{i=2}^5{\cal L}_i[\phi,g_{\mu\nu}]\right\},
\end{eqnarray}
where ${\cal L}_i$ are scalar-tensor Lagrangians given by:
\begin{eqnarray}
{\cal L}_2&=& K , \nonumber \\
{\cal L}_3&=& -G_3 \Box\phi , \nonumber \\
{\cal L}_4&=&  G_4R+G_{4X}\left\{(\Box \phi)^2-\nabla_\mu\nabla_\nu\phi \nabla^\mu\nabla^\nu\phi\right\} , \nonumber \\
{\cal L}_5&=& G_5G_{\mu\nu}\nabla^\mu\nabla^\nu\phi
-\frac{1}{6}G_{5X}\big\{ (\nabla\phi)^3
-3\nabla^\mu\nabla^\nu\phi\nabla_\mu\nabla_\nu\phi\Box\phi 
 \nonumber \\ & & 
+2\nabla^\nu\nabla_\mu\phi \nabla^\alpha\nabla_\nu\phi\nabla^\mu\nabla_\alpha \phi
\big\}  \,. \label{Horndeski}
\end{eqnarray}
This action is known as the Horndeski action and has four free functions, $K$, $G_3$, $G_4$ and $G_5$, which completely characterise this class of theories and depend solely on $\phi$ and $X\equiv-\nabla^\nu\phi\nabla_\nu\phi/2$; subscripts with $X$ and $\phi$ denote derivatives. Notable examples of theories that are encompassed by this action are: quintessence, k-essence, kinetic gravity braiding, galileons, generalised Brans-Dicke, imperfect fluids, cuscuton, metric $f(R)$ and its generalisations \cite{Clifton:2011jh}. Furthermore, this action can be extended by including ``beyond Horndeski'' terms \cite{Zumalacarregui:2013pma,Gleyzes:2014dya,Gleyzes:2014qga}, ``extended scalar-tensor theories'' terms \cite{Langlois:2015cwa,Crisostomi:2016czh,BenAchour:2016fzp} and non-minimal coupling to matter \cite{Gleyzes:2015pma}.

The Horndeski action in eq.~(\ref{eq:L_horndeski}) can be expanded quadratically in linear perturbations (around a spatially flat homogeneous and isotropic background) to obtain an action that should be encompassed by the one in eq.~(\ref{STperts}). Indeed, the two actions are linked via the following relations \cite{Bellini:2014fua}:
\begin{eqnarray}
M^2_S&\equiv&2\left(G_4-2XG_{4X}+XG_{5\phi}-{\dot \phi}HXG_{5X}\right) , \nonumber \\
HM^2_S\hat{\alpha}_M&\equiv&\frac{d}{dt}M^2_S , \nonumber \\
H^2M^2_S\hat{\alpha}_K&\equiv&2X\left(K_X+2XK_{XX}-2G_{3\phi}-2XG_{3\phi X}\right) \nonumber \\ & &
+12\dot{\phi}XH\left(G_{3X}+XG_{3XX}-3G_{4\phi X}-2XG_{4\phi XX}\right) \nonumber \\ & &
+12XH^2\left(G_{4X}+8XG_{4XX}+4X^2G_{4XXX}\right)\nonumber \\ & &
-12XH^2\left(G_{5X}+5XG_{5\phi X}+2X^2G_{5\phi XX}\right)\nonumber \\ & &
+14\dot{\phi}H^3\left(3G_{5X}+7XG_{5XX}+2X^2G_{5XXX}\right)\nonumber , \\
HM^2_S\hat{\alpha}_B&\equiv&2\dot{\phi}\left(XG_{3X}-G_{4\phi}-2XG_{4\phi X}\right) \nonumber \\ & &
+8XH\left(G_{4X}+2XG_{4XX}-G_{5\phi}-XG_{5\phi X}\right) \nonumber \\ & &
+2\dot{\phi}XH^2\left(3G_{5X}+2XG_{5XX}\right) \nonumber , \\ 
M^2_S\hat{\alpha}_T&\equiv&2X\left[2G_{4X}-2G_{5\phi}-\left(\ddot{\phi}-\dot{\phi}H\right)G_{5X}\right] \,,
\end{eqnarray}
where all the functions $K$, $G_3$, $G_4$, $G_5$ and the field $\phi$ with $X$ are evaluated at the background.

We note that both the Horndeski action and the parametrised action in eq.~(\ref{STperts}) have the same number of free parameters, four, and hence the Horndeski quadratic action is as general as the parametrised action in eq.~(\ref{STperts}). This is a remarkable situation as in most cases the parametrised action we find is more general than quadratic expansions of well-known non-linear theories. This is indeed what we show for vector-tensor and tensor-tensor theories in the next subsections, and what has also been previously found for the simple case of a single metric theory \cite{Lagos:2016wyv}.


\subsection{Vector-Tensor Theories}
In this subsection we consider two qualitatively distinct classes of vector-tensor theories. They have a different number of free parameters and encompass different non-linear modified gravity theories. The first one is general vector-tensor theories, where the vector field can be an arbitrary function. This class will encompass Generalised Proca theory. The second one is Einstein-Aether-like vector-tensor theories, where the vector field is a unit time-like vector. This class will encompass generalised Einstein-Aether theory. We note that, contrary to Horndeski and scalar-tensor theories, neither Generalised Proca nor generalised Einstein-Aether are known to be the most general non-linear second-order actions that can be constructed for vector-tensor theories.


\subsubsection{General Vector-Tensor Theories}
\label{Proca}
As before, we can construct the most general linearly diffeomorphism-invariant vector-tensor quadratic action around a spatially flat homogeneous and isotropic background, leading to second-order derivative equations of motion. It is given by:
\begin{align}\label{VectorTensorActionFinal}
 S^{(2)} &= S_{m,\varphi}^{(2)} + \int d^3kdt\; a^3M_V^2\mathcal{L}_{VT}\,,
\end{align}
where $\mathcal{L}_{VT}$ is the vector-tensor Lagrangian given by:
\begin{align}
\mathcal{L}_{VT}= & -\left(1-\frac{\tilde{\alpha}_{\textrm{D}}}{2}\right)\left[3\dot{\Psi}+2\tilde{B}\right]\dot{\Psi}-\left(2-\tilde{\alpha}_{\textrm{D}}\right)\left[3\dot{\Psi}+\tilde{B}\right]H\Phi-2\left(1+\tilde{\alpha}_{\textrm{C}}\right)\frac{k^{2}}{a^{2}}\Phi\Psi\nonumber \\
 & +\left(1+\tilde{\alpha}_{\textrm{T}}\right)\frac{k^{2}}{a^{2}}\Psi^{2}+\frac{1}{6}\tilde{\alpha}_{\textrm{D}}\tilde{B}^{2}-\frac{1}{2}\left[\left(6-3\tilde{\alpha}_{\textrm{D}}-\frac{\tilde{\alpha}_{\textrm{K}}k^{2}}{a^{2}H^{2}}\right)H^{2}-\frac{\dot{\varphi}^{2}}{M_{V}^{2}}\right]\Phi^{2}\nonumber \\
 & +\frac{1}{2}\tilde{\alpha}_{\textrm{K}}\frac{k^{2}}{a^{2}}\dot{{\delta A}_{1}}^{2}+\tilde{\alpha}_{\textrm{K}}\frac{k^{2}}{a^{2}}\Phi\dot{{\delta A}_{1}}+\left(2\tilde{\alpha}_{\textrm{C}}+\tilde{\alpha}_{\textrm{D}}\right)\frac{k^{2}}{a^{2}}\dot{\Psi}{\delta A}_{1}\nonumber \\
 & +\frac{1}{3}\tilde{\alpha}_{\textrm{D}}\tilde{B}\frac{k^{2}}{a^{2}}{\delta A}_{1}+\left(2\tilde{\alpha}_{\textrm{C}}+\tilde{\alpha}_{\textrm{D}}+\tilde{\alpha}_{\textrm{A}}\right)H\frac{k^{2}}{a^{2}}\Phi{\delta A}_{1}+\frac{1}{6}\tilde{\alpha}_{\textrm{D}}\frac{k^{4}}{a^{4}}{\delta A}_{1}^{2}\nonumber \\
 & -\frac{1}{2}\left[2\tilde{\alpha}_{\textrm{C}}\dot{H}-\tilde{\alpha}_{\textrm{V}}H^{2}+\frac{\left(aH\tilde{\alpha}_{\textrm{A}}M_{V}^{2}\right)^{.}}{aM_{V}^{2}}\right]\frac{k^{2}}{a^{2}}{\delta A}_{1}^{2}\nonumber \\
 & +\left(2\dot{H}+\frac{\dot{\varphi}^{2}}{M_{V}^{2}}-\tilde{\alpha}_{\textrm{D}}\dot{H}\right)\left[3\dot{\Psi}+\tilde{B}+3H\Phi-\frac{3}{2}\dot{H}{\delta A}_{0}\right]{\delta A}_{0}\nonumber \\
 & +\left[\frac{\left(a\tilde{\alpha}_{\textrm{C}}M_{V}^{2}\right)^{.}}{M_{V}^{2}}+\left(\tilde{\alpha}_{\textrm{M}}-\tilde{\alpha}_{\textrm{T}}\right)H\right]\left(2\Psi-H{\delta A}_{0}\right)\frac{k^{2}}{a^{2}}{\delta A}_{0}\nonumber \\
 & +\frac{1}{2}\left(2\dot{H}+\frac{\dot{\varphi}^{2}}{M_{V}^{2}}+\tilde{\alpha}_{\textrm{V}}H^{2}\right)\frac{k^{2}}{a^{2}}{\delta A}_{0}^{2}-\tilde{\alpha}_{\textrm{A}}H\frac{k^{2}}{a^{2}}\Phi{\delta A}_{0}\nonumber \\
 & -\tilde{\alpha}_{\textrm{A}}H\frac{k^{2}}{a^{2}}{\delta A}_{0}\dot{{\delta A}_{1}}-\left(\tilde{\alpha}_{\textrm{D}}\dot{H}+\tilde{\alpha}_{\textrm{V}}H^{2}\right)\frac{k^{2}}{a^{2}}{\delta A}_{0}{\delta A}_{1}\,. \label{VTlin}
\end{align}
Here, we have decomposed the gravitational vector field $A^\mu$ as:
\begin{eqnarray}\label{VectorDecompo}
A^{\mu}= & \left(A-\left(\dot{A}+HA\right)\delta A_{0}+A\Phi,\,\partial^{i}\left(\delta A_{1}-B\right)\right),
\end{eqnarray}
 where $A$ is the background value of the field (which does not appear explicitly in the action), and $\delta A_0$ with $\delta A_1$ are the two linear scalar perturbations. We have also introduced the perturbation field $\tilde{B}$ as in eq.~(\ref{Bredefinition}).
 Similarly to scalar-tensor theories, this action depends on one arbitrary background function, the scale factor, and an additional set of $\alpha_X(t)$ functions that completely determine the evolution of the linear perturbations. In particular, we find seven independent free parameters: one mass dimensional parameter $M_V$ and six dimensionless parameters $\tilde{\alpha}_A$, $\tilde{\alpha}_D$, $\tilde{\alpha}_V$, $\tilde{\alpha}_C$, $\tilde{\alpha}_T$, and $\tilde{\alpha}_K$. For simplicity we have again introduced the dependent parameter $\tilde{\alpha}_M$ which is given by the time derivative of $M_V$, analogous to eq.~(\ref{STAlphaM}).
 
 By analysing the different interaction terms of this quadratic action we can associate physical effects to the different free parameters: $M_V$ and $\tilde{\alpha}_M$ characterise the time dependent Newton's constant, $\tilde{\alpha}_K$ is the kineticity term affecting directly the kinetic energy of the field $\delta A_1$, and $\tilde{\alpha}_T$ is the tensor speed excess. These parameters all have an equivalent in scalar-tensor theories. Furthermore, we find $\tilde{\alpha}_V$, the vector mass mixing parameter, which creates mass mixing between the vector perturbations $\delta A_0$ and $\delta A_1$ and hence contributes indirectly to the mass of the vector field. We also find $\tilde{\alpha}_D$, the small scales dynamics parameter, which generates interaction terms of the form $k^4\delta A_1^2$ and which dominate on small scales. In addition, we find $\tilde{\alpha}_C$, the conformal coupling excess parameter. As we will see next, this parameter describes effective deviations from the mass scale $M_V^2$ in a particular family of vector-tensor theories known as Generalised Proca, and hence acts as an additional conformal coupling in their quadratic actions. Due to the particular field decomposition we are employing in eq.~(\ref{VectorDecompo}), it is not straightforward to identify such an effective conformal coupling in the quadratic action given by eq.~(\ref{VTlin}). However, such a feature will be seen clearly when calculating the quadratic action for Generalised Proca. Finally, we also find $\tilde{\alpha}_A$, the auxiliary friction parameter, in general vector-tensor theories. This parameter is the only one that allows a coupling of the form $\delta A_0 \delta \dot{A}_1$ in the action and hence makes the equation of motion for the dynamical field $\delta A_1$ depend on the velocity of the auxiliary field $\delta A_0$. As a result of such a coupling, $\tilde{\alpha}_A$ will indirectly contribute to the kinetic energy of the field $\delta A_1$.
 
 We note that all the previously mentioned parameters affect the evolution of scalar perturbations. As we will see in Section \ref{Sec:Phenomenology}, general vector-tensor theories have an additional parameter $\tilde{\alpha}_{VS}$ that affects cosmological vector perturbations only. All these $\alpha_X$ parameters describe deviations from GR. Indeed, we recover GR when $M_V=M_P$ and $\tilde{\alpha}_K=\tilde{\alpha}_T=\tilde{\alpha}_V=\tilde{\alpha}_D=\tilde{\alpha}_C=\tilde{\alpha}_A=\tilde{\alpha}_{VS}=0$. 
 
 We emphasise that the action presented here has fewer parameters than the one found in \cite{Lagos:2016wyv} for two reasons. Firstly, here we have fixed two of the originally free parameters in order to obtain an action that propagates one scalar instead of two (as it would happen in completely generic, including sick, quadratic vector-tensor actions)\footnote{Two comments related to the above: Firstly, there is more than one parameter choice that leads to a vector-tensor action propagating only one scalar DoF. Here we have chosen one in which the perturbation field $\delta A_0$ appears as an explicit auxiliary variable (without time derivatives), which is the structure of the healthy vector-tensor theory known as Generalised Proca \cite{DeFelice:2016yws}. Secondly, while the `completely generic' actions discussed propagate two scalars, if two scalars are indeed present, one of them is typically a ghost. This is analogous to the situation in massive and bi-gravity.}. Secondly, here we have also imposed gauge invariance under vector perturbations and, even though not shown explicitly here, they lead to one additional parameter constraint, hence reducing by one the total number of free parameters (see \cite{Tattersall:2017eav} for an explicit calculation of the parametrised gauge-invariant action for vector perturbations).

The parametrised gravitational action in eq.~(\ref{VTlin}) propagates only one physical scalar degree of freedom. This can explicitly be seen from the action, as $B$, $\Phi$ and $\delta A_0$ appear as auxiliary fields that can be integrated out, leading to a gravitational action with only three explicit fields $E$, $\Psi$ and $\delta A_1$, two of which can be gauge fixed leaving one physical propagating degree of freedom. 

Similarly to the previous subsection, the action presented here is the most general quadratic action for vector-tensor gravity theories that are linearly invariant under general coordinate transformations and have second-order derivative equations of motion, and they are expected to encompass quadratic expansions of non-linear theories with the same properties. At present, the most general vector-tensor theory known to be diffeomorphism invariant and leading to second-order derivative equations of motion is the Generalised Proca theory \cite{Heisenberg:2014rta} given by:
\begin{equation}
S=\int d^4x\; \sqrt{-g}\left\{ \sum_{i=2}^6\mathcal{L}_i[A_\alpha, g_{\mu\nu}]\right\}, 
\end{equation}
where $\mathcal{L}_i$ are gravitational vector-tensor Lagrangians given by:
\begin{align}\label{VT}
&\mathcal{L}_2=G_2(X,F,Y),\\
&\mathcal{L}_3=G_3(X)\nabla_{\mu}A^\mu,\\
&\mathcal{L}_4=G_4(X)R+G_{4X}\left[ (\nabla_{\mu}A^\mu)^2 - \nabla_{\mu}A^\nu\nabla_{\nu}A^\mu \right],\\
&\mathcal{L}_5=G_5(X)G_{\mu\nu}\nabla^{\mu}A^\mu-\frac{1}{6}G_{5X}\left[ (\nabla_{\mu}A^\mu)^3-3(\nabla_{\alpha}A^\alpha)(\nabla_{\mu}A^\nu\nabla_{\nu}A^\mu) +2 \nabla_{\mu}A^\alpha \nabla_{\nu}A^\mu \nabla_{\alpha}A^\nu \right]\nonumber\\
& - g_5(X)\tilde{F}^{\alpha\mu}\tilde{F}_{\beta\mu}\nabla_{\alpha}A^\beta,\\
&\mathcal{L}_6=G_6(X)L^{\mu\nu\alpha\beta}\nabla_\mu A_\alpha \nabla_\alpha A_\beta +\frac{1}{2}G_{6X}\tilde{F}^{\alpha\beta}\tilde{F}^{\mu\nu}\nabla_{\alpha}A_\mu \nabla_{\beta}A_\nu, 
\end{align}
which are written in terms of 6 free functions $G_2$, $G_3$, $G_4$, $G_5$, $G_6$ and $g_5$. We can define the following tensors:
\begin{align}
& F_{\mu\nu}=\nabla_\mu A_\nu -\nabla_\nu A_\mu ,\\
& \tilde{F}^{\mu\nu}=\frac{1}{2}\epsilon^{\mu\nu\alpha\beta}F_{\alpha\beta},\\
& L^{\mu\nu\alpha\beta}=\frac{1}{4}\epsilon^{\mu\nu\rho\sigma} \epsilon^{\alpha\beta\gamma\delta}R_{\rho\sigma\gamma\delta},
\end{align}
where $R_{\rho\sigma\gamma\delta}$ is the Riemann tensor and $\epsilon^{\mu\nu\alpha\beta}$ is the Levi-Civita antisymmetric tensor. The 5 free parameters $G_i$ are free functions of the following scalar quantities of the previously defined tensors:
\begin{align}
& X = -\frac{1}{2}A_\mu A^\mu,\\
& F =- \frac{1}{4}F^{\mu\nu}F_{\mu\nu},\\
& Y = A^\mu A^\nu F_\mu{}^\alpha F_{\nu\alpha}.
\end{align}
We note that, as in the scalar-tensor case, the subscripts $X$ denotes derivatives with respect to the quantity $X$. While this action in general describes interactions of a massive spin-1 particle with a massless graviton, for special values of the free functions it can also describe the theory of massless spin-1 particles, such as the Einstein-Maxwell action, vector Galileons \cite{Deffayet:2013tca}, theories with spontaneously broken abelian symmetry \cite{Gripaios:2004ms,Tasinato:2014mia} and Tensor-Scalar-Vector theories \cite{Zlosnik:2006sb}. 
We find that the quadratic expansion of Generalised Proca around a spatially flat homogeneous and isotropic background is included in the parametrised action of eq.~(\ref{VTlin}). Explicitly, we find : 
\begin{align}
M_{V}^{2}= & 2G_{4}-A^{2}\left(2G_{4X}-AHG_{5X}\right),\label{MassVT}\\
\tilde{\alpha}_{\textrm{K}}M_{V}^{2}= & A^{2}\left(G_{2F}+2A^{2}G_{2Y}-4g_{5}HA\right)+2A^{2}H^{2}\left(G_{6}+\frac{1}{3}A^{2}G_{6X}\right),\\
\tilde{\alpha}_{\textrm{C}}= & \frac{2G_{4}}{M_{V}^{2}}-1,\label{AlphaCVT}\\
\tilde{\alpha}_{\textrm{T}}= & \tilde{\alpha}_{\textrm{M}}+\tilde{\alpha}_{\textrm{C}}\left(1+\tilde{\alpha}_{\textrm{M}}-\frac{\tilde{\alpha}_{\textrm{A}}}{\tilde{\alpha}_{\textrm{K}}}\right)+\frac{\dot{\tilde{\alpha}}_{\textrm{C}}}{H},\\
\tilde{\alpha}_{\textrm{A}}= & \frac{\dot{A}}{HA}\tilde{\alpha}_{\textrm{K}},\\
\frac{\tilde{\alpha}_{\textrm{K}}}{\tilde{\alpha}_{\textrm{A}}}\tilde{\alpha}_{\textrm{V}}= & \tilde{\alpha}_{\textrm{A}}-\frac{A^{3}}{M_{V}^{2}H}\left(G_{3X}-H^{2}G_{5X}+4AHG_{4XX}-A^{2}H^{2}G_{5XX}\right),\\
\tilde{\alpha}_{\textrm{D}}=&0. \label{AlphaDProca}
\end{align}
Even though the Generalised Proca theory has a similar structure to the Horndeski action, the latter has been shown to be the most general possible scalar-tensor diffeomorphism invariant action with second order derivative equations of motion, while the former one has not. For this reason, it is not surprising that the quadratic Generalised Proca action has fewer free functions than the parametrised action we present in this paper. Indeed, we find that $\tilde{\alpha}_D=0$ in Generalised Proca. It is then possible that new non-linear interactions are allowed in vector-tensor theories but they have yet to be discovered. Such work is beyond the scope of this paper, but a starting point to find them would be to look for the quadratic interactions that $\tilde{\alpha}_D$ generates and reconstruct healthy non-linear terms from there. We note though that it could also happen that $\tilde{\alpha}_D$ is not associated to any healthy theory and that any consistent non-linear completion of the quadratic actions considered here enforces $\tilde{\alpha}_D = 0$.
%

Finally, we mention that from $\mathcal{L}_4$ one might be tempted to identify the function $G_4$ as the mass scale of the full non-linear action. However, once we calculate the quadratic action for perturbations, we find an effective mass scale given by eq.~(\ref{MassVT}), and hence we identify $G_4$ with a function carrying an additional conformal coupling. For this reason, the parameter $\tilde{\alpha}_C$ in eq.~(\ref{AlphaCVT}) is called conformal coupling excess.


\subsubsection{Einstein-Aether Theories}
We now consider the most general linearly diffeomorphism-invariant vector-tensor theories with second-order derivative equations of motion, where the gravitational vector field $A^\mu$ is time-like (often dubbed the ``aether'') such that:
\begin{equation}
A^\mu A_\mu +1=0. \label{AEconstraint}
\end{equation}
We can write an action for such a field by adding a Lagrange multiplier to the vector-tensor action; in doing so, the resulting theory is not strictly vector-tensor as there is an additional scalar field -- the Lagrange multiplier -- which, however, has no dynamics \cite{Jacobson:2000xp}. After eliminating the Lagrange multiplier we find the most general quadratic action satisfying eq.~(\ref{AEconstraint}) to be:
\begin{align}\label{EinsteinAetherActionFinal}
 S^{(2)} &= S_{m,\varphi}^{(2)}+ \int d^3kdt\; a^3M_A^2\mathcal{L}_{EA}\,,
\end{align}
where $\mathcal{L}_{EA}$ is the Lagrangian for an aether vector, given by:
\begin{align}\label{EAlin}
\mathcal{L}_{EA}= & -\left(1-\frac{\check{\alpha}_{\textrm{D}}}{2}\right)\left[3\dot{\Psi}+2\tilde{B}\right]\dot{\Psi}-\left(2-\check{\alpha}_{\textrm{D}}\right)\left[3\dot{\Psi}+\tilde{B}\right]H\Phi-2\left(1+\check{\alpha}_{\textrm{C}}\right)\frac{k^{2}}{a^{2}}\Phi\Psi\nonumber \\
 & +\left(1+\check{\alpha}_{\textrm{T}}\right)\frac{k^{2}}{a^{2}}\Psi^{2}+\frac{1}{6}\check{\alpha}_{\textrm{D}}\tilde{B}^{2}-\frac{1}{2}\left[\left(2-\check{\alpha}_{\textrm{D}}\right)\left(\dot{H}+3H^{2}\right)-\frac{\check{\alpha}_{\textrm{K}}k^{2}}{a^{2}}\right]\Phi^{2}\nonumber \\
 & +\frac{1}{2}\check{\alpha}_{\textrm{K}}\frac{k^{2}}{a^{2}}\dot{{\delta A}_{1}}^{2}+\check{\alpha}_{\textrm{K}}\frac{k^{2}}{a^{2}}\Phi\dot{{\delta A}_{1}}+\left(2\check{\alpha}_{\textrm{C}}+\check{\alpha}_{\textrm{D}}\right)\frac{k^{2}}{a^{2}}\dot{\Psi}{\delta A}_{1}+\frac{1}{3}\check{\alpha}_{\textrm{D}}\tilde{B}\frac{k^{2}}{a^{2}}{\delta A}_{1}\nonumber \\
 & +\left(2\check{\alpha}_{\textrm{C}}+\check{\alpha}_{\textrm{D}}\right)H\frac{k^{2}}{a^{2}}{\delta A}_{1}\Phi+\frac{1}{6}\check{\alpha}_{\textrm{D}}\frac{k^{4}}{a^{4}}{\delta A}_{1}^{2}-\frac{1}{2}\left(2\check{\alpha}_{\textrm{C}}+\check{\alpha}_{\textrm{D}}\right)\dot{H}\frac{k^{2}}{a^{2}}{\delta A}_{1}^{2}\,,
\end{align}
where we have decomposed the gravitational vector field as:
\begin{equation}
A^\mu=\left(1 + {\delta A}_{0}, \partial^i{\delta A}_{1}-\partial^iB\right).
\end{equation}
In addition, we have introduced the perturbation $\tilde{B}$ as in eq.~(\ref{Bredefinition}). We note that, after solving the Lagrange constraint, the perturbation $\delta A_0$ of the time component of the vector field does not appear explicitly in this action as it is constrained via eq.~(\ref{AEconstraint}), which gives $\delta A_0=-\Phi$, and the dependence in the multiplier vanishes. 

The parametrised action in eq.~(\ref{EAlin}) depends on the background scale factor, in addition to three free functions that completely determine the dynamics of the linear perturbations: one mass dimensional parameter $M_A$, and two dimensionless parameters $\check{\alpha}_C$, and $\check{\alpha}_K$. For simplicity, we have also introduced two extra dependent parameters $\check{\alpha}_D$ and $\check{\alpha}_T$ given by:
\begin{equation}\label{DependentAlphasEA}
\check{\alpha}_D\equiv 2+ \left(\rho + P\right)/(M_A^2 \dot{H}), \quad \check{\alpha}_T\equiv \check{\alpha}_\textrm{C} + \left(1 + \check{\alpha}_\textrm{C}\right) \check{\alpha}_\textrm{M} + \check{\alpha}^\prime_\textrm{C} ,
\end{equation}
where $\check{\alpha}_\textrm{M}$ was defined analogously to eq.~(\ref{STAlphaM}), and where $\rho$ and $P$ are the energy density and pressure, respectively, of the external matter (a minimally coupled scalar field in this case). Here, primes represent derivatives with respect to the number of e-foldings ($\ln a$). We note that we have defined these two parameters in analogy to $\tilde{\alpha}_D$ and $\tilde{\alpha}_T$ for general vector-tensor theories; they induce similar interaction terms in the action but for Einstein-Aether theories these parameters are not independent. 

Similarly, as before, we can give a physical interpretation to the free parameters of this action. $M_A$ gives a time dependent Newton's constant, $\check{\alpha}_K$ is the kineticity term, and $\check{\alpha}_C$ is the conformal coupling excess. These parameters describe deviations with respect to GR. Indeed, we recover GR when $M_A=M_P$ and $\check{\alpha}_K=\check{\alpha}_C=0$.

This action depends explicitly on five perturbation fields, and only one of those is a physical propagating DoF. This can explicitly be seen from the parametrised action, as $B$ and $\Phi$ appear as auxiliary fields that can be integrated out, leading to a gravitational action with only three explicit fields $E$, $\Psi$ and $\delta A_1$, two of which are gauge freedoms (that still remain after solving the Lagrange constraint) and one physical propagating one. 

We now proceed to compare this parametrised action with quadratic expansions of well-known non-linear theories that are also linearly diffeomorphism invariant with second-order derivative equations of motion. A possible, restricted, non-linear action for an aether vector is given by the generalised Einstein-Aether theory \cite{Zlosnik:2006zu}:
\begin{equation}
S= \frac{1}{2}M_{P}^2\int d^4x \sqrt{-g}\left[R+{\cal F}(\mathcal{K})+\lambda(A^\mu A_\mu +1)\right], \label{AEaction}
\end{equation}
where $M_P$ is the Planck mass, $R$ is the Ricci scalar, $\lambda$ is a Lagrange multiplier, and {\cal F}($\mathcal{K}$) is an arbitrary function of the following kinetic term:
\begin{equation}
\mathcal{K}=\left(c_1g_{\mu\nu}g_{\alpha\beta}+c_2g_{\mu\alpha}g_{\nu\beta}+c_3g_{\mu\beta}g_{\nu\alpha}\right)\nabla^\mu A^\alpha \nabla^\nu A^\beta,
\end{equation}
where $c_i$ are three free constants. These theories have been extensively studied in \cite{Li:2007vz,Zlosnik:2007bu,Jimenez:2009ai,ArmendarizPicon:2009ai,Zuntz:2010jp}. By calculating the quadratic action of the generalised Einstein-Aether theory, we find it indeed encompassed by the parametrised action in eq.~(\ref{EAlin}), Explicitly, we find the following relations:
\begin{align}
& M_A^2=1+(c_1+c_3)\mathcal{F}',\\
& \check{\alpha}_{K}M_A^2=-c_1{\cal F}',\\
& \check{\alpha}_{C}M_A^2=1-M_A^2,
\end{align}
where $'$ denotes derivatives of the function ${\cal F}$ with respect to the kinetic term $\mathcal{K}$ evaluated at the background level. 
We note that the parameter $\check{\alpha}_C$ does not vanish for this family of theories, even though the action given in eq.~(\ref{AEaction}) does not have a conformal coupling. This is because the Lagrange multiplier induces additional self-interactions for the metric in the quadratic action, leading to similar interaction terms to those with $\tilde{\alpha}_C$ in vector-tensor theories. Similarly to the previous subsection, we can also note that the parametrised general action for aether vectors is more general than the quadratic action of generalised Einstein-Aether. Indeed, all the time evolution of the three free functions is determined a single quantity, $\mathcal{F}'$, in the action for generalised Einstein-Aether. 

 
\subsection{Tensor-Tensor Theories}
We now consider the most general linearly diffeomorphism-invariant bimetric quadratic action, leading to second-order derivative equations of motion. There will be two interacting metrics $g_{\mu\nu}$ and $f_{\mu\nu}$, but only $g_{\mu\nu}$ will be coupled to matter and hence it will describe the physical spacetime. For simplicity, we will restrict ourselves to theories without derivative mixing {\it and} propagating only one scalar DoF. As found in \cite{Lagos:2016gep}, such an action is given by: 
\begin{equation}\label{TotalActionNonDer}
S^{(2)}=S_{m,\varphi}^{(2)} + \int d^3kdt\; \left[a^3M^2_T\mathcal{L}_{T_1}+ a^3M_2^2\mathcal{L}_{T_2}+ a^3M^2_T\mathcal{L}_{T_1T_2} \right],
\end{equation}
where $\mathcal{L}_{T_1}$ and $\mathcal{L}_{T_2}$ are the Lagrangians for the self-interaction terms of the metrics $g_{\mu\nu}$ and $f_{\mu\nu}$, respectively, whereas $\mathcal{L}_{T_1T_2}$ is the Lagrangian for the interaction terms between both metrics (and will have no derivative dependence due to our above restriction). These actions are given by:
\begin{align}
\mathcal{L}_{T_1}&= \left[-3\dot{\Psi}^2-6H\dot{\Psi}\Phi-2k^2\dot{E}(\dot{\Psi}+H\Phi)+2\frac{k^2}{a^2}\dot{\Psi}B +\frac{k^2}{a^2}\left(1+\bar{\alpha}_M\right)\Psi^2-2\frac{k^2}{a^2}\Phi(\Psi-HB) \right. \nonumber\\
& - \left(3H^2 - \frac{\dot{\varphi}^2}{2M_T^2} \right)\Phi^2 + \frac{\bar{\alpha}_{SI}H^2}{2(\bar{\alpha}_{N}+1)}\frac{k^2}{a^2}B^2 + 2H^2\bar{\alpha}_{TI} \Psi\left(\frac{3}{2}\Psi+k^2E\right) \nonumber\\
& \left. +\frac{(\bar{\alpha}_{N}-1)}{1-\bar{\alpha}_{b}}\bar{\alpha}_{SI}H^2\Phi\left(3\Psi+k^2E\right) +k^4\bar{\alpha}_{S}H^2E^2 \right], \label{S1NonDer}\\
%
\mathcal{L}_{T_2}&= \left[-3\frac{\dot{\Psi}_2^2}{\bar{\alpha}_{N}}-6\frac{H\bar{\alpha}_{b}}{\bar{\alpha}_{N}}\dot{\Psi}_2\Phi_2-2\frac{k^2}{\bar{\alpha}_{N}}\dot{E}_2(\dot{\Psi}_2+H\bar{\alpha}_{b}\Phi_2)+2\frac{\dot{\Psi}_2}{\bar{\alpha}_{N}} \frac{k^2}{a^2} B_2 \right. \nonumber\\
& +k^2a^2\frac{\bar{\alpha}_{N}}{\bar{\alpha}_{b}}\left(\bar{\alpha}_{M_2}-\bar{\alpha}_{b}+2\right)\Psi_2^2 -2\frac{k^2}{a^2}\bar{\alpha}_{N}\Phi_2\left(\Psi_2-\frac{H\bar{\alpha}_{b}}{\bar{\alpha}_{N}^2}B_2\right) -3\frac{H^2\bar{\alpha}_{b}^2}{\bar{\alpha}_{N}}\Phi_2^2 \nonumber\\
& + \frac{\bar{\alpha}_{SI}H^2}{2\nu^2_M(\bar{\alpha}_{N}+1)}\frac{k^2}{a^2}B_2^2 + 2\nu_M^{-2}\bar{\alpha}_{TI}H^2 \Psi_2\left(\frac{3}{2}\Psi_2+k^2E_2\right) \nonumber\\
& \left. + \bar{\alpha}_{b}\bar{\alpha}_{SI}H^2\frac{(\bar{\alpha}_{N}-1)}{\nu^2_M(1-\bar{\alpha}_{b})}\Phi_2\left(3\Psi_2+k^2 E_2\right) +\bar{\alpha}_{S}H^2\nu_M^{-2}k^4 E_2^2 \right], \label{S2NonDer}\\
%
\mathcal{L}_{T_1T_2}&=H^2\left[- 2 \bar{\alpha}_{TI}\left( 3\Psi_2\Psi+k^2\Psi_2 E+k^2\Psi E_2 \right) - \bar{\alpha}_{SI} \frac{1}{(\bar{\alpha}_{N}+1)}\frac{k^2}{a^2}B_2B \right. \nonumber\\
& \left. -\bar{\alpha}_{SI} \frac{(\bar{\alpha}_{N}-1)}{(1-\bar{\alpha}_{b})} \left(\Phi\left(3\Psi_2+k^2E_2\right) +\bar{\alpha}_{b}\Phi_2\left(3\Psi+k^2E\right)\right) - 2k^4\bar{\alpha}_S E E_2 \right], \label{ST1T2NonDer}
\end{align}
where we have decomposed the line element of the second metric $f_{\mu\nu}$ as:
\begin{equation}\label{Def4Pert}
ds_f^2=-N^2r^2\left(1+2\Phi_2\right)dt^2+2 r^2\partial_i B_2 dt dx^i +b^2\left[\left(1-2\Psi_2\right)\delta_{ij}+2\partial_i\partial_j E_2\right]dx^i dx^j,
\end{equation}
where $b$ is its scale factor with Hubble rate $H_b=\dot{b}/b$, $N$ is a non-trivial shift function with an associated Hubble rate $H_N=\dot{N}/N$, and $\Phi_2$, $B_2$, $\Psi_2$ with $E_2$ are the four linear scalar perturbations of the metric. We have also introduced the scale factor ratio $r=b/a$, and the mass ratio $\nu_M^2=M_2^2/M_T^2$.

This action depends on only one independent arbitrary background function, the scale factor $a$, in addition to four independent functions that determine completely the evolution of linear scalar perturbations: one mass dimensional parameter $M_T$, and three dimensionless parameters $\bar{\alpha}_S$, $\bar{\alpha}_{SI}$ and $\bar{\alpha}_{TI}$. For simplicity, we have also introduced three dependent parameters: one mass dimensional $M_2$, and two dimensionless $\bar{\alpha}_{N}$ and $\bar{\alpha}_{b}$. They satisfy the following relations:
\begin{align}
& M_T^2\left(\bar{\alpha}_{N}-1\right)\bar{\alpha}_{SI}H^2= \delta_{F} ,\label{DefZ}\\
& 2M_T^2H^3\bar{\alpha}_{TI}(\bar{\alpha}_{b}-1)= \left((3-\bar{\alpha}_{b})H+\frac{\dot{\bar{\alpha}}_N \bar{\alpha}_{b}}{\bar{\alpha}_{N}(1-\bar{\alpha}_{b})}\right)\delta_{F} +\dot{\delta}_{F}\label{DefTildeZ},\\
& 2M_2^2H\bar{\alpha}_{b}\left(\frac{\dot{H}}{H}+\frac{\dot{\bar{\alpha}}_b}{\bar{\alpha}_{b}}-\frac{\dot{\bar{\alpha}}_N}{\bar{\alpha}_{N}} - H(\bar{\alpha}_{b}-1)\right)=-N\delta_{F} \label{L1L2relation},
\end{align} 
where we have defined $\delta_{F}$, which describes deviations from a background Friedmann equation, as:
\begin{equation}
\delta_{F}= \rho+P +2\dot{H}M_T^2.
\end{equation}
We note that the dependent parameters $\bar{\alpha}_{b}$ and $\bar{\alpha}_{N}$ describe the background evolution of the second metric $f_{\mu\nu}$. Indeed, they are defined such that $\bar{\alpha}_{b}=H_b/H$ and $\bar{\alpha}_{N}=N$, and they will in general be determined by the free parameters $\bar{\alpha}_{SI}$ and $\bar{\alpha}_{TI}$ through the constraint equations (\ref{DefZ}) and (\ref{DefTildeZ}). However, when $\bar{\alpha}_{SI}=0$ and $\bar{\alpha}_{TI}=0$ we cannot determine the value of $\bar{\alpha}_{N}$ and $\bar{\alpha}_{b}$. This is because in this case the second metric's perturbations and background decouple from the physical metric $g_{\mu\nu}$ and thus $N$ and $H_b$ become dumb parameters that do not affect the physics (and hence also $M_2$). Furthermore, we have also introduced the dependent parameters $\alpha_M$ and $\alpha_{M_2}$ which describe the running of the two mass scales $M_T$ and $M_2$:
\begin{equation}
\bar{\alpha}_M=\frac{d\ln M_T^2}{d\ln a}, \quad \bar{\alpha}_{M_2}=\frac{d\ln M_2^2}{d\ln a}.
\end{equation}

We can give a physical interpretation of the four relevant independent free parameters of this bimetric action. $M_T$ is the effective Planck mass, $\bar{\alpha}_S$ is the sound speed parameter, which directly determines the sound speed of the gravitational scalar DoF propagating in this family of bimetric theories (see Subsection \ref{StabilityTT} for an explicit expression for this sound speed). We also find $\bar{\alpha}_{SI}$, the scalar-interactions parameter, which is the main parameter determining if there is a mix of scalar perturbations between the two metrics. Indeed, as discussed in \cite{Lagos:2016gep}, when $\bar{\alpha}_{SI}=\bar{\alpha}_S=0$, then there is no scalar interaction (as the dependence on $\bar{\alpha}_{TI}$ in the parametrised action drops out after eliminating all auxiliary variables) and thus we recover the same evolution as GR for scalar perturbations. However, tensor perturbations will still interact through $\bar{\alpha}_{TI}$, the tensor-interaction parameter, which will lead to a modified evolution of gravitational waves (see Section \ref{Sec:Phenomenology}). All these parameters describe deviations from GR. Indeed, we recover linear GR when $M_T=M_P$, and $\bar{\alpha}_S=\bar{\alpha}_{SI}=\bar{\alpha}_{TI}=0$ (as the two metrics decouple from each other in this case). 

The parametrised bimetric action presented here propagates only one physical gravitational scalar DoF. This can be seen as the fields $\Phi_{A}$ and $B_{A}$ appear as auxiliary variables (we use the subscript $A$ to refer to corresponding variables in both metrics simultaneously) and thus they can be integrated out, leading to an action depending only on the fields $\Psi_{A}$ and $E_{A}$. Even though the field $\Psi_2$ appears to be a dynamical field in the original parametrised action, it becomes an auxiliary variable after integrating out $\Phi_{A}$ and $B_{A}$. Therefore, $\Psi_2$ can also be integrated out, leading to an action with three apparent dynamical fields $E$, $\Psi$ and $E_2$. However, two of these fields can be fixed by the gauge freedom to obtain only one physical remaining DoF.

Similarly as the previous subsections, we can now compare this general parametrised action to non-linear gravitational theories. A general non-linear bimetric action with no derivative interactions can be written as:
\begin{eqnarray}
S=\int d^4 x \left[\frac{M_g^2}{2} \sqrt{-g}R_g+\frac{M_f^2}{2} \sqrt{-f} R_f + \sqrt{-g}\,V(g,f) \right] \label{TTfull}
\end{eqnarray}
where the two metrics $g_{\alpha\beta}$ and $f_{\alpha\beta}$ have mass scales $M_g$ and $M_f$ and Ricci curvatures $R_g$ and $R_f$, respectively, and a generic potential interaction $V(g,f)$. As shown in \cite{Boulware:1973my}, generic forms for the potential lead to the presence of ghostly scalar modes that render the theory unstable. For this reason, here we only mention the specific case of dRGT massive bigravity, a ghost-free bimetric action given by \cite{deRham:2010kj,deRham:2010ik,Hassan:2011zd}:
\begin{equation}\label{dRGTaction}
S=\; \int d^4x\; \left[ \frac{M_g^2}{2}\sqrt{-g}R_g +\frac{M_f^2}{2}\sqrt{-f}R_f - m^2M_{g}^2 \sqrt{-g}\sum_{n=0}^4\beta_n e_n\left(\sqrt{g^{-1}f}\right) \right],
\end{equation}
where $\beta_n$ are free dimensionless coefficients, while $m$ is an arbitrary mass scale. The interaction potential is defined in terms of the functions $e_n \left(\sqrt{g^{-1}f}\right)$, which correspond to the elementary symmetric polynomials of the eigenvalues $\lambda_n$ of the matrix $\sqrt{g^{-1}f}$, which satisfies $\sqrt{g^{-1}f}\sqrt{g^{-1}f}=g^{\mu\lambda}f_{\lambda\nu}$. Explicitly, the functions $e_n(\mathbb{X})$ are given by:
\begin{align}
e_0&=1,\nonumber\\
e_1&=[\mathbb{X}],\nonumber\\
e_2&=\frac{1}{2}([\mathbb{X}]^2-[\mathbb{X}^2]),\nonumber\\
e_3&=\frac{1}{6}([\mathbb{X}]^3-3[\mathbb{X}][\mathbb{X}^2]+2[\mathbb{X}^3]),\nonumber\\
e_4&=\det(\mathbb{X})=\frac{1}{24}([\mathbb{X}]^4-6[\mathbb{X}]^2[\mathbb{X}^2]+3[\mathbb{X}^2]^2+8[\mathbb{X}][\mathbb{X}^3]-6[\mathbb{X}^4]),
\end{align}
where $\mathbb{X}$ is a matrix and $[\mathbb{X}]$ stands for the trace of $\mathbb{X}$. From these functions we notice that the terms $\beta_0$ and $\beta_4$ in eq.~(\ref{dRGTaction}) simply describe the presence of cosmological constant terms for each metric, whereas the parameters $\beta_{1,2,3}$ describe genuine interactions between both metrics.
This action is fully diffeomorphism invariant, leads to second-order derivative equations of motion, and propagates one scalar DoF, corresponding to the helicity-0 mode of a massive spin-2 particle, or graviton. More generally, this action describes the interactions of one massless and one massive graviton. Due to the properties of this theory, we expect it to be encompassed by the general parametrised action previously presented above, and indeed it is.

By calculating the quadratic action of dRGT massive bigravity, around a cosmological background, we can find five relations that completely determine the four free parameters of the quadratic action in addition to the background evolution. We find $M_T$, $\bar{\alpha}_S$, $\bar{\alpha}_{SI}$ and $\bar{\alpha}_{TI}$ to have the following expressions in dRGT massive bigravity:
\begin{align}
&\bar{\alpha}_{S}=0, \\
& \bar{\alpha}_{SI}= \frac{m^2}{H^2}r\left(\beta_1+2\beta_2r+\beta_3r^2\right), \\
& \bar{\alpha}_{TI}= \frac{m^2}{H^2}r\left(\beta_1+\beta_2r\left(1+N\right)+\beta_3r^2N\right),
\end{align}
and also $M_T=M_g$ (and $M_2=M_fr$). We notice that the parametrised action in eq.~(\ref{TotalActionNonDer}) is more general than the quadratic action of dRGT massive gravity. Such an additional freedom in the parametrised action could be describing ghostly or even non-fully diffeomorphism-invariant bimetric theories. 

Finally, we comment on the fact that some bimetric theories are not encompassed by the parametrised action in eq.~(\ref{TotalActionNonDer}), such as Eddington inspired Born-Infeld (EiBI) theory \cite{Banados:2008fj}. This is because, in order to obtain this action, we have imposed a specific constraint which is not satisfied in models such as EiBI. As shown in \cite{Lagos:2016gep}, the most general quadratic bimetric action without derivative interactions propagates indeed two scalar DoFs, and in this paper we have imposed an additional constraint to such an action in order to enforce the presence of only one physical scalar DoF. This enforcement procedure can be done in more than one way, and we have chosen one that gives the same structure as dRGT massive gravity, and hence encompasses at least that theory, but it does not encompass EiBI. The explicit parametrised action that encompasses EiBI theory can be found in \cite{Lagos:2016gep}. 


\subsection{Obstacles To Unified Action}\label{sec:unifiedaction}

In the previous subsections we have presented parametrised actions describing different families of gravity theories that introduce additional scalar, vector and tensor fields. Despite their differing field content, all of the examples considered here propagate one new physical scalar degree of freedom. Furthermore, all these models are described by one free background function (the scale factor $a$) as well as a one mass scale and a set of dimensionless $\alpha_X$ parameters that completely determine the evolution of linear scalar cosmological perturbations. In Table \ref{SummaryModels} we compile the free independent parameters found for each model. In all cases, we recover the linearised action for GR when all $\alpha_X$ vanish, the mass scale is the Planck mass. In addition, we recover the background evolution for GR when the scale factor is determined by the Friedmann equation.

\renewcommand{\arraystretch}{1.8}

\begin{table}[h!]
	\centering
	\begin{tabular}{| c | c |}
		\hline
		\rowcolor{gray!30} {\bf Theory} & {\bf Free parameters} \\
		\hline \hline
		ST & $M_S$, $\hat{\alpha}_{K}$, $\hat{\alpha}_{B}$, $\hat{\alpha}_{T}$\\
		\hline
		VT & $M_V$, $\tilde{\alpha}_{K}$, $\tilde{\alpha}_{T}$, $\tilde{\alpha}_{C}$, $\tilde{\alpha}_{D}$, $\tilde{\alpha}_{A}$, $\tilde{\alpha}_{V}$\\
		\hline
		EA & $ M_A$, $\check{\alpha}_{K}$, $\check{\alpha}_{C}$ \\
		\hline
		TT & $M_T$, $\bar{\alpha}_{S}$ , $\bar{\alpha}_{SI}$, $\bar{\alpha}_{TI}$\\
		\hline
	\end{tabular}\caption{\label{SummaryModels} Summary of the free independent parameters found for the four models studied in this paper: scalar-tensor, general vector-tensor, Einstein-Aether and bimetric gravity. In all cases there is one free mass scale, in addition to a number of dimensionless $\alpha_X$ parameters. Furthermore, in all models there is an extra background function, the scale factor $a$, needed to completely determine the evolution of linear cosmological perturbations. We omit this additional background parameter in this table.}
\end{table}

The fact that all different models have been parametrised in a similar way and propagate the same number of scalar DoFs naturally leads to the question: can one construct a unified, local, action that subsumes all of these cases into one set of parameters in order to encompass a larger family of gravity theories propagating one scalar degree of freedom?

Such an action would allow the data to inform us `of its own accord' whether observations are best-fit by a scalar-tensor, vector-tensor, or tensor-tensor theory (if any of these). One could envisage a single multi-dimensional parameter space -- a set of $\alpha_X$ -- with the data (plus some consistency conditions that enforce physical viability, see Section \ref{Sec:constraints}) singling out some preferred patch, a volume in the space of free parameters.

Ideally, we would work with this unified action, however, some obstacles prevent us from writing down such an object. The main obstacle is that the family of models considered here has a set of disjoint free parameters, that is, there are parameters that belong uniquely to each family. However, the unified action would have to include them all and, as result, end up describing theories that can have a mix of these unique parameters. In this case, the action would not correspond to a scalar-tensor, vector-tensor nor tensor-tensor action, and it could not be easily mapped onto any non-linear theory and hence should not be included in any attempts at constraining fundamentally motivated gravitational theories. 

Furthermore, the families of models studied here will have some seemingly common parameters, such as the running of the mass scale $\alpha_M$, but even these common parameters affect each action in a different way and thus they do not have a direct correspondence. For instance, while in tensor-tensor theories $\bar{\alpha}_M$ generates metric self interactions, in scalar-tensor theories $\hat{\alpha}_M$ does not. 

Finally, we also mention that since the four parametrised actions presented above have a different field content, in order to unify them we need to express them in terms of a set of common fields. It would be natural to write the unified action in terms of the single propagating scalar degree of freedom (and maybe a set of common auxiliary fields, and redundant gauge fields). In order to do this, we would need to integrate out all uncommon auxiliary fields, via their own equations of motion, in which case the resulting action in general will include inverse spatial derivative operators (or in Fourier space, wavemodes). This leads to coefficients in the action that are rational polynomials containing inverse powers of the $k$ in the denominator (see Section \ref{Sec:constraints}). As a result, the final parametrised action would be nonlocal defeating the purpose of constructing a unified, local, action.

Even with the obstructions to constructing a unified action discussed here, we will, in Section \ref{Sec:Phenomenology}, focus on the equations of motion and present a set of unified equations that describe the four parametrised models in as standardised a fashion as possible. 


\section{Stability Conditions}\label{Sec:constraints}

Generic modified gravity theories are frequently plagued by instabilities that manifest themselves as growing solutions of linear perturbations around given backgrounds. 
In this section we find conditions on the free parameters shown in the previous section to avoid such instabilities. In order to do this, we first fully reduce all the quadratic actions for scalar perturbations to the two physically propagating degrees of freedom. This is done by first imposing a specific gauge choice and then integrating out the auxiliary variables of the theory. Explicitly, in all cases we first impose the gauge choice $\Psi \to 0, E \to 0$\footnote{We use this gauge because with this choice there is no physical information lost in the action, and the number of dynamical physical DoFs is explicit (see discussions on gauge fixing the action in \cite{PhysRevD.89.024034,Motohashi:2016prk}).} and then subsequently eliminate the auxiliary fields $\Phi$ and $B$, in addition to any extra auxiliary field ($\delta A_0$ for VT theories and $B_2$, $\Phi_2$ with $\Psi_2$ for TT theories). Since all these theories propagate only one scalar gravitational DoF, we will always be able to write the actions for scalar perturbations in Fourier space as:
\begin{equation}\label{ReducedActionGral}
S^{(2)}=\int dtd^3k \left[ \vec{\dot{X}}^t \mathbb{K}\vec{\dot{X}} + \vec{\dot{X}}^t \mathbb{D}\vec{X} + \vec{X}^t \mathbb{M}\vec{X}\right],
\end{equation}
where $\vec{X}$ is a 2 dimensional vector with the gravitational and matter scalar field in its components. In addition, we have that $\mathbb{K}$, $\mathbb{D}$ and $\mathbb{M}$ are $2\times 2$ matrices with each component generically depending on time {\it and} scale $k$. Indeed, as we will find later on, in most cases they will be ratios of polynomials in $k$ with time-dependent coefficients. 

The matrix $\mathbb{K}$ will determine the kinetic energy of these perturbations (it captures all the pure time-derivative interactions of the perturbations $\vec{X}$, including kinetic mixing between the degrees of freedom). The sign of the eigenvalues of this matrix will determine the sign of the kinetic energy of each DoF, and we will diagonalise the kinetic interactions to make this explicit. We will impose positivity on both eigenvalues, thus avoiding ghost instabilities. To be more specific, generically these eigenvalues are $k$-dependent and as such the no-ghost conditions will be as well. In the main body of the paper we will only present no-ghost conditions in the large-$k$ limit (although we also collect results for the small-$k$ limit in Appendix \ref{appendix-smallKghosts}). This is mainly for two reasons: Firstly, theoretically speaking ghosts are only a problem if present in the high energy limit of the theory, since this allows (exponential) run-away behaviour (and the associated decay of the vacuum). If only present at low energies, this run-away behaviour is automatically stopped as soon as the ghost-free high energy regime is reached. If progressing too fast, such a small-$k$ instability may still be worrisome phenomenologically, but whether this is the case will be accurately determined by deriving and simulating the associated cosmology; it is not a fundamental theoretical worry. Another way of understanding this statement is the result of \cite{Gumrukcuoglu:2016jbh}, where it is argued that a small-$k$ ghost is equivalent to the presence of a tachyonic/Jeans instability. Such an instability is of course not only not worrisome, but required in order to reproduce the phenomenology of gravitational collapse. Again, the time-scale of the instability would need to satisfy observational bounds (where $m_i^2$ in \eqref{asymLag} below should be positive or otherwise satisfy $|m_i^2| \lsim H^2$, see \cite{DeFelice:2016ucp} for details), but the mere presence of the tachyonic (or small-$k$ ghost) instability is not a diagnostic signalling sickness of the theory. 

Finally, one may wonder whether this reasoning also extends to large-$k$ ghosts. After all, the theories we consider are effective (field) theories and, even if a ghost is present in the large-$k$ limit of this effective theory, an eventual UV completion can perhaps cure this at even higher energies. 
However, note that if the effective theories we consider are to be valid at and below the Jeans' length, a ghost in the high energy limit of our theory would immediately rule out the theory's phenomenology, since (by the same argument as above) it would generically correspond to a (phenomenologically highly constrained) gradient instability\footnote{Note that this assumes that when taking the large-k limit of the action the mass term will go as $m_i^2\propto k^2$. If, for any reason, in this limit the mass term does not have any $k$ dependence then the high energy ghost would correspond to a tachyon instability and, as before, we would have to check how fast it evolves.}. If these scales are to be within the remit of the theory, this means that requiring the absence of a large-$k$ ghost is a minimal requirement to ensure the theory is a viable candidate for cosmology\footnote{Note that, technically speaking, all of the above reasoning only directly applies to the (reduced) matter field. If $\mathbb{D}$ is order one, then the same reasoning will apply to the `dark energy' field as well, since the dark energy field would feed energy into the matter field via the $\mathbb{D}$-dependent coupling then. If, on the other hand, $\mathbb{D} = 0$, as far as the linear theory is concerned, it does not matter whether the dark energy field has instabilities or not, since it is completely decoupled from matter when $\mathbb{D} = 0$ (at the linear level).}. 

Having ensured the absence of large-$k$ ghosts in this way, we can now also diagonalise $\mathbb{M}$. While $\mathbb{K}$ and $\mathbb{M}$ may not commute and be simultaneously diagonalisable at first sight, it is nevertheless always possible to simultaneously diagonalise kinetic and mass interactions. This is the case, because, upon canonically normalising the fields, $\mathbb{K}$ essentially becomes the identity matrix\footnote{Also note that $\mathbb{M}$ can of course always be written as a symmetric matrix.}. This ensures commutativity of the resulting $\mathbb{K}$ and $\mathbb{M}$ and hence simultaneous diagonalisability. 
This leaves us with the mixed derivative/non-derivative interactions, encoded by the $\mathbb{D}$ matrix, which (after appropriate integration by parts) can always be written as an antisymmetric matrix. In Appendix \ref{appendix-asym} we show that one can then always rewrite eq.~(\ref{ReducedActionGral}) as:
\begin{equation} \label{asymLag}
S^{(2)}=\int dtd^3k \left[ \dot{\tilde{X}}^2_1 + \dot{\tilde{X}}^2_2 + d(t,k)\left(\dot{\tilde{X}}_1\tilde{X}_2-\dot{\tilde{X}}_2\tilde{X}_1\right) -m_1(t,k)^2\tilde{X}_1^2-m_2(t,k)^2\tilde{X}_2^2 \right],
\end{equation}
where $\tilde{X}_1$ and $\tilde{X}_2$ are the new fields after performing the above-mentioned field redefinitions on \eqref{ReducedActionGral}. In principle, one might think that the $d$-dependent mixing term could affect the presence of instabilities. However, while it does affect the precise evolution of the fields, as shown in \cite{DeFelice:2016ucp}, positive-definiteness of the Hamiltonian of such a system (and hence whether it contains instabilities) does not depend on the function $d(t,k)$. Explicitly, the Hamiltonian can be written as
\begin{equation}
H=\dot{\tilde{X}}^2_1 + \dot{\tilde{X}}^2_2 + m_1(t,k)^2\tilde{X}_1^2 + m_2(t,k)^2\tilde{X}_2^2.
\end{equation}
This, of course, is not expressed in canonical Hamiltonian form (we have kept the $\dot{\tilde X}_i$ instead of expressing them in terms of canonical momenta). However, what is important is that the Hamiltonian can be expressed as a quadratic form that is manifestly independent of $d$. Therefore $d$ does not affect the (un-)boundedness of the Hamiltonian, and hence the stability of the perturbations. As such we will be able to ignore $d$ in the instability analysis of this section. 

Finally the matrix $\mathbb{M}$, in its diagonalised form given via $m_1(t,k)^2$ and $m_2(t,k)^2$, encodes all spatial gradient (i.e.~$k$-dependent) interactions and mass terms for both fields. If any $m_i^2$ is negative in the large-$k$ limit, we will say that a gradient instability is present. For this reason, we will also find and impose the conditions ensuring an absence of gradient instabilities here.\footnote{Analogously to the ghost analysis described above, imposing this condition on both fields, i.e.~also on the dark energy field, implicitly assumes that $d$ is not fine-tuned to suppress the interaction between both fields so much, that the observationally relevant matter field does not feel the effect of a gradient instability in the dark energy field.} As mentioned above, we will not require strict positivity of $\mathbb{M}$ in the low-energy limit, as we need to allow tachyonic instabilities in any viable model. As noted above, however, one could (and, indeed, should) impose a ``minimal'' tachyonic ghost-freedom condition along the lines of \cite{DeFelice:2016ucp}, requiring $m_i(t,k)^2 \gtrsim - H^2$ (for negative $m_i^2$ and small $k$). This would not impose the absence of tachyonic instabilities altogether, but would ensure an absence of strong tachyonic instabilities which progress too fast to mimic a Jeans-like instability. 

In what follows we will present explicit constraints on the free parameters of each action presented in the previous section in order to avoid ghosts and gradient instabilities for large $k$. We summarise these results in Tables \ref{SummaryStability} and \ref{GradientStability}. Details on the precise procedure to derive these constraints are collected in Appendix \ref{appendix-stability}. 

\renewcommand{\arraystretch}{1.8}
\begin{table}[h!]
\centering
 \begin{tabular}{| c | c |}
 \hline
 \rowcolor{gray!30} {\bf Theory} & {\bf Ghost Freedom Conditions} \\
 \hline \hline
 ST & $3\hat{\alpha}_B^2 + 2\hat{\alpha}_K > 0$\\
 \hline
VT I & $(\tilde{\alpha}_\textrm{D} - 2)/\tilde{\alpha}_\textrm{D} > 0$\\
 \hline
VT II & $\mathcal{K}_1/\mathcal{K}_2 > 0$\\
 \hline
EA I & $ (\check{\alpha}_D-2)/ \check{\alpha}_\textrm{D} > 0$\\
 \hline
EA II & $\check{\alpha}_K>0$\\
 \hline
TT & $\bar{\alpha}_{SI}(\bar{\alpha}_N-1)(\bar{\alpha}_N^2-\bar{\alpha}_b^2) < 0$\\
 \hline
 \end{tabular}\caption{\label{SummaryStability}Summary of the large-$k$ ghost stability conditions for the four models studied in this paper: scalar-tensor, general vector-tensor, Einstein-Aether and bimetric gravity. VT I and EA I are the general cases whereas VT II and EA II refer to the special case of $\alpha_D = 0$, that corresponds to scenarios encompassing Generalised Proca theories for VT and to imposing a $\Lambda$CDM background for EA. The full expressions for $\mathcal{K}_{1,2}$ are given in eq.~(\ref{KinVTProca}). In the case of scalar-tensor theories there is only one simple condition that ensures ghost freedom for arbitrary $k$.}
\end{table}

\renewcommand{\arraystretch}{2}
\begin{table}[h!]
	\begin{center}
		\begin{tabular}{| c | c |}
			\hline
			\rowcolor{gray!30} {\bf Theory} & {\bf Gradient Instability Freedom Conditions} \\
			\hline \hline
			ST & $\frac{\left( \hat{\alpha}_\textrm{B} + 2 \hat{\alpha}_\textrm{M} - 2 \hat{\alpha}_\textrm{T} + \hat{\alpha}_\textrm{B} \hat{\alpha}_\textrm{T}\right) H^2 M_S^2}{\left(\hat{\alpha}_\textrm{B} - 2\right)} < \frac{2 H M_S^2 \dot{\hat{\alpha}}_\textrm{B} + 2 \left(\hat{\alpha}_\textrm{B} - 2\right) M_S^2 \dot{H} - 2 (\rho+ P)}{\left(\hat{\alpha}_\textrm{B} - 2\right)^2}$ \\
			\hline
			VT I & $ M_{{\rm VT}, k\to \infty}^2 > 0$, cf.~\ref{extraconditions} \\
			\hline
			VT II & $\mu_1 \left(\frac{\mu_1 \mu_2}{\mu_M^2} - 1\right) > 0$ and $M_1 > 0$, cf. \eqref{VTIIgradKlim1} and \eqref{VTIIgradKlim2}\\
			\hline
			EA I & $\dot{\check{\alpha}}_\textrm{C} H + \left(1 + \check{\alpha}_\textrm{C}\right) \left(1 + \check{\alpha}_\textrm{M} -\frac{(2 + 2 \check{\alpha}_\textrm{C})}{\check{\alpha}_\textrm{K}} \right) H^2 < 0$ \\
			\hline
			EA II & 
			$\mu_1 \left(\frac{\mu_1 \mu_2}{\mu_M^2} - 1\right) > 0$ and 
			$\check{\alpha}_\textrm{K} (\rho + P) > 4 H^2 M_A^2$, cf.~\eqref{EAIImus}		\\ 
			\hline
			TT I & $\bar{\alpha}_S<0$ \\
			\hline
			TT II & $ M_{{\rm TT}, k\to \infty}^2 > 0$, cf.~\ref{extraconditionsTT} \\
			\hline
		\end{tabular}\caption{\label{GradientStability}Summary of the large-$k$ gradient instability freedom conditions for the four models studied in this paper: scalar-tensor, general vector-tensor, Einstein-Aether, and bimetric gravity. VT I, EA I, and TT I are the general cases whereas VT II and EA II refer to the special case of $\alpha_D = 0$, that correspond to scenarios encompassing Generalised Proca theories for VT and to imposing a $\Lambda$CDM background for EA. TT II refers to the case $\bar{\alpha}_S=0$, encompassing dRGT massive bigravity. The explicit expressions for $M_{{\rm VT}, k\to \infty}^2$ and $ M_{{\rm TT}, k\to \infty}^2$ are given in Appendix \ref{extraconditions} and \ref{extraconditionsTT}, respectively. The results for VT II and EA II have been given in terms of $\mu_i,M_i$ (see the discussions around equations \eqref{VTIIgradKlim1} and \eqref{EAIIgradKlim}).}
	\end{center}
\end{table}

\subsection{Scalar-Tensor Theories}

In the scalar-tensor case we find the reduced action for the only two physical degrees of freedom as previously explained. Then, we focus on the kinetic matrix and we find that after diagonalising it, the reduced action has the following kinetic terms:
\begin{align}
S^{(2)}_{\rm kin} &= \int d^3k dt a^3 \left[\frac{1}{2} \delta \dot{\tilde{\varphi}}^2 + \frac{\left(3 \hat{\alpha}_\textrm{B}^2 + 2 \hat{\alpha}_\textrm{K}\right) H^2 M_S^2 \dot{\tilde{v}}_X^2}{\left(\hat{\alpha}_\textrm{B} - 2\right)^2}\right],\label{STghost}
\end{align}
where $\delta \tilde{\varphi}$ and $ \tilde{v}_X$ are the two fields that diagonalise the kinetic matrix, and are linear combinations of the matter field $\delta \varphi$ and dark energy field $ v_X$. These kinetic terms are general, i.e.~we have not used any approximation in deriving them and hence these expressions are valid for all wavelengths. Whereas the matter field $\delta \tilde{\varphi}$ is trivially stable, we can read off a non-trivial ghost-freedom condition from eq.~(\ref{STghost}), ensuring that the kinetic term for $\tilde{v}_X$ is positive.

For the mass terms we find that the general expressions are involved, but in the large-$k$ limit they approximate to:
\begin{align}\label{STgradient}
S^{(2)}_{\rm mass, k\rightarrow \infty} = \int d^3k dt a \Big[ &- \frac{1}{2} k^2 \delta \tilde{\varphi}^2 \nn + k^2 {\tilde{v}}_X^2 \left( \frac{\left( \hat{\alpha}_\textrm{B} + 2 \hat{\alpha}_\textrm{M} - 2 \hat{\alpha}_\textrm{T} + \hat{\alpha}_\textrm{B} \hat{\alpha}_\textrm{T}\right) H^2 M_S^2}{\left(\hat{\alpha}_\textrm{B} - 2\right)} \right. \\ &- \left. \frac{2 H M_S^2 \dot{\hat{\alpha}}_\textrm{B} + 2 \left(\hat{\alpha}_\textrm{B} - 2\right) M_S^2 \dot{H} - 2 (\rho+ P)}{\left(\hat{\alpha}_\textrm{B} - 2\right)^2} \right) \Big].
\end{align}
Note that these interactions are automatically diagonalised, i.e.~mass mixing terms drop out in the large-$k$ limit (this would not be the case in the small-$k$ limit). In addition, we mention that kinetic-mass mixing as captured by the ${\mathbb{D}}$ matrix is present and we fix any integration by parts ambiguities by requiring the $\mathbb{D}$ matrix to be antisymmetric\footnote{Note that here, unlike in some of the other cases below, the large $k$ mass matrix is insensitive to how we choose to write the kinetic-mass mixing.}. We will use the same procedure in all the following cases. Similarly as before, the mass term for $\delta\tilde{\varphi}$ is trivially stable, but we can read off a non-trivial expression for the effective speed of sound of $\tilde{v}_X$ in eq.~(\ref{STgradient}). By enforcing this speed to be positive we ensure a gradient-instability free model. We explicitly state the resulting ghost-freedom constraints in Table \ref{SummaryStability} and conditions ensuring the absence of gradient instabilities in Table \ref{GradientStability}, for all models considered in Section \ref{Sec:Models}.

Finally, we discuss the special case of $\hat{\alpha}_B=2$. Looking at the above results, indeed the expressions for both the kinetic and mass terms diverge when $\hat{\alpha}_B=2$ and hence are not valid anymore. In order to obtain the correct expressions for this limit, one has to go back to the original scalar-tensor quadratic action \eqref{STperts}, impose $\hat{\alpha}_B=2$ there directly and gauge fix as before. Crucially, in the resulting action, setting $\hat{\alpha}_B=2$ eliminates the $\Phi B$ mixing term, which turns $B$ into a Lagrange multiplier (while the auxiliary field $\Phi$, which appears quadratically, can be solved for via its own equation of motion). As a result the $\hat{\alpha}_B=2$ scalar-tensor theory propagates no gravitational scalar degree of freedom, i.e. there is no dark energy here\footnote{It is worth emphasising, that it was not necessary to take any limit in $k$ in order to arrive at this conclusion.}. Therefore, such a case will not be of interest in this paper.

\subsection{Vector-Tensor Theories}
As before, we consider the two main sub-classes of vector-tensor theories in turn.
\subsubsection{General Vector-Tensor Theories}

In the general vector-tensor case we proceed as for scalar-tensor theories, except that we now have an additional auxiliary variable (the temporal component $\delta A_0$ of the vector field) to integrate out, which creates non-local terms in the reduced action. 
Compared to scalar-tensor theories, we also have to deal with a new complication: as before, we have two physical degrees of freedom in the reduced action, but their `native' scaling with $k$ is different. While $\delta A_1$ always enters with a $k$ (since these perturbations originally always enter together with a spatial derivative), $\delta\phi$ does not. In order to have a fair counting when taking $k$-limits, we will therefore re-scale $\delta A_1 \to \delta \tilde{A}_1/k$.\footnote{This is analogous to treating e.g.~$\Phi$ and $k^2 E$ on par, when considering scalar perturbations of the metric.} As we will see below, this scaling will be undone in some cases when considering the large-$k$ limit and (canonically) normalising $\delta A_1$ in this limit. 
After diagonalising the kinetic matrix, we then find the following schematic kinetic terms:
\begin{equation}\label{VTGenKin}
	S^{(2)}_{\rm kin} = \int d^3k dt \left[ \frac{\left(A_0+A_2k^2\right)}{\left(B_0+B_2k^2+B_4k^4\right)}\delta \dot{\tilde{A}}_1^2 + \frac{\left(C_0+C_2k^2+C_4k^4\right)}{\left(D_0+D_2k^2+D_4k^4\right)} \delta \dot{\tilde{\varphi}}^2 \right], 
\end{equation} 
where all the coefficients $A_i$, $B_i$, $C_i$ and $D_i$s are functions of time only. 
In the large-$k$ limit, which we are primarily interested in, these greatly simplify. There, we find the following kinetic interactions
\begin{align}\label{VTghostLargek}
	S^{(2)}_{{\rm kin}, k\rightarrow \infty} &= \int d^3k dt \; a^3 
	\left[ \frac{3 \left(\tilde{\alpha}_\textrm{D} - 2\right) H^2 M_V^2}{\tilde{\alpha}_\textrm{D}}\delta \dot{\tilde{A}}_1^2 + \frac{1}{2} \delta \dot{\tilde{\varphi}}^2 \right],
\end{align}
where we have sent $\delta A_1 \to k \delta \tilde{A}_1$ to normalise the k-dependence of all fields (clearly we have not fully canonically normalised at this point, since we want to read off the large-$k$ ghost stability conditions).
We see that $\delta \tilde{\varphi}$ is trivially non-ghostly, whereas $\delta \tilde{A}_1$ imposes a simple but non-trivial ghost-freedom condition, which can be read off from the action \eqref{VTghostLargek}. 

Next, we look at the mass matrix in the large-$k$ limit. Similarly as in the scalar-tensor case, we do not need to explicitly diagonalise this matrix as the off-diagonal terms are subdominant in this limit. We obtain:
\begin{align} \label{VTmass}
	S^{(2)}_{\rm mass,k\rightarrow \infty} &= -\int d^3k \frac{dt}{a} 
	\left[\frac{1}{2} k^2 \delta \tilde{\varphi}^2 a^2 + k^2 M_{{\rm VT}, k\to \infty}^2\delta \tilde{A}_1^2\right].
\end{align}
We find that $\delta \tilde{\varphi}$ is trivially stable but $\delta \tilde{A}_1$ imposes a complicated condition that is explicitly given in Appendix \ref{extraconditions}. 
Kinetic-mass mixing interactions are also present (and are $k$-independent), but have been put into the anti-symmetric form discussed above, which means we can ignore them for the stability analysis. 
Finally note two implicit assumptions we have made: First we have assumed $\tilde{\alpha}_D \neq 0$. What happens when this condition is violated we will discuss below. 
Secondly, in deriving \eqref{VTghostLargek} we have assumed $\tilde{\alpha}_D \neq 2$. If instead $\tilde{\alpha}_D = 2$, note that this does not mean that only one field propagates, but instead a term scaling as $1/k^2$ then becomes the dominant kinetic prefactor for $\delta A$ in the large-$k$ limit (which of course needs to be taken into account when canonically normalising fields). 

Somewhat analogously to the scalar-tensor case, we identify a special case ($\tilde{\alpha}_D=0$), where one of the kinetic coefficients diverges. 
As shown in the previous section (see eq.~(\ref{AlphaDProca})), this case in fact encompasses the whole family of Generalised Proca theories and is therefore of special interest. Going back to the general quadratic vector-tensor theory, we therefore set $\tilde{\alpha}_D=0$, gauge fix and integrate auxiliary fields out as before. Note that, unlike for the scalar-tensor case, the special case in question does not render any auxiliary field into a mere Lagrange multiplier here, so the $\tilde{\alpha}_D=0$ case does still propagate a dynamical `dark energy' helicity-0 degree of freedom. 
Taking the large-$k$ limit eliminates any kinetic mixing (this goes as $1/k$ and is suppressed) and the kinetic terms explicitly become: 
\begin{align}\label{VTghostLargekProca}
	S^{(2)}_{{\rm kin}, k\rightarrow \infty} &= \int d^3k dt a 
	\left[ \frac{{\cal K}_1}{{\cal K}_2}\delta \dot{\tilde{A}}_1^2 + \frac{1}{2} a^2 \delta \dot{\tilde{\varphi}}^2 \right],
\end{align}
where ${\cal K}_1,{\cal K}_2$ are given by
\begin{align}
	{\cal K}_1 &= H^2 M_V^4 \left(2 \tilde{\alpha}_\textrm{A}^2 H^2 M_V^2 + \tilde{\alpha}_\textrm{K} \left({\cal C}-2 (\rho+P) - 4 \dot{H} M_V^2\right)\right), \nn \\
	{\cal K}_2 &= 2 H^2 M_V^2 \left(2 \left(\tilde{\alpha}_\textrm{A} - 1\right) (\rho+P + 2 \dot{H} M_V^2) + {\cal C}\right) -\tilde{\alpha}_\textrm{K} \left(\rho+P + 2 \dot{H} M_V^2\right)^2,
	\label{KinVTProca} 
\end{align}
and where we have used the shorthand notation
\begin{equation}
	{\cal C} \equiv H \left(4 \dot{\tilde{\alpha}}_\textrm{C} + \left(4 \tilde{\alpha}_\textrm{M} + 4 \tilde{\alpha}_\textrm{C} \left(1 + \tilde{\alpha}_\textrm{M}\right) - 4 \tilde{\alpha}_\textrm{T} + \tilde{\alpha}_\textrm{V}\right) H\right) M_V^2.
\end{equation}
This leads to a single non-trivial condition for avoiding ghosts in $\tilde \alpha_D = 0$ theories, namely ${\cal K}_1/{\cal K}_2 > 0$. Note that these theories encompass Generalised Proca theories, but are still much richer, since the additional constraints \eqref{MassVT}-\eqref{AlphaDProca} have not been imposed and we have also not imposed additional Generalised Proca-specific constraints arising from considering which branches of solutions exist or via considering and substituting model-specific background equations of motion \cite{DeFelice:2016yws}. 

From \eqref{KinVTProca} we can already see that the expression for the large-$k$ kinetic terms is significantly more cumbersome for the special $\tilde \alpha_D = 0$ case, than in the general vector-tensor case discussed above. 
Analogously, extracting the mass matrix and resulting gradient stability conditions also requires a little bit more work. 
We now need to maximally diagonalise the full action to pick out the correct eigenmodes, extensively using the results of Appendix \ref{appendix-asym}. Doing so, we use the two general gradient stability conditions (for both modes) from the appendix 
\begin{align}
	\left(1 + c_2^2\right) \mu_1 + c_2^4 \mu_2 + 2 c_2 \mu_M + 2 c_2^3 \mu_M - \left(D + \dot c_2\right) \dot c_2 + c_2^2 \left(\mu_2 - D \dot c_2\right) &> 0, \nn \\
	c_2^4 \mu_1 + \mu_2 - 2 c_2 \mu_M - 2 c_2^3 \mu_M - \left(D + \dot c_2\right) \dot c_2 + c_2^2 \left(\mu_1 + \mu_2 - D \dot c_2\right) &> 0,
	\label{GenGradCon}
\end{align}
where $c_2$ is given by
\be
c_2 = \frac{\mu_2 - \mu_1}{2 \mu_M} \pm \sqrt{\frac{\left(\mu_2 - \mu_1\right)^2}{4 \mu_M^2} + 1},
\ee
and $\mu_1,\mu_2,\mu_M$ are the diagonal and mass-mixing elements in the canonically normalised and kinetically diagonalised action (where kinetic-mass interactions have been put into the anti-symmetric form discussed above). They are defined to be
\begin{align}
	\mu_1 &\equiv \frac{4 K_1 M_1 - 2 K_1 \ddot K_1 + (\dot K_1)^2}{4 K_1^2},
	&\mu_2 &\equiv \frac{4 K_2 M_2 - 2 K_2 \ddot K_2 + (\dot K_2)^2}{4 K_2^2},\nn \\
	D &\equiv \frac{\tilde D}{\sqrt{K_1 K_2}},
	&\mu_M &\equiv \frac{4 K_1 K_2 M_M + \tilde D K_2 \dot K_1 - \tilde D K_1 \dot K_2}{4 K_ 1^{\frac{3}{2}} K_ 2^{\frac{3}{2}}}, 
	\label{VTIIgradKlim1}
\end{align}
where $K_1 = \tfrac{1}{2}a^3, \;  K_2 = a \tfrac{{\cal K}_1}{{\cal K}_2}$ and the expressions for $M_1,M_2,M_M,\tilde D$ are rather cumbersome, so we will not show them here explicitly, but (for reference) they can be found under the following
\href{https://github.com/noller/General-Theory-of-Linear-Cosmological-Perturbations}{link}.
Note, however, that \eqref{GenGradCon} mixes terms with different k-dependencies. Since we are only interested in the large $k$-limit, eliminating all but the leading order terms in $k$ in \eqref{GenGradCon}, we find that the gradient stability conditions reduce to
\begin{align}
	\mu_1 \left(\frac{\mu_1 \mu_2}{\mu_M^2} - 1\right) &> 0,
	&M_1 &> 0,
	\label{VTIIgradKlim2}
\end{align}
where we have assumed a positive scale factor and hence $K_1 > 0$ in the second condition (otherwise it reads $M_1/K_1 > 0$).

This concludes the main analysis for this case, but one may wonder why we encounter this increased complexity, when specialising to $\tilde \alpha_D = 0$, and indeed it is straightforward to understand why this is the case: Going back to the kinetic interactions, contributions that go as $k^4$ are all proportional to the $\tilde \alpha_D$ parameter. 
This makes a lot of sense, when recalling that Generalised Proca theories (a defining feature of which is $\tilde \alpha_D = 0$) do not contain (spatial) higher derivative interactions that would lead to such $k^4$ terms in the (kinetic part of the) linearised action.
Excluding kinetic interaction coefficients that involve $\tilde \alpha_D$, the other $\tilde \alpha$ parameters only contribute to $k$ powers $\leq 2$. Indeed it is straightforward to check that $B_4, C_4, D_4 \propto \tilde \alpha_D$ in \eqref{VTGenKin}, i.e.~all those $k^4$ coefficients vanish when we set $\tilde \alpha_D = 0$. Large-$k$ kinetic terms in the general $\tilde \alpha_D \neq 0$ case are therefore dominated by the small subset of $\tilde \alpha_D$-dependent terms, resulting in very simple and concise expressions. In the $\tilde \alpha_D = 0$ case, on the other hand, a plethora of contributions from other $\tilde \alpha$ parameters survives in the large-$k$ limit, resulting in far more cumbersome expressions. The same logic holds for the large-$k$ mass terms and leads to far more involved diagonal mass terms. In addition, off-diagonal mass-mixing terms are no longer suppressed, when setting $\tilde \alpha_D = 0$. 

\subsubsection{Einstein-Aether Theories}
 Proceeding as in the previous cases, we obtain the reduced action and focus first on the kinetic matrix. We note that, similarly to the ST case, there are no extra auxiliary fields to integrate out in order to obtain the reduced action. Nevertheless, due to the different interaction terms we do obtain non-local terms as in the general VT case. Also, as before, we will re-scale $\delta A_1 \to \delta \tilde{A}_1/k$. 
After diagonalising the kinetic matrix, we find the following schematic structure for the kinetic terms:
\begin{equation}
S^{(2)}_{\rm kin}= \int d^3k dt \; \left[ \frac{A_{0}}{B_0+B_{2}k^2} \delta \dot{\tilde{A}}_1^2 + \frac{C_0+C_{2}k^{2}}{D_0+D_{2}k^{2}}\delta \dot{\tilde{\varphi}}^2\right],
\end{equation}
where all the coefficients $A_i$, $B_i$, $C_i$ and $D_i$s are functions of time only. In the large-$k$ limit the kinetic structure of the theory is then as follows
\begin{align}
S^{(2)}_{\rm kin, k\rightarrow \infty} &= \int d^3k dt a^3 
\left[
\frac{1}{2} \dot{\delta \tilde\varphi}^2 + \frac{ 3M_A^2H^2(\check{\alpha}_D-2)}{ \check{\alpha}_\textrm{D}}\delta \dot{\tilde{A}}_1^2 
 \right], \label{EAkin}
\end{align}
where we 
have sent $\delta A_1 \to k \delta \tilde{A}_1$ to normalise the k-dependence of all fields as above. As in the previous cases, the kinetic terms for $\delta\tilde{\varphi}$ in the large-$k$ limit is always positive, whereas the kinetic term for $\delta\tilde{A}_1$ generates a non-trivial constraint that depends solely on background quantities (recall that $\check{\alpha}_D$ is a dependent parameter given by eq.~(\ref{DependentAlphasEA})). 

Kinetic-mass mixing interactions are present as usual, and (in the large-$k$ limit) they scale $k$-independently and have been brought into anti-symmetric form. Turning our attention to the mass matrix, these potential interactions are diagonal in the large-$k$ limit, since the diagonal terms scale as $k^2$, whereas the mass-mixing term is $k$-independent and therefore automatically drops out. The large $k$ limit mass terms then are
\begin{align}
S^{(2)}_{\rm mass, k\rightarrow \infty} &= \int d^3k dt \; a k^2
\left[ - \frac{1}{2}\delta \tilde{\varphi}^2 a^2 + M_A^2\delta \tilde{A}_1^2 \left(\dot{\check{\alpha}}_\textrm{C} H +\right.\right. \nonumber \\ 
&\left. \left. \left(1 + \check{\alpha}_\textrm{C}\right) \left(1 + \check{\alpha}_\textrm{M} -\frac{(2 + 2 \check{\alpha}_\textrm{C})}{\check{\alpha}_\textrm{K}} \right) H^2\right) \right],
\label{EAmass}
\end{align}
where we have used eq.~\eqref{DependentAlphasEA} to simplify expressions again.
From this action we see that the mass term for $\delta\tilde{\varphi}$ is trivially well behaved while from the mass term for $\delta \tilde{A}_1$ we can extract a positivity condition on the associated speed of sound, i.e.~a condition for the absence of a gradient instability.
As before we have so far assumed $\tilde{\alpha}_D \neq 0$ and will again consider the special case when this condition is violated below. 
Note that we have implicitly assumed that $\tilde{\alpha}_K \neq 0$ in deriving \eqref{EAmass} -- otherwise the expression for the large-$k$ mass matrix changes (and in this way avoids diverging when $\tilde{\alpha}_K = 0$).
\\

This completes the stability analysis for the general Einstein-Aether case, but from the previous results we notice that there is one special case: when $\check{\alpha}_\textrm{D}=0$. This special case is in fact particularly interesting for cosmology as it corresponds to a $\Lambda$CDM background evolution. Setting $\check{\alpha}_\textrm{D}=0$ also changes which terms dominate in the large $k$ limit, so we here re-compute this limit and extract the stability conditions for this special case.
The kinetic matrix now takes on a much simpler form and kinetic interactions are independent of $k$ and given by
\begin{align}
S^{(2)}_{\rm kin, k\rightarrow \infty} &= \int d^3k dt a 
\left[ 
\frac{1}{2} \check{\alpha}_\textrm{K} M_A^2 \dot{\delta \tilde{A}}_1^2
+ \frac{1}{2} \dot{\delta \tilde{\varphi}}^2 a^2 \right].
\end{align}
There is therefore only one simple condition to ensure ghost freedom at any scale, namely $\check{\alpha}_\textrm{K} > 0$. Note that the kinetic interactions are automatically diagonalised (independently of scale) here, since the kinetic mixing term in the full EA theory was proportional to $\check{\alpha}_\textrm{D}$. 

Moving on to the mass matrix, we find that the off-diagonal terms are now no longer subdominant with respect to the diagonal ones in the large-$k$ limit. Explicitly, upon ensuring that the kinetic-mass mixing is written in an anti-symmetric format, we find the following mixed mass interactions
\begin{align}
S^{(2)}_{\rm mass,k\rightarrow \infty} &= - \int d^3k dt \; a \left[
\frac{\left(2 H^2 +\dot{H} \check{\alpha}_\textrm{K} \right)}{4 H^2}k^2 \delta \tilde{\varphi}^2
+ \check{\alpha}_\textrm{C} \dot{H} M_A^2 \delta \tilde{A}_1^2 \right.\nn \\
&+\left. \frac{\dot{\varphi} H \left(\dot{\check{\alpha}}_\textrm{K} - 4 \check{\alpha}_\textrm{C} H\right) - \check{\alpha}_\textrm{K} \left(\dot{H} \dot{\varphi} - H \left(\ddot{\varphi} + \dot{\varphi} H\right)\right)}{4 H^2}k \delta \tilde{A}_1 \delta \tilde{\varphi}\right].
\label{EAII}
\end{align} 
Since this is still in a mixed format, it is important to also keep track of the kinetic-mass mixing, which here takes on the following form
\begin{align}
S^{(2)}_{\rm kinetic-mass,k\rightarrow \infty} &= \int d^3k dt \; a \left(\frac{\check{\alpha}_\textrm{K} \dot{\varphi} k}{4 H}\right) \left(\delta \tilde{\varphi} \delta \dot{\tilde{A}}_1 - \delta \dot{\tilde{\varphi}} \delta \tilde{A}_1\right).
\end{align} 
We now need to maximally diagonalise the full action to pick out the correct eigenmodes, just as for VT II extensively using the results of Appendix \ref{appendix-asym}, especially the general gradient stability condition \eqref{GenGradCon}. In the Einstein-Aether case we now have
\begin{align}
\mu_1 &\equiv \frac{4 K_1 M_1 - 2 K_1 \ddot K_1 + (\dot K_1)^2}{4 K_1^2},
&\mu_2 &\equiv \frac{4 K_2 M_2 - 2 K_2 \ddot K_2 + (\dot K_2)^2}{4 K_2^2},\nn \\
D &\equiv \frac{\tilde D}{\sqrt{K_1 K_2}},
&\mu_M &\equiv \frac{4 K_1 K_2 M_M + \tilde D K_2 \dot K_1 - \tilde D K_1 \dot K_2}{4 K_ 1^{\frac{3}{2}} K_ 2^{\frac{3}{2}}}, \nn \\
K_1 &\equiv \frac{1}{2} a^3,
&K_2 &\equiv \frac{1}{2} \check{\alpha}_\textrm{K} M_A^2 a,\nn \\
M_1 &\equiv \frac{ (\check{\alpha}_\textrm{K} \dot{\varphi}^2 - 4 H^2 M_A^2)}{8 H^2 M_A^2}a k^2 
, &M_2 &\equiv \check{\alpha}_\textrm{C} \dot{H} M_A^2 a,\nn \\
\tilde D &\equiv -\frac{\check{\alpha}_\textrm{K} \dot{\varphi} k a}{4 H},
&M_M &\equiv \frac{\dot{\varphi} H \left(\dot{\check{\alpha}}_\textrm{K} - 4 \check{\alpha}_\textrm{C} H\right) - \check{\alpha}_\textrm{K} \left(\dot{H} \dot{\varphi} - H \left(\ddot{\varphi} + \dot{\varphi} H\right)\right)}{8 H^2}k a.
\label{EAIImus}
\end{align}
In terms of these quantities and in the large-$k$ limit, the general gradient stability conditions \eqref{GenGradCon} for EA II then reduce to
\begin{align}
\mu_1 \left(\frac{\mu_1 \mu_2}{\mu_M^2} - 1\right) &> 0,
&\check{\alpha}_\textrm{K} (\rho + P)  &> 4 H^2 M_A^2,
\label{EAIIgradKlim}
\end{align}
where we have assumed a positive scale factor and ghost-freedom in the second condition.

The analysis for this second Einstein-Aether case is clearly more involved than in the previous examples, primarily since the large $k$ limit in this case did not automatically diagonalise both kinetic and mass interactions. 
This concludes the stability analysis for Einstein-Aether theories and we now move on to tensor-tensor theories. 

\subsection{Tensor-Tensor Theories}\label{StabilityTT}

We proceed as before and analyse the reduced action. We note that, unlike the ST and EA cases, we need to integrate out three extra auxiliary fields from the second metric, namely $\Phi_2$, $B_2$ and $\Psi_2$. This complicates greatly the form of the reduced action and as a result we find a number of non-local terms, as in the VT case. Similarly to the VT case, we rescale the field $E_2$ in order to have a fair counting of the $k$ powers in all terms. We redefine $E_2\rightarrow\tilde{E}_2/k^2$ After diagonalising the kinetic matrix, we find the following schematic form for the kinetic terms in the quadratic action:
\begin{align}
S^{(2)}_{\rm kin} = \int d^3kdt\; & \left[ \frac{\dot{ \tilde{E}}_2^2 \left(A_0 + A_2 k^2 + A_4 k^4 + A_6 k^6+ A_8 k^8+ A_{10} k^{10}+ A_{12} k^{12}+ A_{14} k^{14}\right)}{\left(B_0 + B_2 k^2 + B_4 k^4+ B_6 k^6+ B_8 k^8+ B_{10} k^{10}+ B_{12} k^{12}+ B_{14} k^{14}+ B_{16} k^{16}\right)} \right. \nn \\
&+ \left. \delta\dot{\tilde{\varphi}}^2 \frac{\left(C_0 + C_2 k^2 + C_4 k^4\right)}{\left(D_0 + D_2 k^2 + D_4 k^4\right)} \right],
\end{align}
where all the coefficients $A_i$, $B_i$, $C_i$ and $D_i$s are functions of time only. Here, $\delta \tilde{\varphi}$ and $\tilde{}E_2$ are linear combination of the original matter perturbation $\delta \varphi$ and the metric field $E_2$, chosen in such a way that the mixing kinetic term vanishes. In the large-$k$ limit we find the following leading terms:
\begin{equation}
S^{(2)}_{\rm kin, k\rightarrow \infty} = \int d^3k dt\; \left[\frac{1}{2}a^3\delta \dot{\tilde{\varphi}}^2-\frac{1}{2}\dot{\tilde{E}}_2^2M_T^2a^5H^2\frac{\bar{\alpha}_{SI}(\bar{\alpha}_N-1)(\bar{\alpha}_N^2-\bar{\alpha}_b^2)}{k^2\bar{\alpha}_N^2(1-\bar{\alpha}_b)^2} \right].
\end{equation}

For the mass matrix, we find that when taking the large-$k$ limit the off-diagonal terms are subdominant and hence the mass matrix is diagonal in that limit. Explicitly, we find the following terms:
\begin{equation}
S_{\rm mass,k\rightarrow\infty}^{(2)}=-\int d^3k dt \left[-a^3M_T^2H^2\bar{\alpha}_S \tilde{E}_2^2 +\frac{1}{2}ak^2\delta \tilde{\varphi}^2 \right].
\end{equation}
From here we can easily read off conditions to avoid gradient instabilities, which in this case simply give $\bar{\alpha}_S<0$. 
We emphasise that in the special case $\bar{\alpha}_S=0$ (which includes dRGT massive bigravity), the dominant terms in the mass matrix in the large-$k$ limit are different and hence the gradient condition changes. After canonically normalising the kinetic terms for the fields, we find again that the off-diagonal terms of the mass matrix are subdominant. Explicitly, we then find the following mass terms: 
\begin{equation}
S_{\rm mass,k\rightarrow\infty}^{(2)}=-\int d^3k dt \left[M_{TT,k\rightarrow\infty}^2 \tilde{E}_2^2 +\frac{1}{a^2}k^2\delta \tilde{\varphi}^2 \right],
\end{equation}
where $M_{TT,k\rightarrow\infty}^2$ is given in Appendix \ref{extraconditionsTT}. Here we find only one non-trivial condition to avoid gradient instabilities. We mention that specific branches of massive gravity do present gradient instabilities \cite{Comelli:2012db}, leading to an exponential growth of scalar perturbations at early times \cite{Akrami:2015qga}.


\section{The Quasistatic Approximation}\label{Sec:QSA}
In this section we will focus on the quasistatic approximation (QSA) of the equations of motion of the four actions we have considered in the previous sections. That is to say, we will study the regime where the time evolution of gravitational perturbations is small compared to their spatial evolution, which will be valid for intermediate scales within the sound horizon, $c_s^2k^2/a^2\gg H^2$ \cite{Sawicki:2015zya}. In practice, for a given gravitational field $X$ (metric or additional degree of freedom), we assume that $\abs{\dot{X}} \sim {H} \abs{X}\ll \abs{\partial_i X}$, and hence neglect time derivatives.

In the QSA we can describe the relevant equations of motion in terms of two effective parameters (functions of time). If we choose the Newtonian gauge where $B=E=0$, the first parameter, $\mu(a)$, determines the modified relation between metric potentials and matter density perturbations:
\begin{equation}
\frac{k^2}{a^2}\Phi=-4\pi G\mu(a) \rho\Delta,
\end{equation}
where $\Delta$ is the comoving gauge invariant density perturbation given by $\Delta=\delta \rho/\rho + 3Hv$, with $\delta \rho$ the matter energy density perturbation and $v$ its velocity potential perturbation. 
We also have that $G$ is the Newton's constant, given by $G=1/(8\pi M_P^2)$ where $M_P$ is the Planck mass. From here we can see that $G\mu$ describes an effective Newton's constant, and for $\mu=1$ we recover GR. 
The second, $\gamma(a)$, parametrises the modified relationship between the two metric potentials $\Phi$ and $\Psi$:
\begin{equation}
\gamma(a)=\frac{\Psi}{\Phi},
\end{equation}
which is known as the ``slip'' parameter and describes the presence of an effective gravitational anisotropic stress. For $\gamma=1$ we recover GR with no anisotropic matter stress sources. 
In addition, we can define a third QSA parameter, $\Sigma(a)$, which is not independent of the first two. It parametrises the combination of metric potentials that is probed by gravitational lensing as:
\begin{equation}
\frac{k^2}{a^2}\left(\Phi+\Psi\right)=-8\pi G \,\Sigma(a)\left[\rho \Delta+\frac{3}{2}(\rho+P)\sigma_m\right],
\end{equation}
where $\sigma_m$ is an anisotropic stress source in the matter sector; we will set this to zero hereafter. Because $\Sigma$ is connected to lensing observables, it is often more convenient to discuss constraints on $\Sigma$ and $\mu$ instead of $\gamma$ and $\mu$. However, note that in principle any two of these three functions ($\mu$, $\gamma$, $\Sigma$) can be used as a parametrisation of the quasistatic regime. The three parameters are related by:
\begin{equation}
\Sigma=\frac{1}{2}(1+\gamma)\mu.
\label{sigmugam}
\end{equation}

The QSA is useful, because it leads to simplified equations of motion that, in many cases, are still highly accurate on scales relevant to large-scale structure and weak lensing observations. However, the QSA can break down in some models if fields are rapidly oscillating or accelerating, or their sound speed is very small (for instance in some branches of massive bigravity \cite{Comelli:2012db}). In such cases the full equations must be solved. In addition, in our calculations we will neglect mass terms over spatial derivatives. While such an assumption is valid in many cases (scalar-tensor models \cite{Barreira:2012kk, DeFelice:2010as} and vector-tensor models \cite{DeFelice:2016uil}), it is also known that in some dark energy models, such as $f(R)$ gravity \cite{Amendola:2007nt,Hu:2007nk,Tsujikawa:2007xu}, the mass term of the additional gravitational degree of freedom becomes much larger than $H$ during the early universe, so it cannot be neglected in the QSA. For that model, the results presented here will only be valid at late times.

\subsection{Scalar-Tensor Theories}
To apply the QSA to scalar-tensor theories, we eliminate all time derivatives in the metric field and the extra scalar field, and keep terms with highest powers of $k$. We start from the metric $(0,0)$ and traceless $(i,j)$ equations in addition to the equation for the extra scalar field. We see that these three equations can be written as:
\begin{align}
&A_1 \frac{k^2}{a^2}\Psi+A_2 \frac{k^2}{a^2}v_X+ A_3 \frac{k^2}{a^2}\Phi = -\rho\Delta\label{E00},\\
& B_1\Phi+B_2\Psi+B_3v_X=0\label{Eij},\\
& C_1 \frac{k^2}{a^2}\Psi + C_2 \frac{k^2}{a^2}\Phi+ C_3\frac{k^2}{a^2}v_X=0 \label{EScalar},
\end{align}
where the coefficients $A_i$, $B_i$ and $C_i$ in principle depend on the scale factor and all of the free parameters of the quadratic action.
Combining eq.~(\ref{Eij})-(\ref{EScalar}) we find:
\begin{equation}
\label{gamscalar}
\gamma(a)=\frac{\Psi}{\Phi}=\frac{C_2B_3-B_1C_3}{B_2C_3-C_1B_3},
\end{equation}
Combining eq.~(\ref{E00})-(\ref{Eij}) and using the result for $\gamma$, we also find:
\begin{equation}
4\pi G\mu=\frac{B_2C_3-C_1B_3}{A_1(C_2B_3-B_1C_3)+A_2(B_1C_1-B_2C_2)+A_3(B_2C_3-B_3C_1)}.
\label{muscalar}
\end{equation}
Specifically, in terms of the free parameters, $\gamma$ and $\mu$ are given by:
\begin{align}
&\gamma=\frac{\beta_1+\beta_2}{\beta_3},\\
& \mu= \frac{2\beta_3}{2\beta_1+\beta_2(2-\hat{\alpha}_B)}\frac{M_P^2}{M^2_S},
\end{align}
where we have introduced three new functions $\beta_{i}$ (for $i=1,2,3$) such that:
\begin{align}
&\beta_1={\hat B}-\frac{1}{H^2}\frac{d}{dt}\left(H{\hat \alpha}_B\right),\\
&\beta_2=\hat{\alpha}_B(1+\hat{\alpha}_T)+2(\hat{\alpha}_M-\hat{\alpha}_T),\\
&\beta_3=\beta_1(1+\hat{\alpha}_T)+\beta_2(1+\hat{\alpha}_M),
\end{align}
where have we defined ${\hat B}=(-\dot{\varphi}_*^2+2{\dot H})/H^2$ and $\dot{\varphi}_*^2=\dot{\varphi}^2/M^2_S$. We notice that these results are in agreement with those in \cite{Alonso:2016suf}. Using eq.~(\ref{sigmugam}), from these results we find $\Sigma$:
\begin{equation}
\Sigma=\frac{M_P^2}{M^2_S}\frac{\beta_3+\beta_1+\beta_2}{\left[2\beta_1+\beta_2\left(2-\hat{\alpha}_B\right)\right]}.
\end{equation}
Let us now analyse the set of two parameters ($\mu$, $\Sigma$). We first note that $\hat{\alpha}_K$ does not appear in either of them, and thus does not play a role on quasistatic scales. In addition, it is important to note that, if all the parameters $\hat{\alpha}_X$ vanish and $M_S=M_P$, then we recover GR with $\Sigma=\mu=1$, as expected. Under these circumstances, the equations of motion for sub-horizon perturbations are the same as those of GR, regardless of the background evolution (which could still potentially be different from the $\Lambda$CDM model). 

In general, $\mu$ and $\Sigma$ can be completely arbitrary, but they may satisfy certain relations in some cases. In fact, we see that:
\begin{equation}
\frac{\Sigma-1}{\mu-1}=\frac{(M_P^2/M^2_S)\left[\beta_1(2+\hat{\alpha}_T)+\beta_2(2+\hat{\alpha}_M)\right]-\left[2\beta_1+\beta_2\left(2-\hat{\alpha}_B\right)\right]}{(M_P^2/M^2_S)\left[\beta_1(2+2\hat{\alpha}_T)+\beta_2(2+2\hat{\alpha}_M)\right]-\left[2\beta_1+\beta_2\left(2-\hat{\alpha}_B\right)\right]}.
\end{equation}
Notice that the numerator and denominator above differ only by having different factors of two in a few terms. Some authors have used this to argue that numerator and the denominator should generally have the same sign, and thus, a measurement of different signs for the deviations of $\Sigma$ and $\mu$ from unity would disfavour most Horndeski models \cite{Pogosian:2016ji}. Recently, it has been argued that viable models have $\hat{\alpha}_T=0$ \cite{Baker:2017hug,Ezquiaga:2017ekz,Sakstein:2017xjx,Wang:2017rpx}, and if in addition we assume that we have small $\alpha_X$ (for example variations of the Planck mass are locally constrained to be of order $few\times 10^{-1}$ \cite{PhysRevLett.51.1609}) and therefore $M_P\sim M_S$,
we find that
\begin{equation}
\frac{\Sigma-1}{\mu-1}\simeq \frac{1+\frac{\hat \alpha_B}{\hat \alpha_M}}{2+\frac{\hat \alpha_B}{\hat \alpha_M}}. 
\end{equation}
If $\hat{\alpha}_B$ and $\hat{\alpha}_M$ have different orders of magnitude and hence one of them dominates, then this ratio is always positive and fixed to $1$ or $1/2$. Furthermore, we see that unless $-2<\hat \alpha_B/\hat \alpha_M<-1$, this quantity is always positive. This result reinforces the point made in \cite{Pogosian:2016ji}.

We can also look at the difference between $\Sigma$ and $\mu$, which can be expressed as:
\begin{equation}
\Sigma-\mu=\frac{M_P^2}{M^2_S}\frac{\beta_1+\beta_2-\beta_3}{\left[2\beta_1+\beta_2\left(2-\hat{\alpha}_B\right)\right]}.
\end{equation}
For viable models with $\hat{\alpha}_T= 0$ we see that 
\begin{equation}
\Sigma-\mu = \frac{M_P^2}{M^2_S}\frac{\beta_2\hat{\alpha}_M}{\left[2\beta_1+\beta_2\left(2-\hat{\alpha}_B\right)\right]} ,
\end{equation}
which means that any measurement of $\Sigma \not= \mu$ would signal $\hat{\alpha}_M\not=0$, i.e.~a varying effective mass scale. This was already pointed out in \cite{Pogosian:2016ji}.

\subsection{Vector-Tensor Theories}
We now follow the same procedure as the previous subsection, for vector-tensor theories. As usual, we analyse general vector-tensor and Einstein-Aether theories separately.

\subsubsection{General Vector-Tensor Theories}
In this case we first integrate out the auxiliary field $\delta A_0$ and then apply the QSA. We follow the procedure described above and find the same set of equations (\ref{E00})-(\ref{EScalar}) for the metric potentials and the vector field $\delta A_1$. However, eq.~(\ref{EScalar}) here has a different $k$ dependence:
\begin{equation}
C_1\frac{k^2}{a^2}\Psi+C_2\frac{k^2}{a^2}\Phi+ \left(C_3\frac{k^2}{a^2} + C_4\frac{k^4}{a^4}\right)\delta A_1=0,
\end{equation}
where $C_4\propto \tilde{\alpha}_D$. Hence, when $\tilde{\alpha}_D\not=0$ the term with $C_4$ dominates but if $\tilde{\alpha}_D=0$ (as in Generalised Proca theories) then $C_3$ dominates and the structure of the equations is the same as that for scalar-tensor theories. Let us first give the quasistatic parameters when $\tilde{\alpha}_D\not=0$. 

For these models, we find the following expressions for $\gamma$ and $\mu$:
\begin{align}
&\gamma= \frac{2\tilde{\alpha}_A(1+\tilde{\alpha}_T)\beta_1+(1+\tilde{\alpha}_C)\beta_2}{(1+\tilde{\alpha}_T)^2(2\beta_1^2+\beta_2)},\\
&\mu=-\frac{2(1+\tilde{\alpha}_T)(2\beta_1^2+\beta_2)}{-4\tilde{\alpha}_A(1+\tilde{\alpha}_T)\left[-\tilde{\alpha}_A(1+\tilde{\alpha}_T)+2\beta_1(1+\tilde{\alpha}_C)\right]-2\beta_2(1+\tilde{\alpha}_C)^2+\tilde{\alpha}_{K}(1+\tilde{\alpha}_T)(2\beta_1^2+\beta_2)}\nonumber\\
& \times \frac{M_P^2}{M_V^2},\label{VTmu}\\
\end{align}
where we have defined
\begin{align}
&\beta_1=2\left(\frac{\dot{\tilde{\alpha}}_C}{H} +(\tilde{\alpha}_M+1)\tilde{\alpha}_C+\tilde{\alpha}_M-\tilde{\alpha}_T\right),\\
&\beta_2= 4(\tilde{\alpha}_T+1)\left({\hat B}+\beta_1-\tilde{\alpha}_V\right),
\end{align}
and where, as in the case of scalar-tensor theories, we have defined ${\hat B}=(-\dot{\varphi}_*^2+2{\dot H})/H^2$. The third, dependent, QS parameter is given by
\begin{align}
\Sigma&=-\frac{(1+\tilde{\alpha}_C)\beta_2+2\tilde{\alpha}_A(1+\tilde{\alpha}_T)\beta_1+(1+\tilde{\alpha}_T)^2(2\beta^2_1+\beta_2)}
{-4\tilde{\alpha}_A(1+\tilde{\alpha}_T)\left[-\tilde{\alpha}_A(1+\tilde{\alpha}_T)+2\beta_1(1+\tilde{\alpha}_C)\right]-2\beta_2(1+\tilde{\alpha}_C)^2+\tilde{\alpha}_{K}(1+\tilde{\alpha}_T)(2\beta_1^2+\beta_2)}\nonumber \\
& \times \frac{M_P^2}{M_V^2}.
\end{align}
Again, we look at the set of parameters ($\mu$, $\Sigma$). First of all, we can see that all of the free parameters do contribute, except for $\tilde{\alpha}_D$, and thus it is not possible to constrain this parameter in the QS regime (see the related discussion on limits where $\tilde{\alpha}_D$ dependent-terms drop out in section \ref{Sec:constraints}). Similar to scalar-tensor theories, if all the $\tilde{\alpha}_X$ parameters vanish and $M_V=M_P$, then $\Sigma=\mu=1$; thus at sub-horizon scales the equations of motion for perturbations are the same as in GR, regardless of the background evolution. 

Next, we can look for possible relations between $\mu$ and $\Sigma$. We find that, in general, 
\begin{align}
\frac{\Sigma-1}{\mu-1}=\frac{-(M_P^2/M^2_V)\left[(1+\tilde{\alpha}_C)\beta_2+2\tilde{\alpha}_A(1+\tilde{\alpha}_T)\beta_1+(1+\tilde{\alpha}_T)^2(2\beta^2_1+\beta_2)\right]-D}{-(M_P^2/M^2_V)[2(1+\tilde{\alpha}_T)(2\beta_1^2+\beta_2)]-D},
\end{align}
where we have defined $D$ as the denominator of $\mu$ in the first term of eq.~(\ref{VTmu}). For small $\tilde{\alpha}_X$ (and thus $M_P\sim M_V$) we have that
\begin{align}
\frac{\Sigma-1}{\mu-1}\simeq \frac{(2\tilde{\alpha}_T-3\tilde{\alpha}_C+\tilde{\alpha}_K){\beta^{(1)}}_2+2({\beta^{(1)}}_1-2\tilde{\alpha}_A)(-\tilde{\alpha}_A+{\beta^{(1)}}_1)}{(2{\beta^{(1)}}_1-2\tilde{\alpha}_A)^2+2{\beta^{(1)}}_2(\tilde{\alpha}_T-4\tilde{\alpha}_A-2\tilde{\alpha}_C)},
\end{align}
where ${\beta^{(1)}}_i$ is the leading (linear) term in the Taylor expansion of $\beta_i$ (with $\hat{B}$ also small). This shows that, due to the large number of free parameters, the sign of the ratio of deviations of $\Sigma$ and $\mu$ from unity is undetermined, and additional information is needed in order to see if these models have a specific tendency. Furthermore, we can see that even background deviations from FRW will play a role in determining this sign, through the parameter $\hat{B}$ in ${\beta^{(1)}}_2$. In addition, we look at the difference between these two parameters:
\begin{align}
\Sigma-\mu&=-\frac{(1+\tilde{\alpha}_C)\beta_2+2\tilde{\alpha}_A(1+\tilde{\alpha}_T)\beta_1-(1+\tilde{\alpha}_T)^2(2\beta^2_1+\beta_2)}
{-4\tilde{\alpha}_A(1+\tilde{\alpha}_T)\left[-\tilde{\alpha}_A(1+\tilde{\alpha}_T)+2\beta_1(1+\tilde{\alpha}_C)\right]-2\beta_2(1+\tilde{\alpha}_C)^2+\tilde{\alpha}_{K}(1+\tilde{\alpha}_T)(2\beta_1^2+\beta_2)}\nonumber \\
& \times \frac{M_P^2}{M_V^2}.
\end{align}
Again, due to the large number of parameters, it is not possible to make general statements on the relation between $\Sigma$ and $\mu$ for these models. 

Even though for the case with $\tilde{\alpha}_D=0$ (as in Generalised Proca theories) it will not be possible to obtain general conclusions either (expressions are even more involved than in the $\tilde{\alpha}_D\not=0$ case), for completeness, we leave the expressions of the two QS parameters $\gamma$ and $\mu$ in the following \href{https://github.com/noller/General-Theory-of-Linear-Cosmological-Perturbations}{link}. The derivation and analysis of these QS parameters for Generalised Proca has been performed in \cite{DeFelice:2016uil,deFelice:2017paw,Nakamura:2017dnf}. 

\subsubsection{Einstein-Aether Theories}
We find that the field equations have the same structure as in the general vector-tensor case. Indeed, eq.~(\ref{EScalar}) has the following $k$ dependence:
\begin{equation}
C_1\frac{k^2}{a^2}\Psi+C_2\frac{k^2}{a^2}\Phi+ \left(C_3\frac{k^2}{a^2} + C_4\frac{k^4}{a^4}\right)\delta A_1=0,
\end{equation}
where $C_4\propto \check{\alpha}_D$. Hence, when $\check{\alpha}_D\not=0$ the term with $C_4$ dominates but if $\check{\alpha}_D=0$ (as in a model with a $\Lambda$CDM background evolution) then $C_3$ dominates and the structure of the equations is the same as that for scalar-tensor theories. Let us first give the quasistatic parameters when $\check{\alpha}_D\not=0$. In this case, the parameters $\gamma$ and $\mu$ are given by
\begin{align}
&\gamma=\frac{1+\check{\alpha}_C}{1+\check{\alpha}_T},\nonumber\\
& \mu= \frac{1+\check{\alpha}_T}{\left[2(1+\check{\alpha}_C)^2-\check{\alpha}_K(1+\check{\alpha}_T)\right]}\frac{2M_P^2}{M_A^2},
\end{align}
and then the quasistatic parameter $\Sigma$ is given by:
\begin{equation}
\Sigma = \frac{M_P^2}{M_A^2}\frac{(2+\check{\alpha}_C+\check{\alpha}_T)}{\left[2(1+\check{\alpha}_C)^2-\check{\alpha}_K(1+\check{\alpha}_T)\right]}.
\end{equation}
Similarly as before, we look at the set ($\Sigma$, $\mu$). From here we see that the two free parameters $\check{\alpha}_X$ of these theories (recall $\check{\alpha}_T$ is a derived parameter here) appear in the expressions of $\mu$ and $\Sigma$; thus they are all relevant and, in principle, measurable in this regime. These expressions are valid in the case of an arbitrary background, excluding the $\Lambda$CDM background (where $\check{\alpha}_D=0$). We recover the GR behaviour of perturbations, $\Sigma=\mu=1$ when the two independent $\check{\alpha}_X$ parameters vanish and $M_P=M_A$, regardless of the background evolution. 

Following the same logic as in the scalar-tensor case, we calculate the ratio of $\Sigma-1$ and $\mu-1$ to find:
\begin{equation}
\frac{\Sigma-1}{\mu-1}=\frac{(M_P^2/M_A^2)\left(1+\check{\alpha}_C+\check{\alpha}_T\right)-\left[2(1+\check{\alpha}_C)^2-\check{\alpha}_K(1+\check{\alpha}_T)\right]}{2(1+\check{\alpha}_T)(M_P^2/M_A^2)-\left[2(1+\check{\alpha}_C)^2-\check{\alpha}_K(1+\check{\alpha}_T)\right]}.
\end{equation}
In the approximation of small parameter (and $M_P\sim M_A$), we get:
\begin{equation}
\frac{\Sigma-1}{\mu-1}\simeq \frac{{\check\alpha}_T-3{\check \alpha}_C+{\check\alpha}_K}{2{\check\alpha}_T-4{\check \alpha}_C+{\check\alpha}_K}.
\end{equation}
If we take models with $\check{\alpha}_T\sim 0$ that are in agreement with the latest observation of gravitational waves, then this expression is such that if ${\check \alpha}_C$ and ${\check \alpha}_K$ have different orders of magnitude, then one of them will dominate and in any case we will get that $(\Sigma-1)/(\mu-1)>0$, although the specific ratio will be different, with a ratio of 1 if ${\check \alpha}_K$ dominates and 3/4 if ${\check \alpha}_C$ dominates. A different sign can only happen if $3<{\check \alpha}_K/{\check \alpha}_C<4$. In general then, viable Einstein-Aether models prefer a positive sign, with some exceptional cases. 

We also find that the difference between the two QS parameters, which is given by:
\begin{equation}
\Sigma-\mu= \frac{M_P^2}{M_A^2}\frac{\check{\alpha}_C-\check{\alpha}_T}{\left[2(1+\check{\alpha}_C)^2-\check{\alpha}_K(1+\check{\alpha}_T)\right]}.
\end{equation}
From here we can see that if we take models with $\check{\alpha}_T\sim 0$ that are in agreement with the latest observation of gravitational waves, then any difference between $\Sigma$ and $\mu$ would be an indication of a non-zero $\check{\alpha}_C$, regardless of the value of the kineticity parameter $\check{\alpha}_K$.

Next, we give the QS parameters when $\check{\alpha}_D=0$: 
\begin{align}
& \gamma=\frac{2\dot{H}(1+\check{\alpha}_C)+H\beta (\check{\alpha}_M-\check{\alpha}_T)}{2\dot{H}(1+ \check{\alpha}_T)},\\
& \mu= \frac{2\dot{H}(1+ \check{\alpha}_T)}{\dot{H}\left( 2\check{\alpha}_C(2+\check{\alpha}_C)-\check{\alpha}_K(1+\check{\alpha}_T)+1 \right) +H\beta \left(\check{\alpha}_C(1+\check{\alpha}_M)+\check{\alpha}_M-\check{\alpha}_T \right)}\frac{M_P^2}{M_A^2}, 
\end{align}
where we have defined 
\begin{equation}
\beta \check{\alpha}_C= \dot{\check{\alpha}}_K-2H\check{\alpha}_C+H\check{\alpha}_K(1+ \check{\alpha}_M).
\end{equation}
Therefore, the third QS parameter $\Sigma$ is given by:
\begin{equation}
	\Sigma=\frac{2\dot{H}(2+ \check{\alpha}_T+ \check{\alpha}_C)+ H\beta (\check{\alpha}_M-\check{\alpha}_T)}{\dot{H}\left( 2\check{\alpha}_C(2+\check{\alpha}_C)-\check{\alpha}_K(1+\check{\alpha}_T)+1 \right) +H\beta \left(\check{\alpha}_C(1+\check{\alpha}_M)+\check{\alpha}_M-\check{\alpha}_T \right)}\frac{M_P^2}{M_A^2}.
\end{equation}
If we look at the pair ($\Sigma$, $\mu$) we find that if $\alpha_T=0$ and for small parameters $\check{\alpha}_X$ (and $M_A\sim M_P$) then:
\begin{equation}
\frac{\Sigma-1}{\mu-1}\simeq \frac{H (2 \check{\alpha}_C+ \check{\alpha}_M)\left[ \dot{\check{\alpha}}_K + H(\check{\alpha}_K-2\check{\alpha}_C) \right]+2\check{\alpha}_C\dot{H}(3\check{\alpha}_C-2\check{\alpha}_K)}{H (2 \check{\alpha}_C+ 2\check{\alpha}_M)\left[ \dot{\check{\alpha}}_K + H(\check{\alpha}_K-2\check{\alpha}_C) \right]+2\check{\alpha}_C\dot{H}(4\check{\alpha}_C-\check{\alpha}_K)}.
\end{equation}
Similarly to the case with $\check{\alpha}_D\not=0 $, here we find that if any of the parameters $\check{\alpha}_X$ dominate, then this ratio will always be positive. However, in special situations the sign of the ratio may be negative. 
For the difference between the QS parameters we find, in general, :
\begin{equation}
\Sigma-\mu= \frac{ 2(\check{\alpha}_C-\check{\alpha}_T)\dot{H}+H\beta (\check{\alpha}_M-\check{\alpha}_T) }{ \dot{H}\left( 2\check{\alpha}_C(2+\check{\alpha}_C)-\check{\alpha}_K(1+\check{\alpha}_T)+1 \right) +H \beta \left(\check{\alpha}_C(1+\check{\alpha}_M)+\check{\alpha}_M-\check{\alpha}_T \right)}\frac{M_P^2}{M_A^2}.
\end{equation}
In this case, all the parameters are responsible for this difference and $\check{\alpha}_C$ does not play any special role, as opposed to the general case with $\check{\alpha}_D\not=0$.

\subsection{Tensor-Tensor Theories}
In the bimetric case we first integrate out the auxiliary fields from the second metric, that is, $\Phi_2$, $B_2$ and $\Psi_2$. We identify $E_2$ as the additional gravitational scalar degree of freedom. We follow the QSA and find that the $k$-structure of the quasistatic equations is similar to vector-tensor theories. In this case, the equation for $E_2$ has the following form:
\begin{equation}
 C_1\frac{k^2}{a^2}\Psi_1+C_2\frac{k^2}{a^2}\Phi_1+ \left(C_3\frac{k^2}{a^2}+ C_4\frac{k^4}{a^4}\right)E_2=0,
\end{equation} 
where $C_4\propto \bar{\alpha}_S$ and hence when $\bar{\alpha}_S\not=0$ the term with coefficient $C_4$ dominates but if $\bar{\alpha}_S=0$ (as in massive bigravity) then the term with $C_3$ dominates. First, we give the expressions for the QS parameters when $\bar{\alpha}_S\not=0$. In this case, we find that $\gamma$ and $\mu$ are given by:
\begin{align}
&\gamma= \frac{1}{1+\bar{\alpha}_M},\\
&\mu=(1+\bar{\alpha}_M) \frac{M_P^2}{M_T^2}.
\end{align}
We notice that these expressions are very simple, and when $M_T=M_P$ we find that this model is indistinguishable from GR in this regime, regardless of the background evolution and the value of the three independent parameters $\bar{\alpha}_X$. Furthermore, in general, they satisfy the very simple relation $\gamma \mu=M_P^2/M_T^2$. We emphasise that these results hold only when $\bar{\alpha}_S\not=0$. 

If $\bar{\alpha}_S=0$, as in the specific theory of massive bigravity, the parameters $\gamma$ and $\mu$ are found to be:
\begin{align}
&\gamma=\frac{\bar{\alpha}_{SI}+\bar{\alpha}_N(\beta-\bar{\alpha}_{SI})}{\beta(1+\bar{\alpha}_M)\bar{\alpha}_N-2\bar{\alpha}_{TI}(\bar{\alpha}_b-1)},\\
&\mu=\frac{2\bar{\alpha}_{TI}(\bar{\alpha}_b-1)}{2\gamma \bar{\alpha}_{TI}(\bar{\alpha}_b-1) -\bar{\alpha}_{SI}(\bar{\alpha}_N-1)\left[\gamma(1+\bar{\alpha}_M)-1\right]}\frac{M_P^2}{M_T^2},
\end{align}
where $\beta$ is an additional and lengthy term given in Appendix \ref{app:QSTT}. Hence, we find the derived quasistatic parameter $\Sigma$ to be:
\begin{equation}
	\Sigma=\frac{M_P^2}{M_T^2}\frac{(\gamma+1)\bar{\alpha}_{TI}(\bar{\alpha}_b-1)}{2\gamma \bar{\alpha}_{TI}(\bar{\alpha}_b-1) -\bar{\alpha}_{SI}(\bar{\alpha}_N-1)\left[\gamma(1+\bar{\alpha}_M)-1\right]}.
\end{equation}
 We note that when $\bar{\alpha}_{SI}=\bar{\alpha}_{TI}=0$ and $M_T=M_P$ we recover GR and hence find $\gamma=\mu=\Sigma=1$.
Similarly as before, we focus on the set ($\Sigma$, $\mu$) and calculate the two quantities that are useful to understand the behaviour of the model in the quasistatic regime:
\begin{align}
& \frac{\Sigma-1}{\mu-1}=\frac{(M_P^2/M_T^2)(\gamma+1)\bar{\alpha}_{TI}(\bar{\alpha}_b-1)- \left[2\gamma \bar{\alpha}_{TI}(\bar{\alpha}_b-1) -\bar{\alpha}_{SI}(\bar{\alpha}_N-1)\left[\gamma(1+\bar{\alpha}_M)-1\right]\right] }{(M_P^2/M_T^2)2\bar{\alpha}_{TI}(\bar{\alpha}_b-1)- \left[2\gamma \bar{\alpha}_{TI}(\bar{\alpha}_b-1) -\bar{\alpha}_{SI}(\bar{\alpha}_N-1)\left[\gamma(1+\bar{\alpha}_M)-1\right]\right] },\\
& \Sigma-\mu=\frac{M_P^2}{M_T^2}\frac{(\gamma-1)\bar{\alpha}_{TI}(\bar{\alpha}_b-1)}{2\gamma \bar{\alpha}_{TI}(\bar{\alpha}_b-1) -\bar{\alpha}_{SI}(\bar{\alpha}_N-1)\left[\gamma(1+\bar{\alpha}_M)-1\right]}.
\end{align}
In this case it is not possible to make any general statements due to the complexity of the expressions. An analysis of the quasistatic limit of the specific theory of massive gravity has been performed in \cite{Solomon:2014dua}. However, we point out that some branches of massive gravity present growing modes due to the presence of gradient instabilities, and thus the quasistatic approximation is not valid in those cases.  


\section{Towards Phenomenology}\label{Sec:Phenomenology}

In this section we show how to treat all the models we have looked at in a unified way with the goal of implementing them in numerical Einstein-Boltzmann (EB) solvers. Throughout this section we will use conformal time $\tau$ and primes denote derivatives w.r.t.\ $\tau$. Specifically, we will present a unified set of equations for scalar, vector and tensor perturbations in the presence of a general perfect fluid, such that they describe the four parametrised models discussed in this paper at the same time. In addition, we will explain how to solve the equations and implement them in numerical codes. We shall present our results using the notation and the gauge choice typical of the most widely used EB codes, namely CAMB \cite{Lewis:1999bs} and CLASS \cite{Blas:2011rf}. These codes solve the equations of motion for linear perturbations in conformal time $\tau$, and synchronous gauge. In this case, the metric line element, up to first order, in perturbation theory reads \cite{Ma:1995ey}:
\begin{equation}
	ds^{2}=a^{2}\left[ -d\tau^{2}+\left(\delta_{ij} + \tilde{h}_{ij}\right)dx^{i}dx^{j}\right] \,,\label{eq:metric}
\end{equation}
where in Fourier space
\begin{equation}
	\tilde{h}_{ij}\left(\tau, \vec{k}\right)= \hat{k}_i \hat{k}_j h + 6\left(\hat{k}_i \hat{k}_j - \frac{1}{3}\delta_{ij}\right) \eta + \hat{k}_i F_j + \hat{k}_j F_i + h_{ij}\,.
\end{equation}
Here, $h$ and $\eta$ are the metric scalar perturbations \footnote{Sometimes in the codes an auxiliary variable $\alpha$ is used, such that $2k^2\alpha=h^{\prime}+6\eta^{\prime}$. Here, we report the equations using the original $h$ and $\eta$ variables to avoid confusion.}, $F_i$ is the vector perturbation and $h_{ij}$ is the tensor perturbation. Usually, vector equations are neglected since the amplitude of vector perturbations decays rapidly with the expansion of the universe. However, VT, EA and TT gravity theories propagate a dynamical vector degree of freedom, which may bring non-trivial modifications to the dynamics of vector perturbations, and thus this additional DoF should be properly taken into account.

On top of the metric, matter perturbations are standard in Boltzmann codes. Here we decompose a generic stress-energy tensor into: (i) density $\rho$ and density perturbation $\delta\rho_\textrm{m}$, (ii) pressure $P$ and pressure perturbation $\delta p_\textrm{m}$, (iii) velocity scalar $\theta_\textrm{m}$ and vector $\theta^i_\textrm{m}$ perturbations, and (iv) shear scalar $\sigma_\textrm{m}$ and tensor $\sigma_{\textrm{m}ij}$ perturbations. We will also have the following additional perturbations that enter the equations of motion:
\begin{itemize}
	\item $\chi$: extra scalar propagating degree of freedom common to all the models. It corresponds to $v_X$ in ST, $\delta A_1$ in EA and VT, and the combination $\mathcal{H}\left(E_2+(h+6\eta)/(2k^2)\right)$ in TT;
	\item $\Gamma_{n}$ with $n=0,\ldots,4$: additional auxiliary variables that appear in VT and TT. They correspond to $\Gamma_0=\delta A_0$ in VT and $\Gamma_1=\Phi_2/\mathcal{H}, \, \Gamma_2=B_2, \, \Gamma_3=\Psi_2/\mathcal{H}$ in TT;
	\item $\chi^i$: dynamical vector perturbation for EA, VT and TT. It corresponds to $\delta A_2^i$ in EA and, and $F_2^i-F^i$ in TT;
	\item $\Gamma^i$: auxiliary vector perturbation, which corresponds to $S_2^i+F^{i\prime}$ in TT theories;
	\item $h_{2ij}$: additional dynamical tensor perturbation in the TT case.
\end{itemize}
For simplicity, when we treat vector and tensor perturbations we will drop spatial indices.

Before showing the equations for both the background and the perturbation evolution, it is important to stress that the equations will be presented in a unified form, but it is not correct to constrain the full parameter space of these equations at once. As discussed in Section \ref{sec:unifiedaction}, the full equations presented in this section can describe modes that are neither ST, VT, EA nor TT. For instance, a theory with $\alpha_B\neq 0$ and $\alpha_D\neq 0$ is neither ST nor VT. As a consequence, these equations have to be implemented in EB solvers together with a flag to select separately between ST, EA, VT or TT theories.

\subsection{Background}

Before solving for the evolution of the perturbations, EB solvers compute the background dynamics. In GR, this is done by using the Friedmann equation of motion and by specifying the matter components that fill the universe to obtain the conformal Hubble parameter $\mathcal{H}(\tau)$. In the case of modified gravity theories, since additional DoFs arise, the dynamics of the background is not entirely determined by the matter densities and the Friedmann equation, but it is possible to get additional differential equations that fix the expansion history by solving the dynamics of the new DoFs. However, in the case of the general parametrised models presented in this paper, we do not have a particular non-linear theory, and thus we do not have these additional differential equations. In this framework, the background is arbitrary and it is one of the free parameters of the models. In practice, this means that the time-dependent background 
will need to be parametrised further (e.g.~fitted with some specific functional forms) in order get a background evolution. Furthermore, all of the other free parameters of the quadratic action are in general independent functions that will also have to be parametrised further. The only exception is the matter sector, where we assumed minimal and universal coupling for all matter species. This implies that the matter part at the background level will evolve with the usual scaling relations. Then, the only quantity that has to be specified to completely fix the background expansion history of the universe is $\mathcal{H}(\tau)$. Instead of parametrising directly $\mathcal{H}(\tau)$ it is convenient to introduce a more familiar fluid description, where an effective density $\mathcal{E}$ and pressure $\mathcal{P}$ are added to the Friedmann equations:
\begin{align}
	3M_{P}^2 \mathcal{H}^2&= a^2\left(\rho + \mathcal{E}\right)\label{eq:Friedmann1},\\
	M_{P}^2\left(2\mathcal{H}^\prime+\mathcal{H}^2\right)& = -a^2\left(P+\mathcal{P}\right)\label{eq:Friedmann2}\,.
\end{align} 
Here, $\rho$ and $P$ are the standard energy-density and pressure of the matter components, whereas $\mathcal{E} = \sum_i\mathcal{E}_i$ and $\mathcal{P} = \sum_i\mathcal{P}_i$, where $i=$ ST, EA, VT, TT. Note that, as specified before, it would be inconsistent to have more than one density and pressure of the additional fluid different from zero at the same time. Any code will have to choose one of these four models. From eqs.\ (\ref{eq:Friedmann1}) and (\ref{eq:Friedmann2}) it is possible to derive a conservation equation for the additional DoF:
\begin{align}
	& \mathcal{E}^\prime+3\mathcal{H}\left(\mathcal{E}+\mathcal{P}\right)=0.
\end{align} 
These equations form the complete system one has to implement and solve to compute the expansion history of the universe. We should stress here that, even if these equations may look familiar to the reader, they are completely arbitrary and their scope is just to calculate $\mathcal{H}(\tau)$ from a fluid perspective rather than parametrising it directly. As an example, within our approach it would be perfectly allowed to parametrise the background evolution of the extra fluid as a cosmological constant, i.e.\ imposing $\mathcal{P}=-\mathcal{E}$, or with the well-known equation of state parametrisation $\{w_0,\,w_a\}$. These two choices are common in the literature and they are already implemented in both EFTCAMB \cite{EFTCAMB1} and {\tt hi\_class} \cite{Zumalacarregui:2016pph}. 

\subsection{Stability conditions}

After assuming a background evolution, one has to check if that background is stable against small fluctuations. The issue of stability conditions has been discussed in great detail for this general framework in Section \ref{Sec:constraints}, here we report some considerations that could be useful for the implementation into EB solvers.

In a dynamical system, as our framework is, one has to consider two classes of instabilities, i.e.\ ones of \textit{ghost} and \textit{mass} types. The first kind of instability is related to the sign of the eigenvalues of the kinetic matrix $\mathbb{K}$ in eq.\ (\ref{ReducedActionGral}), while the latter is related to the sign of the eigenvalues of the mass matrix $\mathbb{M}$ (care has to be taken that mixed kinetic-mass interactions have been put into the anti-symmetric form discussed in Section \ref{Sec:constraints}, when deriving these eigenvalues). In general, these eigenvalues will be ratios of polynomials in $k$ with time-dependent coefficients. In principle, any stability condition should be tested for the whole evolution of the universe, and the full expressions for these eigenvalues should be taken into account. Here we mention points that are particularly important when implementing these conditions, and that may help simplifying the process but also to consistently address stability issues:
\begin{enumerate}
	\item EB solvers start integrating at times deep in the radiation-dominated epoch, and it is easy to numerically check stability from those times until today. 
%
%
In order to do this it would be necessary to diagonalise the full action with time and scale dependent coefficients in order to find the general expressions for the eigenvalues of $\mathbb{K}$ and $\mathbb{M}$. This is a very involved process which can, however, frequently be omitted due to the fact that significant instabilities present at these times will lead to unviable cosmologies, and hence such a model will already be (phenomenologically) discarded by codes.\label{item1}
	\item If one nevertheless would like to probe the full expressions for the eigenvalues $k_{i}$ of $\mathbb{K}$ and $m_{i}$ of $\mathbb{M}$, we note that a conservative choice for the stability conditions is $k_i(\tau,k)^2>0$, $m_i(\tau,k)^2>-\mathcal{H}^2$. This would eliminate ghost-like instabilities, but allow for slow tachyonic instabilities, which would make the system evolve mimicking a Jeans-like instability. Such a feature is actually necessary in any viable cosmological model in order to have formation of large-scale structures.
	\item It is crucial to check stability {\it prior} to the time of initial conditions for an EB solver. Otherwise, the initial conditions chosen would be arbitrary and would result in an extremely fine-tuned model, that could lead to a viable cosmology yet with early-time instabilities unprobed by the EB solver. Models with these early-time instabilities would therefore not be discarded by codes and thus we must impose complementary conditions to eliminate them. The relevant set of conditions will be given by the positivity of the large-$k$ limit of the eigenvalues of $\mathbb{K}$ and $\mathbb{M}$. Those conditions were explicitly given in Section \ref{Sec:constraints} for the four models studied in this paper.\label{item2}
  \item Due to points \ref{item1} and \ref{item2}, even if theoretical priors are important to distinguish between healthy and non-healthy theories, it is not always necessary to fully implement them in EB solvers, and it may be enough to take a simplified approach and simply impose positivity conditions in the large-$k$ limit. \label{item3}
  \item 
  For consistency, it is important to remember that the models considered in this paper are effective theories. As such, they are only valid up to some energy scale $\Lambda$, which is unknown in this parametrised linear approach. If the scale $\Lambda$ is at the same energy scale as the one at the time of setting initial conditions, then stability prior to that time can only be assesed within the UV embedding/completion of the models considered here. Therefore, it is important to establish a priori what kind of models we are looking for, that is, what the minimum desired scale $\Lambda$ should be.
  \item Finally, we mention that the conditions given in Section \ref{Sec:constraints} were calculated with a minimally coupled matter scalar instead of a perfect fluid. We conjecture that the same conditions would be valid for a general perfect fluid, but there would be additional stability conditions ensuring stability of the matter DoFs.
\end{enumerate}
This list gives an idea of the issues that need to be addressed to properly implement the stability conditions in a code. We do not propose solutions, but we do recommend that one is conservative and implements only the necessary stability conditions.

\subsection{Perturbations} 

In the parametrised framework of this paper, the evolution of the background quantities is not sufficient to completely fix the evolution of linear perturbations. Indeed, there is some residual freedom that is represented by the set of $\{\alpha_X(\tau)\}$ functions. In a fully covariant theory these functions would be completely determined by the background dynamics, while within our approach they remain unknown free functions of time. This implies that, as for the background, also the $\alpha_X(\tau)$ functions have to be parametrised in some way. For ST theories in both EFTCAMB \cite{EFTCAMB1} and {\tt hi\_class} \cite{Zumalacarregui:2016pph} some parametrisation are already implemented, but in principle there are infinite possible choices. One strategy to identify a meaningful parametrisation is to calculate consistently these functions in known theories using the real background evolution, and then parametrise them to mimic the fully covariant theory we have in mind \cite{Linder:2016wqw,Gleyzes:2017kpi}.

For the sake of clarity, let us summarize what the $\alpha_X(\tau)$ are for each class of models:
\begin{itemize}
	\item Scalar-Tensor: four free independent functions, namely $\left\{ \hat{\alpha}_{\textrm{K}},\,\hat{\alpha}_{\textrm{B}},\,\hat{\alpha}_{\textrm{T}},\,\hat{\alpha}_{\textrm{M}}\,\left(\textrm{or}\,\, M_{S}^{2}\right)\right\}$.
	\item Vector-Tensor: eight independent functions of time, namely $\left\{ \tilde{\alpha}_{\textrm{K}},\,\tilde{\alpha}_{\textrm{T}},\,\tilde{\alpha}_{\textrm{M}}\,\left(\textrm{or}\,\, M_{V}^{2}\right),\right.$ $\left. \tilde{\alpha}_{\textrm{C}},\,\tilde{\alpha}_{\textrm{D}},\,\tilde{\alpha}_{\textrm{A}},\,\tilde{\alpha}_{\textrm{V}},\,\tilde{\alpha}_{\textrm{VS}}\right\}$. Note that the last function, $\tilde{\alpha}_{\textrm{VS}}$, was not included in the previous sections because it only affects the vector dynamics. Such a parameter will be needed in this section to describe the equations for vector perturbations.
	\item Einstein-Aether: three free independent functions. The full list of functions used is\\ $\left\{ \check{\alpha}_{\textrm{K}},\,\check{\alpha}_{\textrm{M}}\,\left(\textrm{or}\,\, M_{A}^{2}\right),\,\check{\alpha}_{\textrm{C}},\,\check{\alpha}_{\textrm{T}},\,\check{\alpha}_{\textrm{D}}\right\}$, where the last two are dependent and obey the following constraint equations:
	\begin{align}
		\check{\alpha}_{\textrm{D}}= & \frac{a^{2}\left(\mathcal{E}_{A}+\mathcal{P}_{A}\right)}{M_{A}^{2}\left(\mathcal{H}^{2}-\mathcal{H}^{\prime}\right)},\\
		\check{\alpha}_{\textrm{T}}= & \check{\alpha}_{\textrm{M}}+\frac{\left(aM_A^{2}\check{\alpha}_{\textrm{C}}\right)^{\prime}}{aM_A^{2}\mathcal{H}}\,.
	\end{align}
	\item Tensor-Tensor: four independent functions of time. The full list of functions used is\\ $\left\{ \bar{\alpha}_{\textrm{M}}\,\left(\textrm{or}\,\, M_{T}^{2}\right),\,\bar{\alpha}_{\textrm{b}},\,\bar{\alpha}_{\textrm{N}},\,\bar{\alpha}_{\textrm{S}},\,\bar{\alpha}_{SI},\,\bar{\alpha}_{TI},\,\bar{\nu}_{\textrm{M}}\right\}$, but the last three are dependent and obey the following constraint equations:
	\begin{align}
		\bar{\alpha}_{SI}= & \frac{a^{2}\left(\mathcal{E}_{T}+\mathcal{P}_{T}\right)}{\left(1-\bar{\alpha}_{\textrm{N}}\right)\mathcal{H}^{2}M_{T}^{2}},\\
		\bar{\alpha}_{TI}= & \frac{\left(aM_T^{2}\mathcal{H}^{2}\left(1-\bar{\alpha}_{\textrm{N}}\right)\bar{\alpha}_{SI}\right)^{\prime}}{2aM_T^{2}\mathcal{H}^{3}\left(1-\bar{\alpha}_{\textrm{b}}\right)}-\frac{1}{2}\bar{\alpha}_{\textrm{b}}\bar{\alpha}_{SI}\frac{\left(1-\bar{\alpha}_{\textrm{N}}\right)}{\left(1-\bar{\alpha}_{\textrm{b}}\right)^{2}}\left(1-\bar{\alpha}_{\textrm{b}}-\frac{\bar{\alpha}_{\textrm{N}}^{\prime}}{\mathcal{H}\bar{\alpha}_{\textrm{N}}}\right),\\
		\bar{\nu}_{\textrm{M}}= & -\frac{\bar{\alpha}_{\textrm{N}}}{\bar{\alpha}_{\textrm{b}}^{2}}\left[\bar{\alpha}_{\textrm{N}}+\left(\frac{\bar{\alpha}_{\textrm{N}}}{\bar{\alpha}_{\textrm{b}}\mathcal{H}}\right)^{\prime}\right]^{-1}\times\frac{a^{2}\left(\mathcal{E}_{T}+\mathcal{P}_{T}\right)}{2M_{T}^{2}\mathcal{H}^{2}}\,.
	\end{align}
\end{itemize}

Next, we proceed to show the equations of motion governing the evolution of linear perturbations in a unified way, for tensor, vector and scalar perturbations. We note that in the equations some free functions, belonging to different models, appear in the same places. For example, the effective Planck mass of ST theories, i.e.\ $M_{S}^{2}$, often multiplies the same perturbation terms as the effective Planck mass of EA, VT or TT theories, i.e.\ $M_{A}^{2}$, $M_{V}^{2}$ and $M_{T}^{2}$, respectively. Then, it makes sense to compress the information and define a ``total'' Planck mass as the sum of the different Planck masses. This can be applied to all the functions which are common to different models. We thus define:
\begin{align}
	M_{*}^{2}\equiv\, & M_{S}^{2}+M_{A}^{2}+M_{V}^{2}+M_{T}^{2},\\
	\alpha_{\textrm{M}}\equiv\, & \check{\alpha}_{\textrm{M}}+\hat{\alpha}_{\textrm{M}}+\bar{\alpha}_{\textrm{M}}+\tilde{\alpha}_{\textrm{M}},\\
	\alpha_{\textrm{K}}\equiv\, & \hat{\alpha}_{\textrm{K}}+\left(\check{\alpha}_{\textrm{K}}+\tilde{\alpha}_{\textrm{K}}\right)\frac{k^{2}}{\mathcal{H}^{2}},\\
	\alpha_{\textrm{D}}\equiv\, & \check{\alpha}_{\textrm{D}}+\tilde{\alpha}_{\textrm{D}},\\
	\alpha_{\textrm{C}}\equiv\, & \check{\alpha}_{\textrm{C}}+\tilde{\alpha}_{\textrm{C}},\\
	\alpha_{\textrm{T}}\equiv\, & \check{\alpha}_{\textrm{T}}+\hat{\alpha}_{\textrm{T}}+\tilde{\alpha}_{\textrm{T}}+\bar{\alpha}_{\textrm{M}}\,.
\end{align}
It should be noted that we defined the ``total'' tensor speed excess $\alpha_{\textrm{T}}$ as the sum of the tensor speed excess of ST, EA and VT theories and the Planck mass run-rate $\bar{\alpha}_{\textrm{M}}$ of TT theories. This comes from the fact that in TT theories there is not an independent tensor speed excess, but it is determined by the time variations of the effective Planck mass. In the following we will consider as ``total'' functions also the $\{\alpha_X(\tau)\}$ that are peculiar to just one class of models, e.g.\ $\alpha_{\textrm{B}}\equiv\, \hat{\alpha}_{\textrm{B}}$, since there can be no confusion.

\subsubsection{Tensor Perturbations}

Tensor modes are dynamical in GR and are also dynamical in this general parametrised framework. The peculiarity here is that in TT theories the physical metric tensor ($h_{ij}$) is coupled to a second dynamical metric tensor ($h_{2ij}$), which implies that we need to solve a second equation. They read:
\begin{align}
	& h^{\prime\prime}+\left(2+\alpha_{\textrm{M}}\right)\mathcal{H}h^{\prime}+\left(1+\alpha_{\textrm{T}}\right)k^{2}h-\left(3\alpha_{\textrm{S}}-\alpha_{TI}\right)\mathcal{H}^{2}\left[h-h_{2}\right]=2\frac{a^{2}\sigma_{\textrm{m}}}{M_{*}^{2}}\label{eq:tensor1},\\
	& h_{2}^{\prime\prime}+\frac{\left(a^{2}M_{*}^{2}\nu_{\textrm{M}}\right)^{\prime}}{a^{2}M_{*}^{2}\nu_{\textrm{M}}}h_{2}^{\prime}+\left[\frac{\left(a^{2}M_{*}^{2}\alpha_{\textrm{N}}\nu_{\textrm{M}}\right)^{\prime}}{a^{2}M_{*}^{2}\mathcal{H}\alpha_{\textrm{N}}\nu_{\textrm{M}}}-\alpha_{\textrm{b}}\right]\frac{\alpha_{\textrm{N}}^{2}}{\alpha_{\textrm{b}}}k^{2}h_{2}+\left(3\alpha_{\textrm{S}}-\alpha_{TI}\right)\frac{\mathcal{H}^{2}}{\nu_{\textrm{M}}}\left[h-h_{2}\right]=0\label{eq:tensor2}\,,
\end{align}
where we omitted the spatial indices in the tensor perturbations and $\sigma_{\textrm{m}ij}$ represents the source for the tensor modes arising from the matter. In order to solve this system one has to:
\begin{itemize}
	\item ST, EA and VT theories: solve only eq.\ (\ref{eq:tensor1}) with appropriate initial conditions for $h_{ij}$.
	\item TT theories: solve both eqs.\ (\ref{eq:tensor1}) and (\ref{eq:tensor2}) with appropriate initial conditions for $h_{ij}$ and $h_{2ij}$.
\end{itemize}

\subsubsection{Vector Perturbations}

Vector perturbations are usually neglected, since their amplitude decays rapidly with the expansion of the universe. However, in EA, VT and TT theories there is an extra propagating vector mode, which may bring non-trivial modifications to the evolution (see for instance \cite{Lagos:2014lca} for massive bigravity). For this reason, we show the complete set of equations here, which read:
\begin{align}
	F^{\prime}= & -2\frac{a^{2}\theta_{\textrm{m}}}{M_{*}^{2}k^{2}}+\alpha_{\textrm{C}}\chi+\frac{2\alpha_{SI}\mathcal{H}^{2}\Gamma}{\left(1+\alpha_{\textrm{N}}\right)k^{2}}\label{eq:vector1},\\
	\left(\check{\alpha}_{\textrm{K}}+\tilde{\alpha}_{\textrm{K}}+\nu_{\textrm{M}}\right)\chi^{\prime\prime}= & -\frac{1}{a^{2}M_{*}^{2}}\left[a^{2}M_{*}^{2}\nu_{\textrm{M}}\left(F^{\prime}+\Gamma\right)\right]^{\prime}+\frac{1}{2}\alpha_{\textrm{C}}k^{2}F^{\prime}\label{eq:vector2}\\
	& -\frac{\left(a^{2}M_{*}^{2}\left(\check{\alpha}_{\textrm{K}}+\tilde{\alpha}_{\textrm{K}}+\nu_{\textrm{M}}\right)\right)^{\prime}}{a^{2}M_{*}^{2}}\chi^{\prime}+\frac{1}{2}\left(\alpha_{\textrm{C}}^{2}-2\alpha_{\textrm{K}}\alpha_{\textrm{VS}}\right)k^{2}\chi\nonumber \\
	& +\left[\left(2\alpha_{\textrm{C}}+\check{\alpha}_{\textrm{D}}\right)\left(1-\frac{\mathcal{H}^{\prime}}{\mathcal{H}^{2}}\right)-\frac{\left(aM_{*}^{2}\mathcal{H}\left(\check{\alpha}_{\textrm{K}}+\tilde{\alpha}_{\textrm{K}}\right)\right)^{\prime}}{aM_{*}^{2}\mathcal{H}^{2}}\right.\nonumber \\
	& \left.\quad-\frac{\left(M_{*}^{2}\mathcal{H}\alpha_{\textrm{A}}\right)^{\prime}}{M_{*}^{2}\mathcal{H}^{2}}+3\alpha_{\textrm{S}}+\alpha_{\textrm{V}}-\alpha_{TI}\right]\mathcal{H}^{2}\chi\nonumber, \\
	F^{\prime}= & -\chi^{\prime}-\left[1+\frac{2\alpha_{SI}\mathcal{H}^{2}}{\nu_{\textrm{M}}\left(1+\alpha_{\textrm{N}}\right)k^{2}}\right]\Gamma\label{eq:vector3}\,,
\end{align}
where we omitted the spatial index in the vector perturbations and $\theta^i_{\textrm{m}}$ describes the vector velocity of all the matter components. To solve this system one has to follow different steps, depending on the model considered:
\begin{itemize}
	\item ST theories: the only relevant equation is eq.\ (\ref{eq:vector1}). Remember that by setting all the EA, VT and TT $\alpha_X(\tau)=0$, the terms proportional to $\chi^i$ and $\Gamma^i$ vanish, and hence eqs.~(\ref{eq:vector2}) and (\ref{eq:vector3}) are automatically satisfied. The solution for $F^{i\prime}$ can be straightforwardly obtained once the evolution of matter perturbations is known.
	\item EA and VT theories: the relevant equations are eqs.\ (\ref{eq:vector1}) and (\ref{eq:vector2}). For these models, the terms involving $\Gamma^i$ vanish. The solution for $F^{i\prime}$ can be obtained through eq.\ (\ref{eq:vector1}) once $\chi^i$ and matter are known. The solution for $\chi^i$ can be obtained by evolving the combination of eqs.\ (\ref{eq:vector2}) and (\ref{eq:vector1}) that eliminates $F^{i\prime}$ with the appropriate initial conditions for $\chi^i$ and $\chi^{i\prime}$.
	\item TT theories: all the equations have to be used. $F^{i\prime}$ can be obtained through eq.\ (\ref{eq:vector1}) once $\chi^i$, $\Gamma^i$ and matter are known. $\Gamma^i$ can be obtained by removing $F^{i\prime}$ with eq.\ (\ref{eq:vector1}) in eq.\ (\ref{eq:vector3}). $\chi^i$ is the solution of the differential equation (\ref{eq:vector2}). To decouple it from $F^{i\prime}$ and $\Gamma^i$, one has to take the time derivative of eqs.\ (\ref{eq:vector1}) and (\ref{eq:vector3}) and solve for $F^{i\prime\prime}$ and $\Gamma^{i\prime}$. Then one has to replace $F^{i\prime}$ and $\Gamma^i$ directly with eqs.\ (\ref{eq:vector1}) and (\ref{eq:vector3}).
\end{itemize}

\subsubsection{Scalar Perturbations}

In this general framework scalar modes are described by the four usual Einstein equations in addition to a dynamical equation for the additional degree of freedom for ST and EA theories. VT theories are described by the same equations plus one constraint, while TT theories involve three additional constraint equations. The Einstein equations, common to all the theories read:
\begin{itemize}
	\item Einstein (0,0)
	\begin{align}
		\left[1-\frac{\alpha_{\textrm{B}}+\alpha_{\textrm{D}}}{2}\right]&\mathcal{H}h^{\prime}= \frac{\delta\rho_{\textrm{m}}}{M_{*}^{2}}+2\left(1+\alpha_{\textrm{C}}\right)k^{2}\eta+\frac{3\left(1-\alpha_{\textrm{N}}\right)\alpha_{SI}}{1-\alpha_{\textrm{b}}}\mathcal{H}^{2}\left[\eta-\mathcal{H}\Gamma_{1}\right]\label{eq:e00}\\
		& -\left(3\alpha_{\textrm{B}}+\alpha_{\textrm{K}}\right)\mathcal{H}^{2}\chi^{\prime}-\left[\alpha_{\textrm{B}}+2\alpha_{\textrm{C}}+\alpha_{\textrm{D}}+\alpha_{\textrm{A}}+\frac{1-\alpha_{\textrm{N}}}{1-\alpha_{\textrm{b}}}\alpha_{SI}\right]\mathcal{H}k^{2}\chi\nonumber \\
		& -\left[\alpha_{\textrm{K}}+3\alpha_{\textrm{B}}\left(2-\frac{\mathcal{H}^{\prime}}{\mathcal{H}^{2}}\right)-3a^{2}\frac{\mathcal{E}_{S}+\mathcal{P}_{S}}{M_{*}^{2}\mathcal{H}^{2}}\right]\mathcal{H}^{3}\chi\nonumber \\
		& -\left[3\tilde{\alpha}_{\textrm{D}}\left(1-\frac{\mathcal{H}^{\prime}}{\mathcal{H}^{2}}\right)-3a^{2}\frac{\mathcal{E}_{V}+\mathcal{P}_{V}}{M_{*}^{2}\mathcal{H}^{2}}-\alpha_{\textrm{A}}\frac{k^{2}}{\mathcal{H}^{2}}\right]\mathcal{H}^{3}\Gamma_{0}\nonumber , 
	\end{align}
	
	\item Einstein (0,i)
	\begin{align}
		2\eta^{\prime}= & \frac{a^{2}\theta_{\textrm{m}}}{M_{*}^{2}k^{2}}-\frac{1}{6}\alpha_{\textrm{D}}h^{\prime}+\alpha_{\textrm{B}}\mathcal{H}\chi^{\prime}+\left[\alpha_{\textrm{B}}-a^{2}\frac{\mathcal{E}_{S}+\mathcal{P}_{S}}{M_{*}^{2}\mathcal{H}^{2}}+\frac{1}{3}\alpha_{\textrm{D}}\frac{k^{2}}{\mathcal{H}^{2}}\right]\mathcal{H}^{2}\chi\label{eq:e0i}\\
		& +\left[\tilde{\alpha}_{\textrm{D}}\left(1-\frac{\mathcal{H}^{\prime}}{\mathcal{H}^{2}}\right)-a^{2}\frac{\mathcal{E}_{V}+\mathcal{P}_{V}}{M_{*}^{2}\mathcal{H}^{2}}\right]\mathcal{H}^{2}\Gamma_{0}+\frac{\alpha_{SI}}{1+\alpha_{\textrm{N}}}\mathcal{H}^{2}\Gamma_{2}\nonumber ,
	\end{align}
	
	\item Einstein (i,j) trace
	\begin{align}
		2\eta^{\prime\prime}= & \frac{a^{2}}{M_{*}^{2}}\left(\delta p_{\textrm{m}}-\sigma_{\textrm{m}}\right)-\frac{1}{6}\alpha_{\textrm{D}}h^{\prime\prime}-2\left(2+\alpha_{\textrm{M}}\right)\mathcal{H}\eta^{\prime}-\frac{1}{6}\frac{\left(a^{2}M_{*}^{2}\alpha_{\textrm{D}}\right)^{\prime}}{a^{2}M_{*}^{2}}h^{\prime}\label{eq:eii}\\
		& -2\mathcal{H}^{2}\alpha_{TI}\left(\eta-\mathcal{H}\Gamma_{1}\right)+\alpha_{\textrm{B}}\mathcal{H}\chi^{\prime\prime}+\frac{1}{3}k^{2}\left[6\alpha_{\textrm{S}}+\frac{\left(a^{2}M_{*}^{2}\alpha_{\textrm{D}}\right)^{\prime}}{a^{2}M_{*}^{2}\mathcal{H}}\right]\mathcal{H}\chi\nonumber \\
		& +\left[\frac{\left(a^{3}M_{*}^{2}\mathcal{H}\alpha_{\textrm{B}}\right)^{\prime}}{a^{3}M_{*}^{2}\mathcal{H}^{2}}-a^{2}\frac{\mathcal{E}_{S}+\mathcal{P}_{S}}{M_{*}^{2}\mathcal{H}^{2}}+\frac{1}{3}\alpha_{\textrm{D}}\frac{k^{2}}{\mathcal{H}^{2}}\right]\mathcal{H}^{2}\chi^{\prime}\nonumber \\
		& +\left[\frac{\left(a^{2}M_{*}^{2}\mathcal{H}^{2}\alpha_{\textrm{B}}\right)^{\prime}}{a^{2}M_{*}^{2}\mathcal{H}^{3}}-a^{2}\frac{\mathcal{P}_{S}^{\prime}+\mathcal{H}\left(\mathcal{E}_{S}+\mathcal{P}_{S}\right)}{M_{*}^{2}\mathcal{H}^{3}}-3\hat{\alpha}_{\textrm{M}}\right]\mathcal{H}^{3}\chi\nonumber \\
		& +\frac{1}{a^{2}M_{*}^{2}}\left[\left(a^{2}M_{*}^{2}\left(\mathcal{H}^{2}-\mathcal{H}^{\prime}\right)\tilde{\alpha}_{\textrm{D}}-a^{4}\left(\mathcal{E}_{V}+\mathcal{P}_{V}\right)\right)\Gamma_{0}\right]^{\prime}-\frac{1-\alpha_{\textrm{N}}}{1-\alpha_{\textrm{b}}}\alpha_{\textrm{b}}\alpha_{SI}\mathcal{H}^{3}\Gamma_{3}\nonumber ,
	\end{align}
	
	\item Einstein (i,j) traceless
	\begin{align}
		h^{\prime\prime}= & -3\frac{a^{2}}{M_{*}^{2}}\sigma_{\textrm{m}}-6\eta^{\prime\prime}-\mathcal{H}\left(2+\alpha_{\textrm{M}}\right)\left[h^{\prime}+6\eta^{\prime}\right]+2\left(1+\alpha_{\textrm{T}}\right)k^{2}\eta\label{eq:eij}\\
		& -2\alpha_{\textrm{C}}k^{2}\chi^{\prime}+2\left[\frac{\left(aM_{*}^{2}\tilde{\alpha}_{\textrm{C}}\right)^{\prime}}{aM_{*}^{2}\mathcal{H}}+\tilde{\alpha}_{\textrm{M}}-\tilde{\alpha}_{\textrm{T}}\right]\mathcal{H}k^{2}\Gamma_{0}\nonumber \\
		& -2\left[\frac{\left(a^{2}M_{*}^{2}\alpha_{\textrm{C}}\right)^{\prime}}{a^{2}M_{*}^{2}\mathcal{H}}-\hat{\alpha}_{\textrm{M}}+\hat{\alpha}_{\textrm{T}}-3\alpha_{\textrm{S}}+\alpha_{TI}\right]\mathcal{H}k^{2}\chi\nonumber .
	\end{align}
\end{itemize}
The last equation in common with all the theories under consideration is a dynamical equation for the additional new degree of freedom
\begin{align}
	\left[\alpha_{\textrm{K}}+\frac{k^{2}}{\mathcal{H}^{2}}\bar{\nu}_{\textrm{M}}\right]&\mathcal{H}^{2}\chi^{\prime\prime}= \frac{1}{2}\alpha_{\textrm{B}}\mathcal{H}h^{\prime\prime}+\frac{1}{2}\mathcal{H}\nu_{\textrm{M}}\left(h^{\prime\prime}+6\eta^{\prime\prime}\right)+\frac{1}{3}\alpha_{\textrm{D}}k^{4}\chi\label{eq:d}\\
	& +2\left(\hat{\alpha}_{\textrm{M}}-\hat{\alpha}_{\textrm{T}}\right)\mathcal{H}k^{2}\eta+\left[3\frac{\left(a^{2}M_{*}^{2}\nu_{\textrm{M}}\right)^{\prime}}{a^{2}M_{*}^{2}\mathcal{H}}+2\alpha_{\textrm{C}}\frac{k^{2}}{\mathcal{H}^{2}}\right]\mathcal{H}^{2}\eta^{\prime}\nonumber \\
	& +\frac{1}{2}\left[\frac{\left(aM_{*}^{2}\mathcal{H}\alpha_{\textrm{B}}\right)^{\prime}}{aM_{*}^{2}\mathcal{H}^{2}}+\frac{\left(a^{2}M_{*}^{2}\nu_{\textrm{M}}\right)^{\prime}}{a^{2}M_{*}^{2}\mathcal{H}}+a^{2}\frac{\mathcal{E}_{S}+\mathcal{P}_{S}}{M_{*}^{2}\mathcal{H}^{2}}-\frac{1}{3}\alpha_{\textrm{D}}\frac{k^{2}}{\mathcal{H}^{2}}\right]\mathcal{H}^{2}h^{\prime}\nonumber \\
	& +\nu_{\textrm{M}}\mathcal{H}k^{2}\left(\Gamma_{2}^{\prime}+\alpha_{\textrm{N}}^{2}\mathcal{H}\Gamma_{3}\right)+\alpha_{\textrm{A}}\mathcal{H}k^{2}\Gamma_{0}^{\prime}+\frac{\left(a^{2}M_{*}^{2}\nu_{\textrm{M}}\right)^{\prime}}{a^{2}M_{*}^{2}}\mathcal{H}k^{2}\Gamma_{2}\nonumber \\
	& +\left[\frac{\left(aM_{*}^{2}\mathcal{H}\alpha_{\textrm{A}}\right)^{\prime}}{aM_{*}^{2}}+\tilde{\alpha}_{\textrm{D}}\left(\mathcal{H}^{2}-\mathcal{H}^{\prime}\right)-\alpha_{\textrm{V}}\mathcal{H}^{2}\right]k^{2}\Gamma_{0}\nonumber \\
	& -\alpha_{\textrm{N}}\left[\frac{\left(a^{2}M_{*}^{2}\alpha_{\textrm{N}}\nu_{\textrm{M}}\right)^{\prime}}{a^{2}M_{*}^{2}\alpha_{\textrm{b}}}-\alpha_{\textrm{N}}\mathcal{H}\nu_{\textrm{M}}\right]\mathcal{H}k^{2}\Gamma_{1}\nonumber \\
	& -\left[\frac{\left(a^{2}M_{*}^{2}\mathcal{H}^{2}\alpha_{\textrm{K}}\right)^{\prime}}{\mathcal{H}^{3}}+\mathcal{H}\left(a^{2}M_{*}^{2}\mathcal{H}^{-2}\nu_{\textrm{M}}\right)^{\prime}\frac{k^{2}}{\mathcal{H}^{2}}\right]\frac{\mathcal{H}^{3}}{a^{2}M_{*}^{2}}\chi^{\prime}\nonumber \\
	& +\left[\frac{\left(a^{2}M_{*}^{2}\mathcal{H}^{-2}\mathcal{H}^{\prime}\nu_{\textrm{M}}\right)^{\prime}}{a^{2}M_{*}^{2}\mathcal{H}}-\frac{\left(M_{*}^{2}\mathcal{H}\left(\alpha_{\textrm{A}}+\alpha_{\textrm{B}}\right)\right)^{\prime}}{M_{*}^{2}\mathcal{H}^{2}}+\left(2\alpha_{\textrm{C}}+\check{\alpha}_{\textrm{D}}\right)\left(1-\frac{\mathcal{H}^{\prime}}{\mathcal{H}^{2}}\right)\right.\nonumber \\
	& \left.\quad-2\left(\hat{\alpha}_{\textrm{M}}-\hat{\alpha}_{\textrm{T}}\right)+3\alpha_{\textrm{S}}+\alpha_{\textrm{V}}-\alpha_{TI}-a^{2}\frac{\mathcal{E}_{S}+\mathcal{P}_{S}}{M_{*}^{2}\mathcal{H}^{2}}\right]\mathcal{H}^{2}k^{2}\chi\nonumber \\
	& -\left[\frac{\left(aM_{*}^{2}\mathcal{H}^{3}\alpha_{\textrm{K}}\right)^{\prime}}{aM_{*}^{2}\mathcal{H}^{4}}+3\frac{\left(\mathcal{H}M_{*}^{2}\left(\mathcal{H}^{2}-\mathcal{H}^{\prime}\right)\alpha_{\textrm{B}}\right)^{\prime}}{M_{*}^{2}\mathcal{H}^{4}}+3a^{2}\left(1-\frac{\mathcal{H}^{\prime}}{\mathcal{H}^{2}}\right)\frac{\mathcal{E}_{S}+\mathcal{P}_{S}}{M_{*}^{2}\mathcal{H}^{2}}\right]\mathcal{H}^{4}\chi\nonumber .
\end{align}
Finally, in VT and TT theories scalar perturbations satisfy the following additional set of constraint equations:
\begin{itemize}
	\item VT theories
	\begin{align}
		& 3\left(\mathcal{H}^{2}-\mathcal{H}^{\prime}\right)\left[\tilde{\alpha}_{\textrm{D}}\left(1-\frac{\mathcal{H}^{\prime}}{\mathcal{H}^{2}}\right)-a^{2}\frac{\mathcal{E}_{V}+\mathcal{P}_{V}}{M_{*}^{2}\mathcal{H}^{2}}\right]\mathcal{H}^{2}\Gamma_{0}\label{eq:c0}\\
		& -\left[2\left(\tilde{\alpha}_{\textrm{M}}-\tilde{\alpha}_{\textrm{T}}\right)-\tilde{\alpha}_{\textrm{V}}+2\frac{\left(aM_{*}^{2}\tilde{\alpha}_{\textrm{C}}\right)^{\prime}}{aM_{*}^{2}\mathcal{H}}+a^{2}\frac{\mathcal{E}_{V}+\mathcal{P}_{V}}{M_{*}^{2}\mathcal{H}^{2}}\right]\mathcal{H}^{2}k^{2}\Gamma_{0}\nonumber \\
		& +2\left[\tilde{\alpha}_{\textrm{M}}-\tilde{\alpha}_{\textrm{T}}+\frac{\left(aM_{*}^{2}\tilde{\alpha}_{\textrm{C}}\right)^{\prime}}{aM_{*}^{2}\mathcal{H}}\right]\mathcal{H}k^{2}\eta+\left[\tilde{\alpha}_{\textrm{D}}\left(1-\frac{\mathcal{H}^{\prime}}{\mathcal{H}^{2}}\right)-\alpha_{\textrm{A}}-\alpha_{\textrm{V}}\right]\mathcal{H}^{2}k^{2}\chi\nonumber \\
		& -\tilde{\alpha}_{\textrm{A}}\mathcal{H}k^{2}\chi^{\prime}-\frac{1}{2}\left[\tilde{\alpha}_{\textrm{D}}\left(1-\frac{\mathcal{H}^{\prime}}{\mathcal{H}^{2}}\right)-a^{2}\frac{\mathcal{E}_{V}+\mathcal{P}_{V}}{M_{*}^{2}\mathcal{H}^{2}}\right]\mathcal{H}^{2}h^{\prime}=0\nonumber ,
	\end{align}
	
	\item TT theories
	\begin{align}
		& \left[3\frac{1-\alpha_{\textrm{N}}}{1-\alpha_{\textrm{b}}}\alpha_{SI}+2\frac{\alpha_{\textrm{N}}^{2}}{\alpha_{\textrm{b}}}\nu_{\textrm{M}}\frac{k^{2}}{\mathcal{H}^{2}}\right]\mathcal{H}^{4}\Gamma_{1}-\left[\frac{3\alpha_{SI}}{1+\alpha_{\textrm{N}}}+2\nu_{\textrm{M}}\frac{k^{2}}{\mathcal{H}^{2}}\right]\mathcal{H}^{4}\Gamma_{2}\label{eq:c1}\\
		&-\nu_{\textrm{M}}\mathcal{H}^{2}\left[h^{\prime}+6\eta^{\prime}\right]-3\frac{1-\alpha_{\textrm{N}}}{1-\alpha_{\textrm{b}}}\alpha_{SI}\mathcal{H}^{3}\eta\nonumber \\
		& +\left[\alpha_{SI}\frac{1-\alpha_{\textrm{N}}}{1-\alpha_{\textrm{b}}}-2\nu_{\textrm{M}}\frac{\mathcal{H}^{\prime}}{\mathcal{H}^{2}}\right]\mathcal{H}^{2}k^{2}\chi+2\nu_{\textrm{M}}\mathcal{H}k^{2}\chi^{\prime}=0\nonumber, \\
		& 2\alpha_{\textrm{b}}\Gamma_{3}= -\frac{\alpha_{SI}\Gamma_{2}}{\left(1+\alpha_{\textrm{N}}\right)\nu_{\textrm{M}}}-\frac{2}{\mathcal{H}^{2}}\left(\mathcal{H}\Gamma_{1}\right)^{\prime}\label{eq:c2}, \\
		& 2\alpha_{\textrm{S}}\frac{k^{2}}{\mathcal{H}^{2}}\chi-2\alpha_{TI}\left(\frac{\eta}{\mathcal{H}}-\Gamma_{1}\right)-\frac{1-\alpha_{\textrm{N}}}{1-\alpha_{\textrm{b}}}\alpha_{\textrm{b}}\alpha_{SI}\Gamma_{3}-\frac{\left[\frac{a^{2}M_{*}^{2}\mathcal{H}^{2}\alpha_{SI}}{1+\alpha_{\textrm{N}}}\Gamma_{2}\right]^{\prime}}{a^{2}M_{*}^{2}\mathcal{H}^{3}}=0\label{eq:c3}.
	\end{align}
\end{itemize}

This forms the complete set of equations that one has to implement in an EB solver to describe the evolution of scalar perturbations in ST, VT, EA or TT theories. In order to implement these equations in a code we need to diagonalise them. Next, we give the steps necessary to do so. Excluding matter, which has its own conservation equations to satisfy, the full list of perturbations we have in these equations is $\{h^{\prime},\,h^{\prime\prime},\,\eta,\,\eta^{\prime},\,\eta^{\prime\prime},\,\chi,\,\chi^{\prime},\,\chi^{\prime\prime},\,\Gamma_0,\,\Gamma_0^{\prime},\,\Gamma_1,\,\Gamma_1^{\prime},\,\Gamma_2,$ $\Gamma_2^{\prime},\,\Gamma_3\}$ and we need to find them all. As in GR, we will give initial conditions for $\eta$, although it is not a dynamical variable but for numerical purposes it is convenient in synchronous gauge to calculate it from the integrator rather than from algebraic relations. Furthermore, in all these modified gravity models we will have the dynamical variable $\chi$, and thus we will obtain $\chi$ and $\chi^\prime$ from the integrator as well. This reduces the number of perturbations from 15 to 12, which can all be obtained from algebraic relations following these steps in order:
\begin{enumerate}
	\item TT: obtain $\Gamma_2$ from the time derivative of eq.\ (\ref{eq:c1}). One has to remove the set $\{h^{\prime},\,\eta^{\prime},\,\chi^{\prime\prime},\,\Gamma_1,\,\Gamma_3,\,\Gamma_2^{\prime}\}$ by using eqs.\ (\ref{eq:e00}), (\ref{eq:e0i}), (\ref{eq:d}), (\ref{eq:c1}), (\ref{eq:c2}) and (\ref{eq:c3}). Naively, one would also generically expect to have a term with $\Gamma_1^{\prime}$. However, after replacing these constraints, the dependence on $\Gamma_1^{\prime}$ vanishes leaving only $\Gamma_2$, matter perturbations and the integrator variables $\{\eta,\,\chi,\,\chi^\prime\}$.
	\item TT: obtain $\Gamma_2^\prime$ from the time derivative of the equation calculated in step 1. Eqs.\ (\ref{eq:e00}), (\ref{eq:e0i}), (\ref{eq:eii}), (\ref{eq:eij}) and (\ref{eq:d}) have to be used to eliminate the variables $\{h^{\prime},\,\eta^{\prime},\,h^{\prime\prime},\,\eta^{\prime\prime},\,\chi^{\prime\prime}\}$.
	\item TT: obtain $\Gamma_1$ from eq.\ (\ref{eq:c1}). $h^\prime$ and $\eta^\prime$ have to be removed with eqs.\ (\ref{eq:e00}) and (\ref{eq:e0i}).
	\item TT: obtain $\Gamma_1^\prime$ from eq.\ (\ref{eq:c2}).
	\item TT: obtain $\Gamma_3$ from eq.\ (\ref{eq:c3}).
	\item VT: obtain $\Gamma_0$ from eq.\ (\ref{eq:c0}). $h^\prime$ has to be removed with eq.\ (\ref{eq:e00}).
	\item VT: obtain $\Gamma_0^\prime$ from the time derivative of eq.\ (\ref{eq:c0}). $\chi^{\prime\prime}$ has to be removed using eq.\ (\ref{eq:d}), $h^{\prime\prime}$ using eq.\ (\ref{eq:eij}), $h^\prime$ with eq.\ (\ref{eq:e00}) and $\eta^\prime$ with eq.\ (\ref{eq:e0i}).
	\item ALL: obtain $h^\prime$ from eq.\ (\ref{eq:e00}).
	\item ALL: obtain $\eta^\prime$ from eq.\ (\ref{eq:e0i}).
	\item ALL: obtain $\eta^{\prime\prime}$ from eq.\ (\ref{eq:eii}). $\chi^{\prime\prime}$ has to be removed using eq.\ (\ref{eq:d}) and $h^{\prime\prime}$ with eq.\ (\ref{eq:eij}).
	\item ALL: obtain $h^{\prime\prime}$ from eq.\ (\ref{eq:eij}).
	\item ALL: obtain $\chi^{\prime\prime}$ from eq.\ (\ref{eq:d}).
\end{enumerate}

As a final remark, it is important to note that the constraint eq.\ (\ref{eq:c0}) is singular if the VT parameters $\alpha_X$ vanish. This implies that in order to implement these equations, a flag that excludes this equation when dealing with non-VT theories is needed. In the same way one has to add a flag to exclude eqs.\ (\ref{eq:c1}), (\ref{eq:c2}) and (\ref{eq:c3}) in non-TT theories. 


\section{Discussion}\label{Sec:discussion}
In this paper we have analysed the dynamics of linear cosmological perturbations of the most general modified gravity theories that are linearly diffeomorphism invariant, have second-order derivative equations of motion, and whose field content is given by a spacetime metric in addition to either a scalar, a vector or a tensor field. We describe all these models in terms of free parameters that characterise different possible gravitational interactions that can be present and can modify general relativity. In particular, we use the formalism developed in \cite{Lagos:2016wyv,Lagos:2016gep} and consider four families of parametrised models: scalar-tensor, general vector-tensor, Einstein-Aether, and tensor-tensor gravity. This parametrised approach allows us to analyse, in a unified and efficient manner, the cosmological predictions of all these models, which encompass specific non-linear theories such as Horndeski, Generalised Proca, generalised Einstein-Aether and massive bigravity. 

We have standardised the way modifications to gravity are parametrised. In particular, we write all models in such a way that the dynamics of linear perturbations is completely determined by one background parameter (the scale factor), one mass scale, and a set of dimensionless parameters $\alpha_X$. In all cases, we recover GR when all $\alpha_X$ vanish, the mass scale is the Planck mass, and the evolution of the scale factor is determined by the Friedmann equation. We summarise the number of free parameters describing scalar cosmological perturbations in each case in Table \ref{SummaryModels}. We have restricted ourselves to theories with minimal coupling to the matter sector and we have not included theories which invoke degeneracies in the action to cancel higher order degrees of freedom (such as, for example ``beyond Horndeski" theories of \cite{Gleyzes:2014dya,Zumalacarregui:2013pma,Langlois:2017mxy}); this can be done following the results found in \cite{Lagos:2016wyv,Lagos:2016gep}.

Modified gravity theories can be plagued by instabilities that manifest themselves as growing solutions of the linear cosmological perturbations. These end up breaking the perturbative validity of the calculations, and lead to unviable models. For this reason, in this paper we analyse the stability of the four families of parametrised models. We find general conditions that the free parameters need to satisfy in order to avoid ghost and gradient instabilities in the large-$k$ limit. Our results are summarised in Tables \ref{SummaryStability} and \ref{GradientStability}.

In addition, we analyse the behaviour of all the models in the quasistatic limit. This allows us to determine how the $\alpha_i$ parameters relate to quantities influencing cosmological observables, such as the effective Newton's constant and the effective shear. While for general vector-tensor and bimetric theories it is not possible to make general qualitative statements about their behaviour in this regime, we find that most scalar-tensor and Einstein-Aether models have a positive sign for the ratio of $(\Sigma-1)/(\mu-1)$, where $\Sigma$ describes a modified lensing potential and $\mu$ an effective Newton's constant. We also find that for models with tensor modes propagating at the speed of light, evidence for $\Sigma\not=\mu$ would signal the presence of a running of the Planck mass in scalar-tensor theories, or an effective conformal coupling in Einstein-Aether models.

Finally, we present a set of unified equations of motion that determine the evolution of linear cosmological perturbations in all four families of models. We present these equations in synchronous gauge, so that they can be implemented in numerical Einstein-Boltzmann codes for modified gravity, in a manner similar to EFTCAMB or HiCLASS. Such codes can be used for analysing data from future large-scale structure surveys such as Euclid, SKA, LSST and WFIRST, allowing us to test and compare the performance of theories. We also discuss considerations to take into account when implementing stability conditions in these codes.

A key aspect that we have not addressed is the theoretical priors (and non-cosmological constraints) that we should impose on the free parameters $\alpha_X$. As was shown in \cite{Baker:2017hug,Creminelli:2017sry,Sakstein:2017xjx,Ezquiaga:2017ekz}, constraints on the gravitational waves from GW170817 and its electromagnetic counterpart GRB170817A severely limit the values of some of $\alpha_X$: in practice we have that $\alpha_T\simeq0$. But given the underlying structure of the $\alpha_X$ as functions of some background fields, it remains to be explored how other $\alpha_X$ might also be constrained. Combined with the stability constraints from Section \ref{Sec:constraints} one expects that only a reduced subspace of the $\alpha_X$ needs to be explored in any attempts at parameter estimation. We believe that, now that we have a unified approach to tackling this broad class of theories, the focus should be on identifying an efficient yet complete parametrisation of the $\alpha_X$. With this in hand, it will be possible to completely characterise a vast space of gravitational theories on cosmological scales.


\begin{acknowledgments}
JN acknowledges support from Dr.~Max Rössler, the Walter Haefner Foundation and the ETH Zurich Foundation. TB is supported by All Souls College, University of Oxford. EB is supported by the ERC and BIPAC. PGF acknowledges support from STFC, BIPAC, the Higgs Centre at the University of Edinburgh and ERC. ML is supported by the Kavli Institute for Cosmological Physics through an endowment from the Kavli Foundation and its founder Fred Kavli. 
\end{acknowledgments}


\appendix

\section{Stability Conditions} \label{appendix-stability}

\subsection{Diagonalising Interactions} \label{appendix-asym}

Here we wish to prove, that one can always express the reduced Lagrangian for the two dynamical scalar degrees of freedom in the form \eqref{asymLag}. To prove this, we start with the following Lagrangian (in Fourier space) for the two fields $\tilde{X}_1$ and $\tilde{X}_2$:
\begin{equation}
L= \dot{\tilde{X}}^2_1 + \dot{\tilde{X}}^2_2 + D\left(\dot{\tilde{X}}_1\tilde{X}_2-\dot{\tilde{X}}_2\tilde{X}_1\right) -\mu_1 \tilde{X}_1^2-\mu_2 \tilde{X}_2^2 - 2 \mu_M \tilde{X}_1 \tilde{X}_2,
\label{appL-v0}
\end{equation}
where $D,\mu_i$ are all functions of $t$ and $k$. This form can always be straightforwardly achieved, by diagonalising the kinetic interactions, canonically normalising the fields and integrating by parts in order to put the kinetic-mass-mixing term into the above antisymmetric form. Note that we have assumed that we have two `correct'-sign kinetic terms, i.e~no ghosts, here. We now perform the following field re-definitions
\begin{align}
\tilde{X}_1 &\to \tilde{X}_1 + c_1 \tilde{X}_2, &\tilde{X}_2 &\to \tilde{X}_2 + c_2 \tilde{X}_1,
\end{align}
where the $c_i$ can be (and generally are) functions of $t$ and $k$ as well. After removing terms of the form $f(t,k)\tilde X \dot{\tilde X}$ via integration by parts, the resulting action can be written
\begin{align}
L &= (1+c_2^2)\dot{\tilde{X}}^2_1 + (1+c_1^2)\dot{\tilde{X}}^2_2 +(D - c_1 c_2 D + 2 \dot c_2)\dot{\tilde{X}}_2\tilde{X}_1 -(D - c_1 c_2 D - 2 \dot c_1)\dot{\tilde{X}}_1\tilde{X}_2
\nn \\
&-\left[\mu_1+c_2^2 \mu_2- D\dot c_2 + c_2 (2\mu_M+\ddot c_2)\right] \tilde{X}_1^2
-\left[\mu_2+c_1^2 \mu_1 + D \dot c_1 + c_1 (2\mu_M+\ddot c_1)\right] \tilde{X}_2^2 
\nn \\
&-\left[2 \mu_M + c_2 (2 \mu_2 + D \dot c_1)+ c_1 (2 \mu_1- D \dot c_2) + 2 c_1 c_2 \mu_3\right] \tilde{X}_1 \tilde{X}_2 + (2 c_1 + 2 c_2) \dot{\tilde{X}}_1 \dot{\tilde{X}}_2.
\label{appL-v1}
\end{align}
In order for us to be able to write this in the form of equation \eqref{asymLag}, we first need to avoid re-introducing kinetic mixing. This amounts to setting
\be
c_2 = - c_1,
\ee
which has the immediate nice effect, that it also guarantees that the anti-symmetric structure of the kinetic-mass mixing terms is not altered. This is because, as a consequence of imposing no kinetic mixing, we now also obtain the desired anti-symmetric form of the kinetic-mass mixing terms
\begin{align}
L_{\rm KM-mix} = (D + c_2^2 D + 2 \dot c_2)(\dot{\tilde{X}}_2\tilde{X}_1 - \dot{\tilde{X}}_1\tilde{X}_2)
\end{align}
With an eye on the form of \eqref{asymLag}, we then (canonically) normalise the kinetic terms as follows
\begin{align}
\tilde{X}_1 &\to \frac{\tilde{X}_1}{\sqrt{1+c_2^2}}, &\tilde{X}_2 &\to \frac{\tilde{X}_2}{\sqrt{1+c_2^2}}.
\end{align}
The argument of the square-root is trivially positive-definite, so the kinetic terms manifestly remain free of ghostly instabilities under this replacement. After some integration-by-parts, the resulting Lagrangian can be written
\begin{align}
L &= \dot{\tilde{X}}^2_1 + \dot{\tilde{X}}^2_2 +\frac{\left(D + c_2^2 D + 2 \dot c_2\right)}{1 + c_2^2}(\dot{\tilde{X}}_2\tilde{X}_1 -\dot{\tilde{X}}_1\tilde{X}_2)
\nn \\
&-\frac{\left(\left(1 + c_2^2\right) \mu_1 + c_2^4 \mu_2 + 2 c_2 \mu_M + 2 c_2^3 \mu_M - \left(D + \dot c_2\right) \dot c_2 + c_2^2 \left(\mu_2 - D \dot c_2\right)\right)}{\left(1 + c_2^2\right)^2} \tilde{X}_1^2 \nn \\
&-\frac{\left(c_2^4 \mu_1 + \mu_2 - 2 c_2 \mu_M - 2 c_2^3 \mu_M - \left(D + \dot c_2\right) \dot c_2 + c_2^2 \left(\mu_1 + \mu_2 - D \dot c_2\right)\right)}{\left(1 + c_2^2\right)^2} \tilde{X}_2^2 \nn \\
&-\frac{2 \left(c_2 \left(\mu_2 - \mu_1\right) + \mu_M - c_2^2 \mu_M\right)}{1 + c_a^2} \tilde{X}_1 \tilde{X}_2.
\end{align}
From this Lagrangian we can then see, that the final condition we need to impose (corresponding to requiring a diagonalised mass matrix), now reduces to the algebraic relation (the derivative terms contributing to the mass-mixing in \eqref{appL-v1} cancel with the above condition implemented)
\be
 c_2^2+ c_2 \frac{\left(\mu_1 - \mu_2\right)}{\mu_M} = 1,
\ee
which can be solved for $c_2$ to give
\be \label{c2sol}
c_2 = \frac{\mu_2 - \mu_1}{2 \mu_M} \pm \sqrt{\frac{\left(\mu_2 - \mu_1\right)^2}{4 \mu_M^2} + 1}.
\ee
With these choices of $c_1$ and $c_2$, the resulting form of the Lagrangian now is 
\begin{align}
L &= \dot{\tilde{X}}^2_1 + \dot{\tilde{X}}^2_2 + 
\tilde D(\dot{\tilde{X}}_2\tilde{X}_1 - \dot{\tilde{X}}_1\tilde{X}_2)
-\tilde \mu_1 \tilde{X}_1^2
 -\tilde \mu_2 \tilde{X}_2^2,
 \label{appL-v3}
\end{align}
where the anti-symmetric kinetic-mass mixing coefficient $\tilde D$ and diagonal mass coefficients $\tilde \mu_1, \tilde \mu_2$ are given in terms of $D, \mu_1, \mu_2,\mu_M$ from \eqref{appL-v0} and $c_2$ as defined in \eqref{c2sol} via
\begin{align}
\tilde D &= \frac{\left(D + c_2^2 D + 2 \dot c_2\right)}{1 + c_2^2}, \nn \\
\tilde \mu_1 &= \frac{\left(\left(1 + c_2^2\right) \mu_1 + c_2^4 \mu_2 + 2 c_2 \mu_M + 2 c_2^3 \mu_M - \left(D + \dot c_2\right) \dot c_2 + c_2^2 \left(\mu_2 - D \dot c_2\right)\right)}{\left(1 + c_2^2\right)^2}, \nn \\
\tilde \mu_2 &= \frac{\left(c_2^4 \mu_1 + \mu_2 - 2 c_2 \mu_M - 2 c_2^3 \mu_M - \left(D + \dot c_2\right) \dot c_2 + c_2^2 \left(\mu_1 + \mu_2 - D \dot c_2\right)\right)}{\left(1 + c_2^2\right)^2}.
\end{align}
The final Lagrangian \eqref{appL-v3} is now precisely of the form \eqref{asymLag}, as required. As we will explicitly see below, this form guarantees that the theory is free of instabilities, as long as $\tilde \mu_1, \tilde \mu_2 > 0$.

\subsection{Bounded Hamiltonian}
Let us start with the following Lagrangian in Fourier space for two fields $\tilde{X}_1$ and $\tilde{X}_2$:
\begin{equation}\label{LagrangianKineticMassMix}
L= \dot{\tilde{X}}^2_1 + \dot{\tilde{X}}^2_2 + d(t,k)\left(\dot{\tilde{X}}_1\tilde{X}_2-\dot{\tilde{X}}_2\tilde{X}_1\right) -m_1(t,k)^2\tilde{X}_1^2-m_2(t,k)^2\tilde{X}_2^2,
\end{equation}
where the $d$, $m_1$ and $m_2$ are generic functions of time and the wavenumber $k$. In order to explore the effect of the mixing coefficient $d$, we compute the Hamiltonian of this coupled system, which results in the following expression:
\begin{equation}\label{HamiltonKineticMassMix}
H =\frac{1}{2}\left[ P_1^2 + P_2^2 + d \left( P_1 Q_2 - P_2 Q_1 \right) + m_1^2 Q_1^2 + m_2^2 Q_2^2 \right].
\end{equation}
We notice that from here it is not clear whether the mixing terms can make the Hamiltonian unbounded from below and hence allow for infinitely large growing modes. However, if the write $H$ in terms of the Lagrangian variables $\dot{\tilde{X}}_{1,2}$ and $\tilde{X}_{1,2}$ it becomes:
\begin{equation}
H=\dot{\tilde{X}}^2_1 + \dot{\tilde{X}}^2_2 + m_1(t,k)^2\tilde{X}_1^2 + m_2(t,k)^2\tilde{X}_2^2,
\end{equation}
from where we explicitly see that the Hamiltonian is positive definite. We emphasise that the fact that $H$ does not depend on $d$ in this last expression does not mean that the physics of this field is independent of $d$, as here the Hamiltonian is written in terms of $\dot{\tilde{X}}_{1,2}$ and $\tilde{X}_{1,2}$, which are not the appropriate variables to derive the Hamiltonian equations. Indeed, in the same way that the equations of motion from the Lagrangian in eq.~(\ref{LagrangianKineticMassMix}) depend on $d$, the Hamiltonian equations from eq.~(\ref{HamiltonKineticMassMix}) will also depend on $d$.

\subsection{Small-$k$ Ghost Conditions} \label{appendix-smallKghosts}

Having discussed why low-energy ghosts are harmless in general in section \ref{Sec:constraints}, it is still interesting to list conditions for the presence of such a small-$k$ ghost, since it is one way for a (potentially desired) low-energy tachyonic instability to enter. In any case, they are also associated with phenomenological effects, so here we list the conditions for the existence of low-energy ghosts for the theories discussed in Section \ref{Sec:constraints}, which are compiled in Table \ref{SummaryStabilityII}.

\begin{table}[h!]
	\centering
	\begin{center}
		\begin{tabular}{| c | c | }
			\hline
			\rowcolor{gray!30} {\bf Theory} & {\bf Ghost Freedom Conditions (small k)}\\
			\hline \hline
			ST & $3\alpha_B^2 + 2\alpha_K > 0$\\
			\hline
			VT I \& II & $\tilde{\alpha}_K>0 \text{ and } \left(2 (\dot{H} + 3 H^2) M_V^2 + \rho+P \right) > 0$\\
			\hline
			EA I & $\tilde{\alpha}_K>0 \text{ and }\left(\check{\alpha}_\textrm{D} \dot{H} + 6 H^2\right)> 0$\\
			\hline
			EA II & $\check{\alpha}_K>0$ \\
			\hline
			TT & $(\nu_M^{-2}\bar{\alpha}_N+1)>0$ and $[2(\dot{H}+3H^2)+\nu_M^{-2}\bar{\alpha}_N(\bar{\alpha}_{SI}(\bar{\alpha}_N-1)+6H^2)]>0$ \\
			\hline
		\end{tabular}\caption{\label{SummaryStabilityII} Summary of the small-$k$ ghost stability conditions for the four models studied in this paper: scalar-tensor, general vector-tensor, Einstein-Aether, and bimetric gravity. VT I is the general vector-tensor case and VT II refers to the special $\check{\alpha}_\textrm{D} = 0$ case (incorporating Generalised Proca) - note that both cases give rise to identical conditions in the limit shown here. EA I is the general Einstein-Aether case and EA II refers to the special $\check{\alpha}_\textrm{D} = 0$ case that corresponds to having a $\Lambda$CDM background solution. In the case of scalar-tensor theories the condition is the same as for large $k$, since there is only a single condition that ensures ghost freedom in any energy regime for those theories.}
	\end{center}
\end{table}

\noindent {\bf Scalar-Tensor}: The conditions we gave in Section \ref{Sec:constraints} were $k$-independent for scalar-tensor theories, so we still have the straightforward 
\be
3\alpha_B^2 + 2\alpha_K > 0
\ee
as the only ghost-freedom condition here. Scalar-tensor theories of this type can therefore only have a small-$k$ ghost, if there is a large-$k$ ghost as well.
\\

\noindent {\bf Vector-Tensor I and II}: In the small-$k$ limit, kinetic interactions are automatically diagonalised here (the mixing term to leading order goes as $k$) and identical for both VT I and VT II, since the general kinetic term small-$k$ limit for VT I is already independent of $\tilde \alpha_D$. We find the following set of interactions
\begin{align}
S^{(2)}_{\rm kin, k\rightarrow 0} = \int d^3k dt
\left[\frac{1}{2} \tilde{\alpha}_\textrm{K} M_V^2 a \dot{\tilde{\delta A}}_1^2 + \frac{3 H^2 M_V^2 a^3}{\rho+P+ 2 \left(\dot{H} + 3 H^2\right) M_V^2}\delta \dot{\tilde{ \varphi}}^2
 \right].
\end{align}
Ghost-freedom conditions are therefore rather straightforward and read
\begin{align}
\check{\alpha}_\textrm{K} &> 0, \nn \\
\rho+P + 2 \left(\dot{H} + 3 H^2\right) M_V^2 &> 0.
\end{align}
Note that the second condition is effectively just a background-dependent condition on how negative $\dot H$ is allowed to be. 
\\

\noindent {\bf Einstein-Aether I}: In the small-$k$ limit, kinetic interactions are automatically diagonalised now (the mixing term to leading order goes as $k$) and we find the following set of interactions
\begin{align}
S^{(2)}_{\rm kin, k\rightarrow 0} &= \int d^3k dt a 
\left[
\frac{1}{2} \check{\alpha}_\textrm{K} M_A^2 \dot{\delta \tilde{A}}_1^2
+
\frac{3 H^2 M_A^2a^2 }{ \left(\check{\alpha}_\textrm{D} \dot{H} + 6 H^2\right) M_A^2}\delta \dot{ \tilde{\varphi}}^2 
 \right].
\end{align}
from which we can easily read off the stability conditions
\begin{align}
\check{\alpha}_\textrm{K} &> 0, \nn \\
 \left(\check{\alpha}_\textrm{D} \dot{H} + 6 H^2\right) &> 0.
\end{align}
\\

\noindent {\bf Einstein-Aether II}: 
As we saw in Section \ref{Sec:constraints}, kinetic interactions for this class of theories are automatically diagonalised and $k$-independent. The kinetic interactions for small $k$ are therefore the same as for large-$k$ and we still have 
\be
\check{\alpha}_\textrm{K} > 0
\ee
as the only ghost-freedom condition. Just as discussed for scalar-tensor theories above, this class of Einstein-Aether theories can therefore only have a small-$k$ ghost, if there is a large-$k$ ghost as well.
\\

\noindent {\bf Tensor-Tensor}:
In the small-$k$ limit, after diagonalising the kinetic matrix, we find the following kinetic interaction terms:
\begin{equation}
S^{(2)}_{\rm kin, k\rightarrow 0} = \int d^3k dt\; \left[\delta \dot{\tilde{\varphi}}^2\frac{3H^2a^3(\nu_M^{-2}\bar{\alpha}_N+1)}{2(\dot{H}+3H^2)+\nu
	_M^{-2}\bar{\alpha}_N(\bar{\alpha}_{SI}(\bar{\alpha}_N-1)+6H^2)}+\frac{M_T^2}{3}\dot{\tilde{E}}_2^2\frac{a^3}{(\nu_M^{-2}\bar{\alpha}_N+1)}\right],
\end{equation}
where we are using the mass scales ratio $\nu_M^2=M_2^2/M_T^2$. From here we can easily read two nontrivial ghost conditions. We note that these expressions hold for any value of $\bar{\alpha}_S$.

\section{Additional Stability Conditions}
In this section we collect some of the unwieldy stability conditions suppressed in the main text.

\subsection{Vector-Tensor Theories}\label{extraconditions}
For the vector-tensor case, from \eqref{VTmass} we find one of the gradient stability condition to be 
\be
M_{{\rm VT}, k\to \infty}^2 > 0,
\ee
where $M_{{\rm VT}, k\to \infty}^2$ is given by
\begin{align}
\frac{{\mathcal A} M_{{\rm VT}, k\to \infty}^2}{M_V^4 a^2} &= \frac{H^2 \left(\left(8 \tilde{\alpha}_\textrm{C} + 4 \tilde{\alpha}_\textrm{C}^2 - 2 \left(\tilde{\alpha}_\textrm{K} + \tilde{\alpha}_\textrm{K} \tilde{\alpha}_\textrm{T} -2\right)\right)(\rho+P) + \left(\tilde{\alpha}_\textrm{K} + \tilde{\alpha}_\textrm{K} \tilde{\alpha}_\textrm{T}-4 \tilde{\alpha}_\textrm{C} - 2 \tilde{\alpha}_\textrm{C}^2 \right) \tilde{\alpha}_\textrm{V} H^2 M_V^2\right)}{M_V^2} \nn \\
&+2 \left(4 \tilde{\alpha}_\textrm{C}^2 \tilde{\alpha}_\textrm{T} - 2 \tilde{\alpha}_\textrm{K} \left(1 + \tilde{\alpha}_\textrm{M}\right) \tilde{\alpha}_\textrm{T} + 2 \tilde{\alpha}_\textrm{C} \left(4 + 2 \tilde{\alpha}_\textrm{A} - \tilde{\alpha}_\textrm{K} \left(1 + \tilde{\alpha}_\textrm{M}\right)\right) \tilde{\alpha}_\textrm{T} - \tilde{\alpha}_\textrm{V}\right) H^4 \nn \\
&+2 \left(2 \tilde{\alpha}_\textrm{K} \tilde{\alpha}_\textrm{M}^2 + 2 \tilde{\alpha}_\textrm{C}^2 \tilde{\alpha}_\textrm{K} \tilde{\alpha}_\textrm{M} \left(2 + \tilde{\alpha}_\textrm{M}\right) + 2 \tilde{\alpha}_\textrm{C} \tilde{\alpha}_\textrm{K} \tilde{\alpha}_\textrm{M} \left(3 + 2 \tilde{\alpha}_\textrm{M}\right) + \left(2 + \tilde{\alpha}_\textrm{A}\right)^2 \tilde{\alpha}_\textrm{T}\right) H^4 \nn \\
&-4 \left(2 + 6 \tilde{\alpha}_\textrm{C} + 6 \tilde{\alpha}_\textrm{C}^2 + 2 \tilde{\alpha}_\textrm{C}^3 + 2 \tilde{\alpha}_\textrm{A} \left(1 + \tilde{\alpha}_\textrm{C}\right)^2 - \tilde{\alpha}_\textrm{K}\right) \tilde{\alpha}_\textrm{M} H^4 \nn \\
&+2 \left(\tilde{\alpha}_\textrm{A}^2 - 4 \tilde{\alpha}_\textrm{A} \tilde{\alpha}_\textrm{C} \left(1 + \tilde{\alpha}_\textrm{C}\right) - 2 \tilde{\alpha}_\textrm{C} \left(1 + \tilde{\alpha}_\textrm{C}\right) \left(2 + 2 \tilde{\alpha}_\textrm{C} - \tilde{\alpha}_\textrm{K}\right)\right) H^4 \nn \\
&+4 \left(-2 \tilde{\alpha}_\textrm{C}^2 - 2 \tilde{\alpha}_\textrm{C} \left(2 + \tilde{\alpha}_\textrm{A} - \tilde{\alpha}_\textrm{K} \left(1 + \tilde{\alpha}_\textrm{M}\right)\right) + \tilde{\alpha}_\textrm{K} \left(1 + 2 \tilde{\alpha}_\textrm{M} - \tilde{\alpha}_\textrm{T}\right)\right) \dot{\tilde{\alpha}}_\textrm{C} H^3 \nn \\
&-4 H^2 \left(\tilde{\alpha}_\textrm{K} \left(- \dot{\tilde{\alpha}}_\textrm{C}^2 + \dot{H} + \tilde{\alpha}_\textrm{T} \dot{H}\right) - 2 \left(\left(1 + \tilde{\alpha}_\textrm{C}\right)^2 \dot{H} - \left(1 + \tilde{\alpha}_\textrm{A}\right) \dot{\tilde{\alpha}}_\textrm{C} H\right)\right)
\end{align}
where, in order to simplify the expression, we have defined 
\be
{\mathcal A} \equiv \tilde{\alpha}_\textrm{K} \left(2(\rho+P) + \left(4 \dot{H} - H \left(4 \dot{\tilde{\alpha}}_\textrm{C} + \left(4 \tilde{\alpha}_\textrm{M} + 4 \tilde{\alpha}_\textrm{C} \left(1 + \tilde{\alpha}_\textrm{M}\right) - 4 \tilde{\alpha}_\textrm{T} + \tilde{\alpha}_\textrm{V}\right) H\right)\right) M_V^2\right) -2 \tilde{\alpha}_\textrm{A}^2 H^2 M_V^2.
\ee

\subsection{Tensor-Tensor Theories}\label{extraconditionsTT}
For the tensor-tensor case, we find one of the gradient stability conditions to be:
\be
M_{{\rm TT}, k\to \infty}^2 > 0.
\ee
where $M_{{\rm TT}, k\to \infty}^2$ is given by
\begin{align}
M_{{\rm TT}, k\to \infty}^2& =-\frac{M_{T}^2a^5}{4\bar{\alpha}_{b_-}^4\bar{\alpha}_{b}\bar{\alpha}_{N}^2} [(2\bar{\alpha}_{SI}\bar{\alpha}_{N_-}\bar{\alpha}_{b_-}H\bar{\alpha}_{b}^2(\bar{\alpha}_{N}^2-\bar{\alpha}_{b})\ddot{H}-2\bar{\alpha}_{SI}\bar{\alpha}_{N_-}\bar{\alpha}_{b_-}\bar{\alpha}_{b}^2(\bar{\alpha}_{N}^2-2\bar{\alpha}_{b})\dot{H}^2\nn\\
&+2\bar{\alpha}_{b}(-\bar{\alpha}_{N_-}((\bar{\alpha}_{b}+1)\bar{\alpha}_{N}^2-2\bar{\alpha}_{b}^2)\bar{\alpha}_{SI} \dot{\bar{\alpha}}_{b} +(\bar{\alpha}_{SI}^2 \nu_M^{-2}\bar{\alpha}_{N}(\bar{\alpha}_{N}+\bar{\alpha}_{b})\bar{\alpha}_{N_-}^3\nn\\
&-\bar{\alpha}_{SI}(\bar{\alpha}_{N}+\bar{\alpha}_{b})(\bar{\alpha}_{SI}\bar{\alpha}_{b_-} \nu_M^{-2}\bar{\alpha}_{N}-2\bar{\alpha}_{b})\bar{\alpha}_{N_-}^2+(1/2( \nu_M^{-2}\bar{\alpha}_{N}(-\bar{\alpha}_{b}^2-2\bar{\alpha}_{b}+2\bar{\alpha}_{N}\bar{\alpha}_{b}+\bar{\alpha}_{N}^3)\bar{\alpha}_{SI}\nn\\
&-4\bar{\alpha}_{b_-}\bar{\alpha}_{b}(\bar{\alpha}_{N}+\bar{\alpha}_{b})))\bar{\alpha}_{SI}\bar{\alpha}_{N_-}+2\bar{\alpha}_{b_-}\bar{\alpha}_{b}\bar{\alpha}_{TI}(\bar{\alpha}_{N}^2-\bar{\alpha}_{b}))\bar{\alpha}_{b_-}H)H\dot{H}\nn\\
&+\bar{\alpha}_{N_-}(- \nu_M^{-2}(2\bar{\alpha}_{N}-1-\bar{\alpha}_{b})(\bar{\alpha}_{N_-}-\bar{\alpha}_{b_-})(\bar{\alpha}_{N}+\bar{\alpha}_{b})\bar{\alpha}_{SI} \dot{\bar{\alpha}}_{b} \nn\\
&+(\bar{\alpha}_{N_-} \nu_M^{-2}\bar{\alpha}_{b}(\bar{\alpha}_{N}+\bar{\alpha}_{b})(\bar{\alpha}_{N_-}-\bar{\alpha}_{b_-})\dot{\bar{\alpha}}_{SI} +(((\bar{\alpha}_{N}^3-(1/2)\bar{\alpha}_{N}^2-(1/2)\bar{\alpha}_{b}^2) \nu_M^{-4}\bar{\alpha}_{SI}\nn\\
&+\bar{\alpha}_{b_-}\bar{\alpha}_{b}(1+\bar{\alpha}_{M}))\bar{\alpha}_{N}\bar{\alpha}_{SI}\bar{\alpha}_{N_-}-\bar{\alpha}_{b}( \nu_M^{-2}((\bar{\alpha}_{M_2}+2\bar{\alpha}_{b}-\bar{\alpha}_{M}-4)\bar{\alpha}_{N}^3\nn\\
&+(-\bar{\alpha}_{M_2}-\bar{\alpha}_{b}+\bar{\alpha}_{M}+3)\bar{\alpha}_{N}^2-\bar{\alpha}_{b}^2(\bar{\alpha}_{M_2}-\bar{\alpha}_{M}-2)\bar{\alpha}_{N}\nn\\
&+(\bar{\alpha}_{M_2}-\bar{\alpha}_{b}-\bar{\alpha}_{M}-1)\bar{\alpha}_{b}^2)\bar{\alpha}_{SI}+2\bar{\alpha}_{TI}(-\bar{\alpha}_{N}^2 \nu_M^{-2}+\bar{\alpha}_{b}^2 \nu_M^{-2}\nn\\
&+2\bar{\alpha}_{b_-}\bar{\alpha}_{N})\bar{\alpha}_{b_-}))H)\bar{\alpha}_{b_-})\bar{\alpha}_{N}\bar{\alpha}_{SI}H^3)],
\end{align}
where we have defined
\be
\bar{\alpha}_{b_-}=\bar{\alpha}_{b}-1; \quad \bar{\alpha}_{N_-}=\bar{\alpha}_{N}-1.
\ee

\section{Tensor-Tensor Quasistatic Limit}\label{app:QSTT}
In this section we define the quantity $\beta$ that appears in the expressions for the quasi static parameters in the case of general parametrised bimetric theories with $\bar{\alpha}_S=0$. 
\begin{align}
\beta&= \{4 \nu_M^{4} \bar{\alpha}_{SI}\bar{\alpha}_b^2(\bar{\alpha}_b-1)(\bar{\alpha}_N-1)[ (\bar{\alpha}_N^2-\bar{\alpha}_b)\ddot{H}H-\dot{H}^2(\bar{\alpha}_N^2-2\bar{\alpha}_b)]\nonumber\\
& + [ -4 \nu_M^{2} H \bar{\alpha}_{SI} (\bar{\alpha}_{N}-1) (-2 \bar{\alpha}_{b}^2+\bar{\alpha}_{N}^2+\bar{\alpha}_{b} \bar{\alpha}_{N}^2) \dot{ \bar{\alpha}}_{b} + (-8H^2 \nu_M^{2} \left( (\bar{\alpha}_{N}-1) \bar{\alpha}_{SI}+\bar{\alpha}_{TI} \right) \bar{\alpha}_{b}^3\nonumber\\
& + (-2 \bar{\alpha}_{N}  (\bar{\alpha}_{N}-1) H^2\bar{\alpha}_{SI}^2 + 2 \nu_M^{2} H^2 \bar{\alpha}_{TI} (\bar{\alpha}_{N}^2+1) ) \bar{\alpha}_{b}^2 - 8\bar{\alpha}_{N} (-1/2  (\bar{\alpha}_{N}-1)^2 H^2\bar{\alpha}_{SI}^2 \nonumber\\
& -\bar{\alpha}_{N} \nu_M^{2} H^2 (\bar{\alpha}_{N}-1) \bar{\alpha}_{SI}+H^2 \bar{\alpha}_{N} \nu_M^{2} \bar{\alpha}_{TI} ) \bar{\alpha}_{b} + 2 \bar{\alpha}_{N}^4 H^2 \bar{\alpha}_{SI}^2 (\bar{\alpha}_{N}-1)) (\bar{\alpha}_{b}-1) ] \bar{\alpha}_{b} \nu_M^{2} \dot{H}\nonumber\\
& + (\bar{\alpha}_{N}-1) \bar{\alpha}_{N} \bar{\alpha}_{SI} \left[ -2(2\bar{\alpha}_{N}-1- \bar{\alpha}_{b}) (\bar{\alpha}_{N}+\bar{\alpha}_{b}) \nu_M^{2} \bar{\alpha}_{SI} H^3 (\bar{\alpha}_{N}-\bar{\alpha}_{b}) \dot{ \bar{\alpha}}_{b}  \right. \nonumber\\
& + ( 2 H \nu_M^{2} \bar{\alpha}_{b} (\bar{\alpha}_{N}-1) (\bar{\alpha}_{N}-\bar{\alpha}_{b}) (\bar{\alpha}_{N}+\bar{\alpha}_{b}) H(2\dot{H}\bar{\alpha}_{SI}+H\dot{\bar{\alpha}}_{SI}) - 2 \nu_M^{2} ( (-\bar{\alpha}_{b}- \bar{\alpha}_{N} \bar{\alpha}_{M_2}+ \bar{\alpha}_{M_2} ) \bar{\alpha}_{SI}\nonumber\\
& +2\bar{\alpha}_{TI} \bar{\alpha}_{b}) H^4 \bar{\alpha}_{b}^3 + 2 ( ( -(2+ \bar{\alpha}_{M}) \bar{\alpha}_{N}+1+ \bar{\alpha}_{M}) \bar{\alpha}_{SI} + 2\bar{\alpha}_{TI} ) \nu_M^{2} H^4 \bar{\alpha}_{b}^3\nonumber\\
& + (- M_{T}^2 (\bar{\alpha}_{N}-1) H^4\bar{\alpha}_{SI}^2\bar{\alpha}_{b} - 2 ((\bar{\alpha}_{M_2}+2\bar{\alpha}_{b} ) \bar{\alpha}_{N}-\bar{\alpha}_{M_2}-\bar{\alpha}_{b}) \bar{\alpha}_{N} \nu_M^{2} H^4 \bar{\alpha}_{SI}\nonumber\\
& +4 H^4 \bar{\alpha}_{N} \nu_M^{2} \bar{\alpha}_{TI} \bar{\alpha}_{b} ) \bar{\alpha}_{N} \bar{\alpha}_{b} - \bar{\alpha}_{N}^2 \nu_M^{2} H^4 2 ( ( -(4+\bar{\alpha}_{M}) \bar{\alpha}_{N}\nonumber\\
& \left. + \bar{\alpha}_{M}+3 ) \bar{\alpha}_{SI} + 2\bar{\alpha}_{TI} ) \bar{\alpha}_{b} + (\bar{\alpha}_{N}-1) (-1+2\bar{\alpha}_{N}) \bar{\alpha}_{N}^3 M_{T}^2H^4 \bar{\alpha}_{SI}^2 ) (\bar{\alpha}_{b}-1)\right]\}\nonumber\\
& /[4 \nu_M^{4} \bar{\alpha}_N^3H^4\bar{\alpha}_b(\bar{\alpha}_b-1)^3\bar{\alpha}_{TI}].
\end{align}

\bibliographystyle{apsrev4-1}
\bibliography{RefModifiedGravity}

\end{document}